\documentclass[useAMS,usenatbib]{mnras}
\usepackage[british]{babel}             
\usepackage{newtxtext}               
\usepackage[slantedGreek]{newtxmath}    
\let\la=\lesssim 
\usepackage[T1]{fontenc}   
\usepackage{ae,aecompl}
\usepackage{graphicx}
\usepackage{graphics}
\usepackage{caption}
\usepackage{gensymb}
\usepackage{hyperref}
\usepackage{tikz}
\usepackage{pifont}% http://ctan.org/pkg/pifont

\graphicspath{./{Figs/}}
\DeclareGraphicsExtensions{.pdf}

\newcommand{\Hii}{H{\sc ii}\ }

\newcommand{\mstar}{M$_{\star}$}

\newcommand{\oiii}{[\ion{O}{iii}]$\lambda5007$}
\newcommand{\oii}{[\ion{O}{ii}]$\lambda3727,29$}
\newcommand{\sii}{[\ion{S}{ii}]$\lambda6717,31$}
\newcommand{\nii}{[\ion{N}{ii}]$\lambda6584$}

\newcommand{\oiiihb}{[\ion{O}{iii}]$\lambda5007$/H$\upbeta$}
\newcommand{\niiha}{[\ion{N}{ii}]$\lambda6584$/H$\upalpha$}
\newcommand{\siiha}{[\ion{S}{ii}]$\lambda\lambda6717,31$/H$\upalpha$}
\newcommand{\sigha}{$\sigma_{\text{H}\alpha}$}
\newcommand{\ewha}{EW[$\text{H}\alpha$]}

\newcommand{\delU}{$\upDelta$log([\ion{O}{iii}]/[\ion{O}{ii}])}
\newcommand{\delNO}{$\upDelta$log([\ion{N}{ii}]/[\ion{S}{ii}])}
\newcommand{\delNOii}{$\upDelta$log([\ion{N}{ii}]/[\ion{O}{ii}])}

\newcommand{\delsfr}{$\upDelta$log(SFR)}

\newcommand{\niibpt}{[\ion{N}{ii}]-BPT}
\newcommand{\siibpt}{[\ion{S}{ii}]-BPT}

\newcommand{\cmark}{\ding{51}}%
\newcommand{\xmark}{\ding{55}}%

\title[What drives the scatter in the BPTs?]{What drives the scatter of local star-forming galaxies in the BPT diagrams? A Machine Learning based analysis}	
\author[M. Curti et al.]{Mirko Curti,$^{1,2}$ \thanks{E-mail: mc2041@cam.ac.uk}
Connor Hayden-Pawson,$^{1,2}$
Roberto Maiolino,$^{1,2,3}$
Francesco Belfiore,$^{4}$ \newauthor
Filippo Mannucci,$^{4}$
Alice Concas,$^{5,6}$ 
Giovanni Cresci,$^{4}$
Alessandro Marconi,$^{5,4}$ \newauthor
and Michele Cirasuolo$^{6}$
\\
\\
$^{1}$Cavendish Laboratory, University of Cambridge, 19 J. J. Thomson Ave., Cambridge CB3 0HE, UK\\
$^{2}$Kavli Institute for Cosmology, University of Cambridge, Madingley Road, Cambridge CB3 0HA, UK\\
$^{3}$ Department of Physics and Astronomy, University College London, Gower Street, London WC1E 6BT, UK \\
$^{4}$INAF - Osservatorio Astrofisico di Arcetri, Largo E. Fermi 5, 50125, Firenze, Italy\\
$^{5}$Dipartimento di Fisica e Astronomia, Universitá di Firenze, Via G. Sansone 1, 50019, Sesto Fiorentino (Firenze), Italy\\
$^{6}$ European Southern Observatory, Karl-Schwarzschild-Strasse 2, D-85748 Garching bei Muenchen, Germany\\
}

\begin{document}

	\date{Accepted XXX. Received XXX}
	\pagerange{\pageref{firstpage}--\pageref{fig:lastfig}} \pubyear{2021}
	\maketitle
	\label{firstpage}

%%%%%%%%%%%%%%%%%%%%% ABSTRACT %%%%%%%%%%%%%%%%%%%
\begin{abstract}
We investigate which physical properties are most predictive of the position of local star forming galaxies on the BPT diagrams, by means of different Machine Learning (ML) algorithms. Exploiting the large statistics from the Sloan Digital Sky Survey (SDSS), we define a framework in which the deviation of star-forming galaxies from their median sequence can be described in terms of the relative variations in a variety of observational parameters. We train artificial neural networks (ANN) and random forest (RF) trees to predict whether galaxies are offset above or below the sequence (via classification), and to estimate the exact magnitude of the offset itself (via regression). We find, with high significance, that parameters primarily associated to variations in the nitrogen-over-oxygen abundance ratio (N/O) are the most predictive for the \niibpt\ diagram, whereas properties related to star formation (like variations in SFR or EW(H$\upalpha$)) perform better in the \siibpt\ diagram. 
We interpret the former as a reflection of the N/O-O/H relationship for local galaxies, while the latter as primarily tracing the variation in the effective size of the S$^{+}$ emitting region, which directly impacts the [\ion{S}{ii}] emission lines.
This analysis paves the way to assess to what extent the physics shaping local BPT diagrams is also responsible for the offsets seen in high redshift galaxies or, instead, whether a different framework or even different mechanisms need to be invoked.

\end{abstract}
%Artificial Neural Networks and Random Forest trees are implemented to solve both classification (i.e., to describe whether galaxies are offset above/below the sequence) and regression (i.e, to predict the exact magnitude of the offset from the sequence itself). We achieve a high accuracy on the test sample in both classification and regression tasks (AUC$>95$ per cent, RMSE$\sim 0.035$), with no overfitting. 
%However, we also find that the relative impact of some parameters change as we consider different regions separately within the diagrams. 
%We interpret the former as a reflection of the N/O-O/H relationship for local galaxies, while the latter as primarily tracing the variation in the effective size of the S$^{+}$ emitting region, which directly impacts the [S II] emission lines.


%%%%%%
\section{Introduction}
\label{sec:Intro}

Rest-frame optical emission lines provide a wealth of information about the physics of gas and stars in star-forming galaxies. The relative intensity of both collisionally excited and recombination lines indeed reflects the properties of the ionising radiation source, dust content, as well as density, temperature, chemical abundances and kinematics of the gas within the emitting HII regions \citep{kewley_understanding_2019}. Classical diagnostic diagrams based on optical emission lines, such as the \oiiihb\ versus \niiha\ \citep{baldwin_classification_1981} and \oiiihb\ versus \siiha\  \citep{veilleux_spectral_1987}, also known as the `BPT' diagrams, have been widely used in the literature to discriminate between different ionising sources and excitation mechanisms in galaxies, in order to separate, for instance, galaxies ionised by star formation processes from those whose spectra are dominated by the presence active galactic nuclei (AGNs).
Different classification schemes to separate star-forming galaxies from AGNs are provided in literature, some based on the predictions from photoionization models \citep{kewley_theoretical_2001, stasinska_bpt_class_2006}, while some others like \cite{kauffmann_host_2003} are more empirically-based.
Star-forming galaxies in the local universe are observed to follow a remarkably tight sequence in these diagrams, which is generally interpreted as a result of the correlation between metallicity and ionization parameter (U) \citep{mc_call_1985, dopita_evans_1986, mingozzi_sdss_2020}. 
Indeed, strong-line metallicity diagnostics widely adopted in large statistical studies are often based on calibrating the position of galaxies on such diagrams against their oxygen abundance \citep[see][for a review]{maiolino_re_2019}. 

In the last decade, the advent of integral field spectroscopic surveys of local galaxies like CALIFA \citep{sanchez_califa_2012}, MaNGA \citep{bundy_overview_2015}, and SAMI \citep{croom_sydney-aao_2012} provided the chance to review the standard classification schemes by leveraging on the information about the spatial variation of emission line ratios across galaxies (see e.g., \citealt{espinosa-ponce_califa_hii_2020}; we also refer to the review by \citealt{sanchez_araa_2020}, and references therein). 
For instance, many studies have shown that spectra from low-ionisation emission line regions
(LINERs) are not necessarily associated to a nuclear origin
\citep[e.g.,][]{cid_fernandes_comprehensive_2011,yan_blanton_liner_2012,belfiore_sdss_2016,hsieh_liners_2017}.
% \citep[e.g.,][]{cid_fernandes_comprehensive_2011,}
% (Sarzi et al.2006;CidFernandesetal.2010,2011;Yan & Blanton2012; Singh et al.2013; Belfiore et al.2016;Hsiehet al.2017)
Various authors have also explored and modelled different multi-dimensional re-projections of the standard line ratio diagnostics to attempt breaking some of the degeneracies in the determination of seyfert-like, shock-like and star-forming spaxels in integral field data \citep[e.g.,][]{dagostino_bpt_2019,ji_yan_bpt_2020,law_bpt_2021}.
% Beside strong-line ratios in the optical, 
% several diagnostics in the rest-frame UV have been shown to be either valuable alternatives or provide complementary information about the excitation sources powering galaxy spectra.
% Several diagnostics in the rest-frame UV (e.g., [\ion{C}{iv}]$\lambda1549$, [\ion{C}{iii}]$\lambda1909$, [\ion{He}{ii}]$\lambda1640$, \ion{N}{v}, \ion{Ne}{v}) has been explored in the literature to separate excitation sources \citep[e.g.,][]{groves_2004, feltre_uv_2016, gutkin_modelling_2016} and could be used to discriminate, for instance, the contribution of photoionisation from that of shock heated gas even in galaxies classified as dominated by star-formation, as fast shock models tracks (like those from \citealt{allen_mappings_2008}) have been shown to overlap  with the upper right boundary of the star-forming sequence on the [\ion{N}{ii}]-BPT diagram.
% Indeed, many UV diagnostic diagrams have been proposed to distinguish between ionisation driven by AGNs, star formation or shocks \citep{groves_2004, feltre_uv_2016, gutkin_modelling_2016} exploiting line ratios of high excitation lines (e.g.,\ion{He}{ii}, \ion{C}{iii}], [\ion{C}{iv}], \ion{N}{v}, \ion{Ne}{v}).
%(Allen et al. 1998, Best et al. 2000, Moy & Rocca-Volmerange 2002, Groves et al. 2004b, Feltre et al. 2016, Jaskot & Ravindranath 2016)
Aside from discriminating between different ionising sources, it is interesting to note that the distribution of star-forming galaxies in the BPT diagrams presents a non-negligible amount of scatter, which is shown to correlate with different physical properties \citep[e.g.,][]{brinchmann_bpt_highz_2008, sanchez_califa_bpt_2015, masters_tight_2016, faisst_bpt_2018}. Therefore, such diagrams are a valuable source of information, as the relative position of sources within the plane can be used to constrain the physical conditions of the gas and of the ionising stellar populations in photoionisation modelling of \Hii regions \citep[e.g.,][]{morisset_bpt_models_2016,baugh_hii_models_2022}. 

Moreover, several lines of evidence indicate that star-forming galaxies at high redshift (i.e., $1<$z$<3$) occupy a slightly different position on the classical BPT diagrams compared to their local counterparts showing, on average, an offset towards higher \oiiihb\ and/or \niiha\ . In general, such deviation is attributed to a combined effect of the evolution in the underlying stellar populations associated to, e.g., a hardening of the far ultraviolet (FUV) ionising spectrum, alpha-enhancement, contribution from binarity and rotation \citep[e.g.,][]{steidel_strong_2014, strom_nebular_2017, topping_mosdef-lris_2020_i, topping_mosdef-lris_2020_ii} and/or in the physical properties of the ISM like density, ionisation parameter and gas chemical abundances \citep[e.g.,][]{brinchmann_bpt_highz_2008, shapley_mosdef_2015, yabe_subaru_2015, masters_tight_2016, kashino_fmos-cosmos_2017}.
Several attempts have been made to theoretically model the emission line ratios in the BPT diagrams and reproduce their variation with cosmic time, by coupling the evolution of galaxy properties from cosmological simulations with state-of-the-art stellar and nebular emission models \citep[e.g.,][]{kewley_cosmic_2013, byler_nebular_2017, hirschmann_synthetic_2017, kaasinen_ionization_2018, xiao_hii_modelling_2018}.
However, although variations in emission line ratios can be theoretically reproduced by means of the interplay of many different parameters, it is often difficult to break the degeneracy and disentangle their true relative contribution, which requires both a large variety of independent observational constraints as well as a careful assessment of the underlying model assumptions.
For instance, photionisation models often assume fixed, underlying trends between different abundance patterns with zero-scatter (e.g., to describe how N/O and C/O varies with oxygen abundance, O/H), and hence struggle to grasp the direct impact of variations of such abundances at fixed metallicity on the modelling of the emission lines.

With this scenario in mind, in this work we present a complementary, self-consistent and fully data-based framework which exploits machine learning algorithms to quantitatively describe how the distribution of local star-forming galaxies across the BPT diagrams is connected to different observational properties, and what we can infer about the relationships between the observed variations in line ratios and the underlying physics of star-forming galaxies.
In particular, we leverage on the large statistics provided by the Sloan Digital Sky Survey \citep[SDSS,][]{york_sloan_2000} to perform a detailed analysis of the dependencies between the deviation of galaxies from the mean star formation (SF) locus and a variety of key observational parameters, directly or indirectly tracing different physical properties of galaxies.

In recent years indeed, machine learning techniques have seen an increasingly significant impact on astronomical studies,
in response to the undergoing rapid growth in size and complexity of datasets as provided by current surveys like SDSS \citep{york_sloan_2000}, MANGA \citep{bundy_overview_2015} or GAIA \citep{gaia_coll_2016}, and in preparation for future large observational campaigns like those provided by DESI \citep{desi_2013}, SKA \citep{dewdney_ska_2009} and LSST \citep{ivezic_lsst_2008}. 
Such algorithms are successfully implemented to solve a variety of different problems, including the classification of galaxy morphological types \citep[e.g.,][]{dela_calleja_ML_class_2004,barchi_ML_class_2020,vavilova_ML_class_2021,reza_ML_class_2021}, the identification of transients \citep[][]{Sooknunan_ML_transients_2021}, or the multi-parametric analysis of very large databases of galaxy properties \citep[e.g.,][]{teimoorinia_artificial_2016,teimoorinia_re-assessment_2021,ho_machine_2019,bluck_what_2019, bluck_are_2019,bluck_how_2020,bluck_ml_causality_2022}. %thanks to their powerful computational algorithms combined with a relatively straightforward implementation.
Inspired especially by the latter works, in this paper we train and test artificial neural networks (ANN) and random forest decision trees (RF) to assess the performance of a set of carefully selected parameters (both individually and as a whole) and identify which properties are the most relevant in predicting the observed deviation of star-forming galaxies from their average sequence in both the [\ion{N}{ii}]- and [\ion{S}{ii}]-BPT diagrams.
In a forthcoming paper of this series, we will expand on the present work by exploiting the information provided by MaNGA in order to compare trends on global/integrated and local/spatially resolved scales.
This approach, if successful in describing what observed in the local Universe, could then be tested on high redshift galaxy samples to assess to what extent the physics that govern the scatter in local BPT diagrams is the same causing the observed evolution in the emission line properties at high-z or, instead, whether a different framework or even different physical mechanisms need to be involved.

The current paper is structured as follows.
In Section~\ref{sec:data} we describe the observational dataset and the sample selection, and we introduce and discuss the set of parameters adopted in the analysis. In Section~\ref{sec:frame} we describe framework and metrics adopted for describing the scatter within the [\ion{N}{ii}]-BPT diagram, whereas in Section~\ref{sec:ML} the proper machine learning analysis is performed.
In Section~\ref{sec:S2_ML}, we repeat the same analysis for the [\ion{S}{ii}]-BPT diagram. We summarise the results and present our conclusions in Section~\ref{sec:summary}.
% Throughout this paper, we assume a \cite{planck_collaboration_planck_2016} $\Lambda$CDM cosmology.
Throughout this paper, we assume a standard $\Lambda$CDM cosmology with with H$_{0}$ = 70 km s$^{-1}$, $\upOmega_{m}$=0.3, and $\upOmega_{\Lambda}$ = 0.7.

\section{Data}
\label{sec:data}

% \subsection{SDSS data}
% \label{sec:sdss}
\subsection{Sample Selection}

The galaxy sample adopted in this work is drawn from the seventh data release (DR7) of the Sloan Digital Sky Survey (SDSS) \citep{abazajian_seventh_2009}, whose galaxy properties and emission line fluxes are provided by the MPA/JHU catalog\footnote{available at \href{http://www.mpa-garching.mpg.de/SDSS/DR7/}{http://www.mpa-garching.mpg.de/SDSS/DR7/}}.
We selected galaxies classified as star-forming according to their position on the [\ion{N}{ii}]-BPT diagram, following the more conservative classification scheme by \citealt{kauffmann_dependence_2003}, and further requiring the equivalent width of the H$\upalpha$ to be higher than $6\AA$, in order to set a more stringent limit to contributions from low ionisation gas powered by different types of stellar populations \citep[e.g.,][]{cid_fernandes_comprehensive_2011,zhang_sdss-iv_2017,lacerda_dig_2018}.
We applied a redshift cut on $\text{z}>0.035$ in order to ensure the presence of the [\ion{O}{ii}]$\lambda3727$ emission line  within the wavelength coverage of the SDSS spectrograph and sample at least the inner $2$ kpc of each galaxy.
% However, such a low redshift cut would imply including galaxies with very small sampling inside the SDSS fiber, as $3\arcsec$ are equivalent to a projected physical distance of only $\sim 1.6$ kpc at z$=0.027$, thus sampling only the most inner regions.
% On one side, this would introduce more uncertainties when applying the aperture corrections for the SFR (see below) and, on the other side, it would make the metallicity measured within the fibre less representative of the global galaxy metallicity, being more sensitive to the presence of metallicity gradients.
% To mitigate this problem, we decided to include in our analysis only those galaxies with a covering factor of at least $10\%$, as inferred from the fraction of the total light that goes into the fibre. 
 %This criterion removes $12,252$ galaxies.
In addition, we discarded all galaxies whose catalogue flags indicates unreliable stellar mass and star-formation rate (SFR) estimates.
Moreover, we applied a signal-to-noise (S/N) threshold\footnote{applying the re-scaled uncertainties provided by the MPA/JHU group, which include both the uncertainties on the spectrophotometry and continuum subtraction} of $5$ on H$\upalpha$, and of $3$ on all the other main emission lines involved in the analysis, namely H$\upbeta$, [\ion{O}{ii}]$\lambda3726,3729$, [\ion{O}{iii}]$\lambda5007$, [\ion{N}{ii}]$\lambda6584$ and [\ion{S}{ii}]$\lambda6718,6732$.
All emission lines were corrected for reddening, where required, from the measured Balmer Decrement (assuming an intrisinc value of H$\upalpha$/H$\upbeta$=2.87, as given by the case B recombination) and adopting the \cite{cardelli_relationship_1989} extinction law.
% We then also discarded from the analysed sample all galaxies characterized by extreme extinction, i.e. with values of E(B-V) higher than $0.8$.
%as inferred from the Balmer decrement assuming the case B recombination (H$\upalpha$/H$\upbeta$=2.87), and adopting the \cite{Cardelli:1989aa} MW extinction law.
Finally, we removed galaxies affected by poor photometric deblending by selecting on the $\rm{DEBLEND\_NOPEAK}$ and $\rm{DEBLEND\_AT\_EDGE}$ flags, as well as 
galaxies whose aperture correction factors are lower than $1$ (e.g., where the stellar mass derived from the total photometry is lower that the stellar mass derived within the fibre).
% In addition, we have also visually inspected all the objects with log(\mstar) < 8.6 and manually removed the residual poorly deblended systems, which account for another $3\%$.
After applying all these criteria, the total analysed sample is reduced to 128,120 galaxies. 

\subsection{Observational parameters and physical properties}
\label{sec:parameters}

In this work we aim at quantitatively assessing which physical properties are most connected with the position of galaxies in the BPT diagrams.
Therefore, we consider direct measurements of physical quantities, as well as a variety of different observational proxies, as the main parameters in our analysis.
% For physical parameters derived from emission lines, with We prefer to stick with direct observational tracers to reduce the systematics introduced by the choice of different calibrations.
The full list of involved parameters is described in the following, and also reported in Table~\ref{tab:nii_parameters} and~\ref{tab:sii_parameters}.

The galaxy stellar masses are provided by the MPA/JHU catalog and have been estimated from fits to the photometry, following the prescription of \cite{kauffmann_stellar_2003} and \cite{salim_uv_2007}. 
Star formation rates adopted in this work are derived from the extinction corrected H$\upalpha$ luminosity inside the fibre, adopting the calibration proposed by \cite{kennicutt_star_2012}. 
We decide to adopt such fibre-based SFR in our fiducial analysis for consistency with the other emission line properties computed within the SDSS fibre.
However, we also apply the aperture corrections provided by the MPA/JHU catalog, which build on the work of \cite{salim_uv_2007} to improve those originally provided by \cite{brinchmann_physical_2004}, to compute the total SFR for our galaxies. We stress that adopting the total SFR instead of the fibre SFR does not change any of the main conclusions of the paper; nonetheless, part of the analysis including the total SFR is presented in Appendix \ref{sec:appendix_A}.
Both stellar masses and SFRs estimates are re-scaled to a common \cite{chabrier_galactic_2003} IMF.

Spectral indices like D$_{\text{N}}$(4000) and EW(H$\alpha$) are provided for SDSS galaxies by the MPA/JHU catalog too. %and are related to the age of the underlying stellar populations.
In particular, EW(H$\alpha$) is a model-independent tracer of the specific star formation rate (sSFR$=$M$_{\star}$/SFR), quantifying the relative contribution of recent star formation on the integrated star formation history (SFH) of galaxies, whereas D$_{\text{N}}$(4000) is a sensitive probe of the overall ageing of the stellar population.
We also consider the central velocity dispersion of the Balmer lines (e.g., \sigha) as a tracer of the gas kinematics and potentially revelatory of non-virial motions (as shocks are known to produce kinematic components with velocity dispersion significantly larger than those of HII regions, see e.g., \citealt[][]{rich_shocks_2010, dagostino_bpt_2019, law_bpt_2021}).
%and collisionally excited emission lines.

In terms of properties derived from emission line ratios, we measure the gas-phase metallicity exploiting the calibrations presented in \cite{curti_new_2017, curti_massmetallicity_2020} (which are built on electron temperature-based abundance measurements).
We refer to \cite{curti_massmetallicity_2020} for a detailed description of the procedure, where metallicity is constrained by simultaneously adopting several emission line ratios in order to minimise the degeneracies and biases intrinsic to each individual calibration. 

The [\ion{N}{ii}]/[\ion{O}{ii}] and [\ion{N}{ii}]/[\ion{S}{ii}] ratios are taken as observational proxies of the nitrogen-over-oxygen (N/O) abundance and, more generally, of nitrogen-over-$\upalpha$ elemental abundances, which are key diagnostics of the chemical enrichment timescales and of the evolutionary stage of galaxies \citep{edmunds_nitrogen_1978, vila_costas_nitrogen-ratio_1993, thuan_heavy_1995, perez-montero_deriving_2014, berg_direct_2012,berg_chemical_2019,belfiore_sdss_2017,hayden_pawson_NO_klever_2021_arxiv}.
More specifically, [\ion{N}{ii}]/[\ion{O}{ii}] primarily traces the N$^{+}$/O$^{+}$ ionic abundance ratio, which closely matches the total N/O abundance ratio because of the similar ionization structures of the two elements (i.e., the ionisation correction factors are small, \citealt{garnett_nitrogen_1990, amayo_ICFs_2020}). 
The [\ion{N}{ii}]/[\ion{S}{ii}] ratio is another commonly adopted diagnostics of N/O, exploiting the similar ionisation potential of S$^{+}$ and O$^{+}$ ions and the proximity of the nitrogen and sulfur emission lines in the optical waveband \citep{perez-montero_impact_2009}.
However, this ratio is sensitive to variations in the sulphur-over-oxygen abundance (S/O), which is however observed as fairly constant with metallicity \citep{amayo_ICFs_2020}, as well as to differential depletion of O and S onto dust grains, which is hard to constrain though \citep{jenkins_depletion_2009,laas_sulphur_depletion_2019}. Moreover, recent evidence that a non-negligible fraction of sulphur is produced by Type Ia Supernovae \citep{kobayashi_origin_2020, palla_SNIa_2021} questions the effectiveness of using such element as a probe of alpha-enhancement. 
Finally, it has been observed that [\ion{N}{ii}]/[\ion{S}{ii}] presents some degree of correlation with other parameters (like specific star-formation rate and gas velocity dispersion) in SDSS galaxies at fixed [\ion{N}{ii}]/[\ion{O}{ii}] \citep{hayden_pawson_NO_klever_2021_arxiv}.
Therefore, interpreting [\ion{N}{ii}]/[\ion{S}{ii}] as a primary tracer of, specifically, N/O should necessarily keep all these assumptions and systematics in mind.
% Both line ratios can be converted to the N/O abundance following a variety of different calibrations (e.g., \citealt{hayden_pawson_NO_klever_2021_arxiv}, based on the SDSS stacked spectra described in \citealt{curti_new_2017}, or \citealt{perez-montero_impact_2009}).

The ionisation parameter U is instead mapped on the [\ion{O}{iii}]$\lambda5007$/[\ion{O}{ii}]$\lambda3727,29$ and [\ion{Ne}{iii}]$\lambda3869$/[\ion{O}{ii}]$\lambda3727,29$ line ratios, which can be converted to U following the calibration relations presented in \cite{kewley_understanding_2019}.
The former index \citep{aller_ion_par_1942,diaz_hii_regions_2000, dors_teff_2003} is physically motivated (as involves emission lines from the same atomic species in different ionization states), and is easily observed across the full sample of selected star-forming galaxies, but it is largely affected by extinction, and presents a secondary dependence on metallicity \citep{kewley_using_2002, kobulnicky_metallicities_2004}, as well as on the softness of the ionising radiation \citep{morisset_bpt_models_2016}; the latter \citep{levesque_ion_par_2014} instead is mildly affected by dust extinction, but requires the detection of the rather faint [\ion{Ne}{iii}]$\lambda3869$ emission line in individual sources and depends also on the neon-over-oxygen abundance pattern.
Forthcoming analysis based on MaNGA data (Curti et al., in preparation) will exploit the more straight and independent \ion{S}{iii}$\lambda9068$/\ion{S}{ii}$\lambda6718,32$ ratio as a tracer \citep{diaz_hii_1991, kewley_understanding_2019} of the ionisation parameter in galaxies.

Finally, the electron density of the gas (N$_{\text{e}}$) is traced by the observed ratio between the lines of the sulfur doublet, i.e. [\ion{S}{ii}]$\lambda6718$/[\ion{S}{ii}]$\lambda6732$, which is a widely adopted diagnostic in star-forming \Hii regions as it is highly sensitive to N$_{\text{e}}$ in the regime between the critical densities of the two lines \citep{osterbrock_astrophysics_2006}.
% From atomic physics, ....

%separating shocks from star-forming regions can also be accomplished using high spectral resolution integral field spectroscopy. This method takes advantage of the impact of shocks on the gas kinematics. Gas excited by thermal shocks produces emission lines with velocity dispersions (i.e., emission-line widths in velocity units) around the mean shock velocity. These shocked com- ponents often present as separate kinematic components in integral field data, and show as broad lines, usually underlying the narrow lines typical of star-forming regions.  Both galactic wind shocks and merger-induced shocks are known to produce separate kinematic components to the emission lines that have velocity dispersions of 150–500 km s−1 , which is significantly larger than the veloc- ity width of Hii regions or gas ionized by evolved stellar populations (both typically <40 km s−1).
%The correlation between velocity dispersion and line ratios such as [Nii]/Hα or [Sii]/Hα (see Rich et al. 2010) is critical for unambiguous identification of shocks because neither beam smearing nor aged stellar populations can produce correlations between velocity dispersion and emission- line ratios. High spectral resolution of 30–50 km s−1 at Hα is typically required to identify shocks with this method. 

% \subsection{MaNGA data}
% \label{sec:manga}

\section{Framework}
\label{sec:frame}

\subsection{Variation of physical properties across the [N II]-BPT star-forming sequence}
\label{sec:nii_params_sequence}

\begin{figure*}
    \centering
\includegraphics[width=0.95\textwidth]{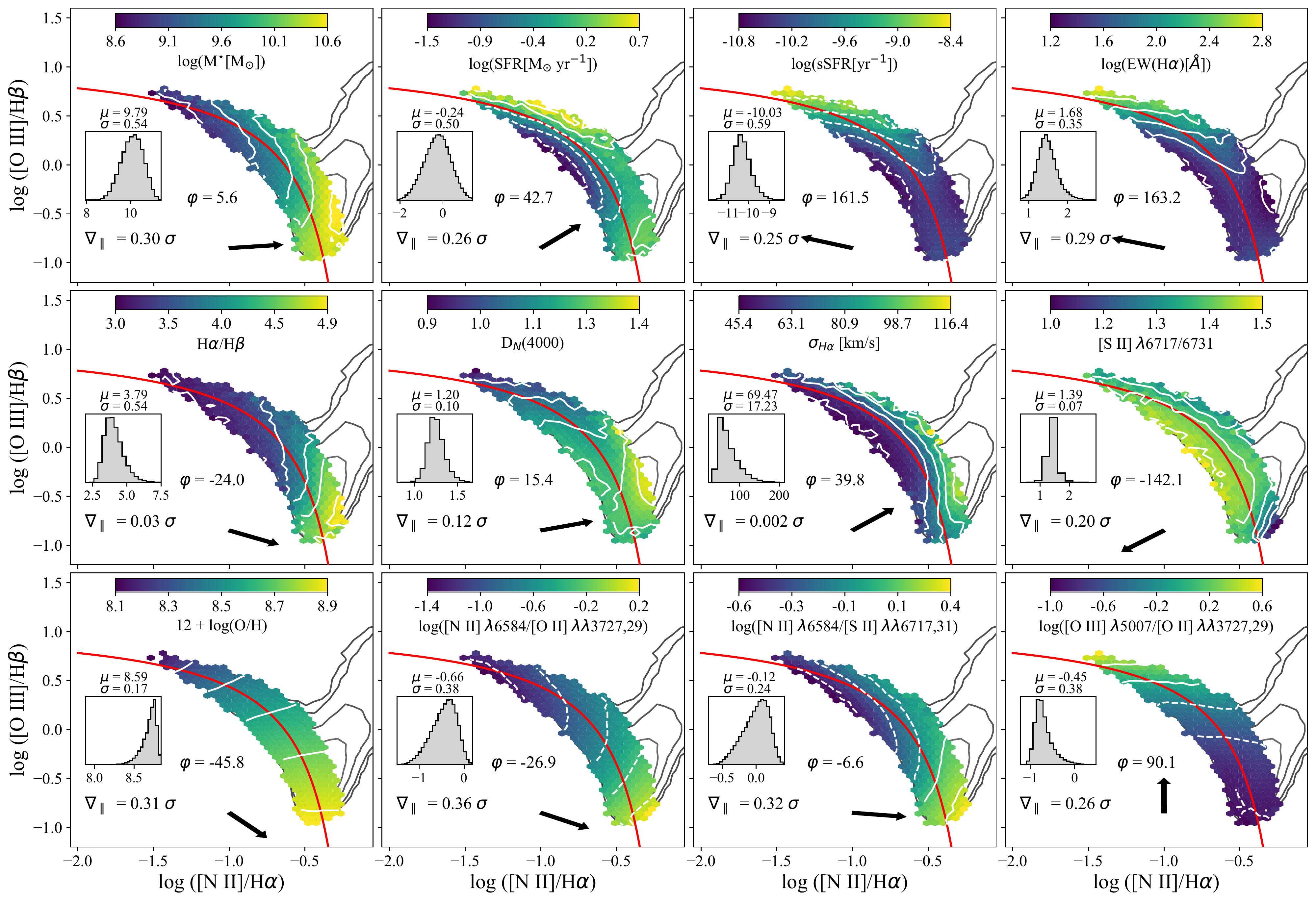}
    \caption{Distribution of the selected star-forming galaxies from the SDSS in the [\ion{N}{ii}]-BPT diagram. In each panel, the colour-coding reflects the average value computed in small hexagonal bins for the following parameters (moving along rows from the upper left to the bottom right): log(M$_{\star}$), log(SFR), log(sSFR), log(EW[H$\upalpha$]), H$\upalpha$/H$\upbeta$, Dn4000, $\sigma_{\text{H}\alpha}$, [\ion{S}{ii}]$\lambda6717/\lambda6731$,12+log(O/H),
    log([\ion{N}{ii}]$\lambda6584$/[\ion{O}{ii}]$\lambda\lambda3727,29$), log([\ion{N}{ii}]$\lambda6584$/[\ion{S}{ii}]$\lambda\lambda6717,31$), log([\ion{O}{iii}]$\lambda5007$/[\ion{O}{ii}]$\lambda\lambda3727,29$). White contours indicate lines of constant values in each parameter (with dashed lines marking negative values). The best-fit to the median SF locus from Eq.~\ref{eq:nii_sequence} (\citealt[][]{kewley_theoretical_2013}) is indicated by the red curve. In each panel, we also plot an histogram of the distribution of values for the reference parameter, reporting its average and standard deviation, and the $\nabla_{\parallel}$ statistics, which quantifies the magnitude of the gradient in that parameter along the best-fit curve, in units of standard deviation.
    Finally, in each panel we show the vector representative of the direction of the steepest average gradient across the diagram in the reference parameter, whose angle is computed (positive counter-clock wise from the x-axis) from the ratio of the partial correlation coefficients between that given parameter and each individual BPT-axis, i.e. with \oiiihb\ at fixed \niiha, and viceversa.}
    \label{fig:Nii_properties}
\end{figure*}

%description of the SDSS BPT diagram, parametrisation of the SF local sequence best-fit
We will initially focus our investigation on the [\ion{N}{ii}]-BPT diagram, where star forming galaxies in the local Universe are observed to form quite a tight sequence. 
In this work, we adopt a polynomial representation of the locus of highest density of star-forming galaxies along the sequence (hereinafter, SF sequence, or SF locus), as originally provided by \cite{kewley_cosmic_2013} and given by
\begin{equation}
\text{log}([\ion{O}{iii}]/\text{H}\beta) = 0.61/(\text{log}([\ion{N}{ii}]/\text{H}\alpha)+0.08) + 1.1 \ .
\label{eq:nii_sequence}
\end{equation}

In Figure \ref{fig:Nii_properties}, the selected star-forming SDSS galaxies are plotted in the [\ion{N}{ii}]-BPT diagram and colour-coded in each panel by the different galaxy properties and parameters described in Section~\ref{sec:parameters}. 
To aid the visualisation of the underlying trends, galaxies are binned in small hexagons, and the average value of each parameter in such bins is considered. In addition, lines of constant value (i.e., iso-contours) in each parameter are marked in white (with dashed lines associated to negative values), while the polynomial fit to the SF sequence of equation \ref{eq:nii_sequence} is shown by the red curve.
Finally, the histogram of values in each given parameter is shown, together with the mean value and standard deviation of the distribution, within its reference panel.

By visually inspecting Fig.~\ref{fig:Nii_properties}, it is readily evident that the relative position of galaxies on the [\ion{N}{ii}]-BPT diagram is strongly correlated with different physical properties. 
In particular, moving along the sequence of star-forming galaxies (e.g., from the bottom-right to the upper-left) we can recognise clear trends in stellar mass, specific star formation rate (or, equivalently, EW(H$\upalpha$)), gas-phase metallicity, \nii/\sii\ and \nii/\oii\ (both tracing primarily N/O), \oiii/\oii\ (tracing primarily U) and H$\upalpha$/H$\upbeta$ (tracing dust extinction).
% More subtle is the dependence shown, for instance, on SFR, as variations in such quantity appear correlated with crossing the sequence rather than moving along it.
% Far less prominent instead appears the dependence of BPT-location on the electron density. 

However, a more careful assessment provides deeper insights about the nature of such dependencies.
In particular, the distribution of star-forming galaxies in the diagram form a very smooth sequence in oxygen abundance, with variations in log(O/H) closely following the shape of the SF locus, and lines of constant metallicity which are, instead, orthogonal to the sequence almost everywhere along the best-fit line. Not surprisingly, the [\ion{N}{ii}]-BPT has been modelled and calibrated for a long time against oxygen abundance to serve as a metallicity diagnostic, and likewise have been the individual line ratios upon which the diagram is built \citep[e.g.,][]{pettini_oiiinii_2004, maiolino_amaze_2008}.
Different properties (and their tracers), which are nonetheless physically connected to metallicity, like M$_{\star}$, N/O and U, although characterised by the presence of a strong gradient along the sequence do show iso-contours at various levels of inclination with respect to the best-fit line. 
For instance, lines of constant \oiii/\oii\ (i.e., closely tracing lines of constant U) are almost everywhere parallel to the x-axis (hence spanning a variety of different inclinations from the best-fit line of the SF of equation~\ref{eq:nii_sequence}), suggesting a correlation between the ionisation parameter and \oiii/H$\upbeta$ \citep{kewley_understanding_2019}.
For parameters like SFR and \sigha\ instead, it is more difficult to identify a clear trend along the SF sequence, whereas clear segregation in such parameters can be seen between galaxies lying leftmost and rightmost the best-fit curve.

We can try to quantitatively estimate the amplitude of the variation in each parameter along the SF sequence as described below.
First, we perform a bi-variate spline interpolation over the underlying (binned) distribution of star-forming galaxies, so to infer the values assumed by each parameter at any discrete (sampling) point along the best-fit curve of the SF locus.
From the array of values assumed by each parameter on the SF sequence best-fit curve, we can then compute a `gradient array' (from second order accurate central differences in the interior points of the original array), whose amplitude (i.e., the square root of the sum of its elements, taken in quadrature) is reported in each panel as $\nabla_{\parallel}$: such statistics are useful to quantify how strongly each parameter varies as we move along the the best-fit line of the SF locus or, in other words, to what extent the sequence of star-forming galaxies in the \niibpt diagram can be interpreted as a sequence in a given physical property.
This quantity is further normalised by the 1$\sigma$ dispersion of values in each parameter, in order to account for the different dynamic ranges and allow a meaningful comparison between different quantities.

The highest $\nabla_{\parallel}$ values ($>0.30 \sigma$) are found for M$_{\star}$, O/H, [\ion{N}{ii}]/[\ion{O}{ii}] and [\ion{N}{ii}]/[\ion{S}{ii}], confirming that the sequence of star-forming galaxies in the [\ion{N}{ii}]-BPT is primarily a sequence in stellar mass, metallicity and nitrogen-over-oxygen abundance (i.e., the variation in such parameters along the SF sequence is relatively large compared to the overall distribution of values within the entire diagram).
Relatively high scores in $\nabla_{\parallel}$ are marked also by [\ion{O}{iii}]/[\ion{O}{ii}], (s)SFR and \ewha.
However, we note that although for some properties like O/H a large $\nabla_{\parallel}$ is actually associated with a smooth and monotonic variation along the sequence, for others (like e.g., SFR) it is the result of the best-fit line crossing more irregular patterns within the diagram, hence varying `rapidly' but not necessarily monotonically along the curve.

Another potentially interesting aspect to consider is how much each parameter is correlated individually with the line ratios of the \niibpt, i.e., with the two axis of the diagram, if taken separately. 
To estimate this we compute, for any given parameter, the Spearman partial correlation coefficients with both \niiha\ and \oiiihb; a partial correlation coefficient quantifies the strength of correlation between two variables while keeping fixed the third (and/or further variables in case of higher dimensionality problems), and is defined as  
\begin{equation}
    \rho_{AB,C} = \frac{ \rho_{AB} - \rho_{AC}\cdot \rho_{BC} }{\sqrt{1-\rho_{AC}^{2}} \sqrt{1-\rho_{BC}^{2}}} \,
\end{equation}
where $\rho_{AB}$ indicates, in general, the Spearman correlation coefficient between the two variables A and B. 
We then follow, e.g., \cite{bluck_are_2019,baker_almaquest_2022}, in defining the vector representing the direction of the steepest average gradient in a given parameter (i.e., which is the average direction one should follow across the diagram in order to maximise the variation in that given parameter), hence constraining the relative role of the quantities on the x- and y-axis in driving the coloured coded quantity.
The inclination of such vector with respect to the horizontal axis can be derived from the arctangent of the ratio of its two components, i.e., from the ratio of the partial correlation coefficients of its specific reference parameter with the individual BPT axis:
\begin{equation}
    \varphi = \tan^{-1} \bigg( \frac{\rho_{Yp,X}}{\rho_{Xp,Y}} \bigg) \,
\end{equation}
where \textit{p} is any of the parameters in our set and Y, X are the log(\oiiihb), log(\niiha) line ratios, respectively.
Such `correlation vector' and its associated $\varphi$ angle are shown for all parameters in the corresponding panel of Fig.~\ref{fig:Nii_properties}.
This analysis is performed on the hexagon binned data, in order to avoid biases introduced by the non homogeneous density distribution of individual galaxies across the diagram, as well as to remove strong outliers.

The introduction of the `correlation vector' further confirms that the direction of preferred variation in metallicity is closely aligned with the shape of the SF sequence ($\varphi=-46\degree$, pointing from the upper-left to the bottom-right in the diagram), being positively correlated with the x-axis while negatively with the y-axis. 
We further note that for N/O tracers, the gradient vector is more inclined towards the x-axis for [\ion{N}{ii}]/[\ion{O}{ii}] than it is for metallicity, whereas it has almost a flat inclination ($\varphi=-6.6\degree$) for [\ion{N}{ii}]/[\ion{S}{ii}], as possibly driven by the secondary, additional dependencies of such line ratio (for instance with the (s)SFR, see discussion in Section~\ref{sec:parameters}, and also \citealt{hayden_pawson_NO_klever_2021_arxiv}).
For ionisation parameter tracers like log(\oiii/\oii) instead, the vector is almost perfectly vertical ($\varphi=90.1\degree$), showing that such parameter in star-forming galaxies is almost entirely (positively) correlated with the y-axis and basically uncorrelated with the x-axis (when the other axis is fixed) in the [\ion{N}{ii}]-BPT. 
For the other parameters, the direction of the gradient vector presents different levels of inclination with respect to the SF locus: D$_{\text{N}}$(4000) is well correlated with \niiha\ (low $\varphi$ values), whereas \sigha\ and SFR present gradient vectors whose inclination (close to $\varphi \sim45\degree$) suggests a comparable level of correlation with both [\ion{N}{ii}]-BPT axis. 

\subsection{Metrics: $\upDelta$log(p), distance \textbf{D} and angle $\theta$}
\label{sec:metrics}

Following the observations and the analysis presented in the previous Section, we now take a step further and try to build a relatively straightforward modelling of the observed scatter in the BPT diagrams, which is based on the two main assumptions described below:
\newline \textbf{i)} galaxies which are shifted from the best-fit curve (i.e., which do not follow the bulk of galaxy distribution along the SF sequence) experience an offset which we describe to occur \textit{orthogonal to the curve} at any point (hence, we account for the minimum possible offset);
\newline \textbf{ii)} such an offset correlates with \textit{relative variations} in, either one or more, physical parameters, when compared to the average values pertaining to galaxies which closely follow the SF sequence.

The aim of the subsequent analysis is therefore to connect the offset from the best-fit line of the SF sequence with the observed variation in different physical parameters, quantify the amount of underlying correlation and assess which parameters are the most useful in predicting the observed deviation of galaxies from the median loci across the diagram.
For each galaxy in the sample, we thus introduce the $\upDelta \text{log}(p)$ metric, defined as the difference between the (logarithm of the) value assumed by a given galaxy in the parameter \textit{p} and the average value assumed by galaxies which lie on the closest point along the best-fit curve of the SF sequence (i.e., assuming a purely orthogonal offset):
\begin{equation}
   \upDelta \text{log}(p) =  \text{log}(p)\ - <\text{log}(p)>_{\text{SF locus}} .
   \label{eq:delta_prop}
\end{equation}
We note here that considering the logarithm of each quantity makes the comparison between different parameters more meaningful and straightforward.

\begin{figure*}
    \centering
\includegraphics[width=0.95\textwidth]{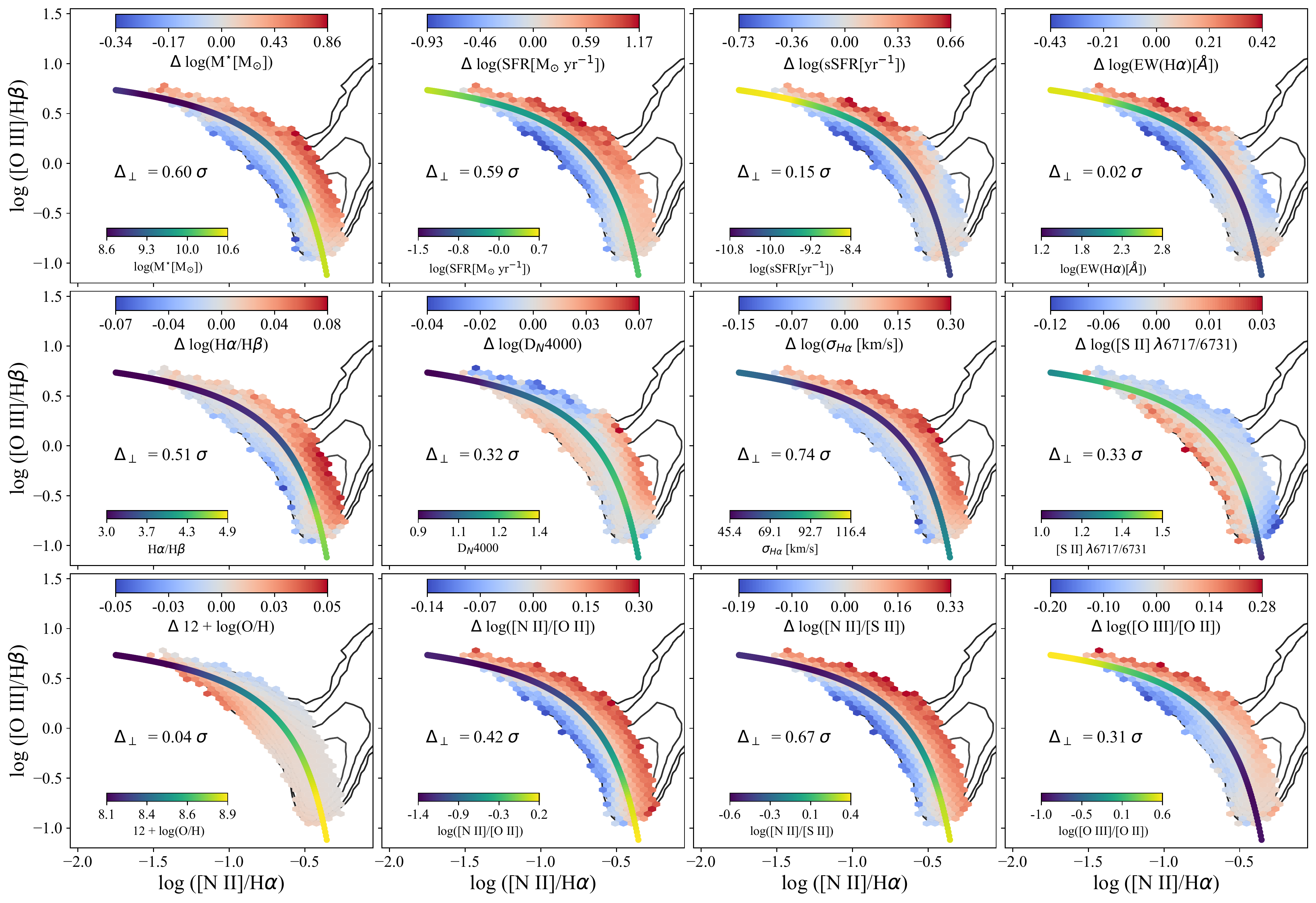}
    \caption{In this figure, the panels are organised as in Fig.~\ref{fig:Nii_properties}, but now the hexagonal bins are colour-coded by the average $\upDelta$log(p), as defined for equation~\ref{eq:delta_prop} for each parameter and assuming a purely orthogonal `offset vector' from the best-fit curve of the SF locus. Within each panel, such curve is colour-coded by the average value assumed in that parameter by galaxies lying exactly on the SF sequence, at any given point. We also report the $\upDelta_{\perp}$ statistics, which quantifies the difference in $\upDelta$log(p) between galaxies lying above and below the SF locus curve, reported in units of standard deviation.}
    \label{fig:nii_delta_prop}
\end{figure*}

In Fig.~\ref{fig:nii_delta_prop} we replicate the scheme of Fig.~\ref{fig:Nii_properties}, but in this case the small hexagons in each panel are colour-coded by the average variation in the logarithm of the relative parameter (i.e., the average $\upDelta \text{log}(p)$ in the bin), as defined in equation~\ref{eq:delta_prop}; the best-fit line of the SF sequence is instead coloured according to the average value assumed by each parameter at any given point along the curve, as inferred from interpolating over the underlying galaxy distribution (we refer to the previous subsection for more details). 
The colour scheme (centred on zero) helps to identify trends between the relative location of galaxies and the magnitude of variations in the various parameters: for instance, whether a galaxy occupies the region above or below the best-fit curve is visually seen to correlate overall very well with different properties, e.g., with both N/O tracers, \delU, and $\upDelta$log($\sigma_{\text{H}\alpha}$), whereas the strength of the correlation with variations in other parameters like SFR or EW(H$\upalpha$) appears more limited to specific regions of the diagram.

We note here that an important corollary following directly from our framework and main assumptions is that the location of any given galaxy lying on the best-fit curve can be, in principle, predicted by the only knowledge of its gas-phase oxygen abundance, whereas any offset can be considered to occur at fixed O/H, as iso-contours in this quantity appear orthogonal to the SF sequence (see Fig.~\ref{fig:Nii_properties}). Indeed, the amplitude of $\upDelta$log(O/H) is basically zero (or very small) almost everywhere across the diagram, meaning that for any given point on the SF sequence, all galaxies located along an orthogonal line originating from that point can be assumed to have, at the first order, the same metallicity.

We first attempt to quantify the amount of variation in each parameter across the SF sequence by introducing the $\upDelta_{\perp}$ statistics, defined as the difference between the average $\upDelta$log(p) computed in galaxies lying in the regions upward and downward the best-fit line, normalised by the standard deviation of (the logarithm of) values spanned by each parameter; this quantity is reported for each quantity within the corresponding panel of Fig.~\ref{fig:nii_delta_prop}. 
In terms of absolute dynamic range, the amplitude of $\upDelta$log(p) is maximum for \sigha (0.74 $\sigma$), and relatively high also for M$^{\star}$, SFR and both N/O tracers, whereas it is minimum (as expected, given what we discussed slightly above) for metallicity.
We note here that $\upDelta$log(p) statistics grasp well what could be already inferred by visually inspecting Fig.~\ref{fig:Nii_properties} and ~\ref{fig:nii_delta_prop}, quantifying how the average orthogonal variation in a given parameter with respect to the SF sequence best-fit compares with the `width' of the overall distribution of values in that parameter (i.e., with $\sigma$).
However, we also stress that a high value in $\upDelta$log(p) does not necessarily imply a stronger \textit{causal} connection with the offset from the SF sequence, as some parameters might be intrinsically more connected with it even if characterised by a lower dynamical range in their logarithmic variation.
In order to properly ascertain which parameters in our set are intrinsically of most impact on the level of scatter in the diagram, we will therefore exploit a number of machine learning (ML) techniques, as outlined in the following Sections.  
The full set of parameters and the associated properties and statistics discussed in this Section are summarised in Table~\ref{tab:nii_parameters}.

\begin{figure*}

\includegraphics[width=0.85\textwidth]{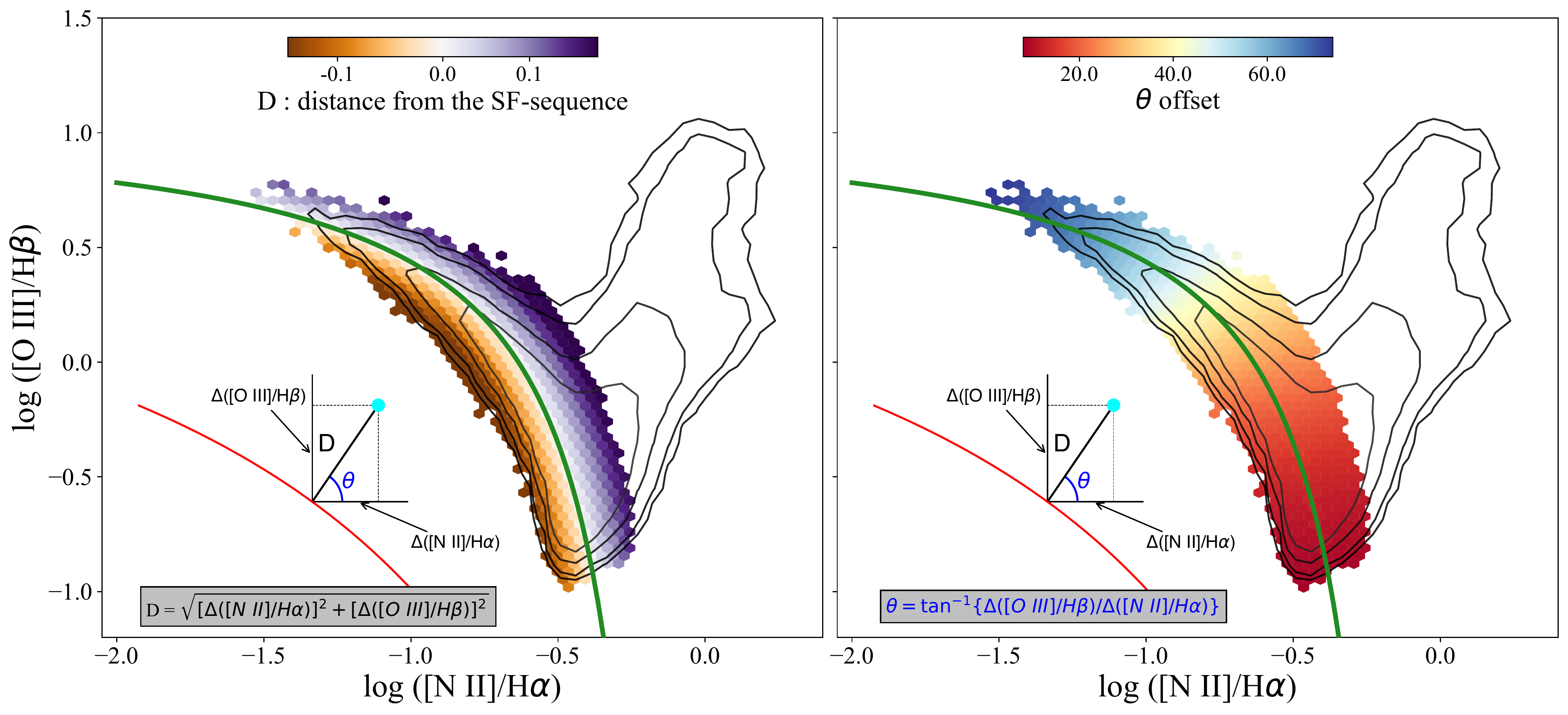}

\caption{The distribution of SDSS star forming galaxies in the [\ion{N}{ii}]-BPT diagram is colour coded by the distance metric defined by equation~\ref{eq:nii_distance} (\textit{left panel}) and by the corresponding angle $\theta$ of the `offset vector' with respect to the horizontal axis (\textit{right panel}), according to equation~\ref{eq:theta_angle}. The `offset vector' is assumed orthogonal to the best-fit curve of the median SF sequence at any given point.}
\label{fig:Nii_bpt_metrics}

\end{figure*}

\begin{table*}
    \centering
    \begin{tabular}{lccccc}
        \hline
        \hline

Parameter & Physical property & $\nabla_{\parallel}$ & $\varphi$ & $\upDelta_{\perp}$ & multi-parameter set \\

\hline

log(M$^{\star}$[M$_{\odot}$]) & Stellar mass & $0.3\sigma$ & $5.6\degree$ & $0.6\sigma$ & \cmark \\
log(SFR[M$_{\odot}$ yr$^{-1}$]) & Star formation rate & $0.26\sigma$ & $42.7\degree$ & $0.59\sigma$ & \cmark \\
log(sSFR[yr$^{-1}$]) & Specific SFR & $0.25\sigma$ & $161.5\degree$ & $0.15\sigma$ & \xmark \\
log(EW[H$\alpha$]) & Specific SFR & $0.29\sigma$ & $163.2\degree$ & $0.02\sigma$ & \cmark \\
H$\alpha$/H$\beta$ & Dust extinction & $0.03\sigma$ & $-24.0\degree$ & $0.51\sigma$ & \xmark \\
D$_{\text{N}}$(4000) & Age of stellar populations & $0.12\sigma$ & $15.4\degree$ & $0.32\sigma$ & \cmark \\
$\sigma_{H\alpha}$ [km/s] & Gas velocity dispersion & $0.002\sigma$ & $39.8\degree$ & $0.74\sigma$ & \cmark \\
$[\ion{S}{ii}] \lambda 6717/6731$ & Gas density & $0.2\sigma$ & $217.9\degree$ & $0.33\sigma$ & \cmark \\
12 + log(O/H) & Oxygen abundance & $0.31\sigma$ & $-45.8\degree$ & $0.04\sigma$ & \xmark \\
log([\ion{N}{ii}] $\lambda 6584$/[\ion{O}{ii}] $\lambda\lambda 3727,29$) & N/O abundance & $0.36\sigma$ & $-26.9\degree$ & $0.42\sigma$ & \xmark \\
log([\ion{N}{ii}] $\lambda 6584$/[\ion{S}{ii}] $\lambda\lambda 6717,31$) & N/O abundance & $0.32\sigma$ & $-6.6\degree$ & $0.67\sigma$ & \cmark \\
log([\ion{O}{iii}] $\lambda 5007$/[\ion{O}{ii}] $\lambda\lambda 3727,29$) & Ionisation parameter (U) & $0.26\sigma$ & $90.1\degree$ & $0.31\sigma$ & \cmark \\
%log([\ion{Ne}{iii}] $\lambda$3869/[\ion{O}{ii}] $\lambda\lambda$3727,29) & Ionisation parameter (U) & $0.32\sigma$ & $55.9\degree$ & $0.38\sigma$ & \cmark \\
         \hline
    \end{tabular}
     
    \caption{Full list of the parameters of interest included in the analysis of the [\ion{N}{ii}]-BPT diagram. For each quantity, we report the values of the statistics introduced in Secion~\ref{sec:frame} and reported in each panel of Fig.~\ref{fig:Nii_properties} and~\ref{fig:nii_delta_prop}. In the last column, we mark the parameters which are included in the set for the multi-parametric ANN and RF analysis; for more details about the justification of such selection, we refer to Section~\ref{sec:ML}.}
    \label{tab:nii_parameters}
\end{table*}

More specifically, the ML analysis will be targeted at reproducing (with the highest possible accuracy) the direction and amplitude of the offset of galaxies from the SF sequence in the BPT diagrams. 
We can thus introduce a few more parameters, whose definitions are based on the framework described above, which will help us in identifying the target labels for the ML algorithms; such quantities are here described for the \niibpt, but are defined equivalently for the \siibpt, as reported in Section~\ref{sec:S2_ML}.
First, for each galaxy in the diagram we can define the \textit{offset vector} as the vector pointing to that galaxy and originating from the closest point on the best-fit line of the SF sequence (i.e., the vector is orthogonal to the curve at any point and has the minimum possible amplitude). 
Its length \textbf{D} is then simply given by the Euclidean distance of the galaxy from the best-fit curve defined by equation~\ref{eq:nii_sequence} in the [\ion{N}{ii}]-BPT diagram parameter space, and can be written, in terms of its components, as : 
\begin{equation}
    \textbf{D} = \sqrt{ \sum_{i} (\upDelta q_{i})^{2} } \ ,
\label{eq:nii_distance}    
\end{equation}
where $\upDelta$\textit{q} is the difference between the \textit{q}-coordinate of a given galaxy in the diagram and the \textit{q}-coordinate of the nearest point on the best-fit curve, with \textit{q} $\in$ [log(\niiha), log(\oiiihb)] for the [\ion{N}{ii}]-BPT.
In the left-hand panel of Fig.~\ref{fig:Nii_bpt_metrics}, the distribution of star-forming galaxies in the diagram is now colour-coded by \textbf{D} (again, averaged in small hexagonal bins to aid visualisation): galaxies lying below the SF locus best-fit are assigned a negative value of \textbf{D}, in order to distinguish them from galaxies located above. This quantity represents one of the target labels for the machine learning analysis presented in Section~\ref{sec:ML}, but can be also simply used to identify whether a galaxy is located above or below the SF sequence.

In the right-hand panel of Fig.~\ref{fig:Nii_bpt_metrics} instead, each hexagonal bin is colour-coded according to the (average) angle formed by the offset vector of a galaxy with the horizontal axis (increasing positive counterclockwise), as given by :  
\begin{equation}
    \theta = \tan^{-1}\Big( \frac{\upDelta\ \text{log}([\ion{O}{iii}]/\text{H} \beta)}{\upDelta\ \text{log}([\ion{N}{ii}]/\text{H} \alpha)} \Big) .
    \label{eq:theta_angle}
\end{equation}
This parameter is useful to quantify which is the predominant component of the offset vector, i.e. whether the offset occurs preferentially along the [\ion{N}{ii}]/H$\upalpha$- or the [\ion{O}{iii}]/H$\upbeta$-axis.
Given the shape of the distribution of star forming galaxies within the [\ion{N}{ii}]-BPT, offsets in the bottom-right part of the sequence occur preferentially along [\ion{N}{ii}]/H$\alpha$ (i.e., low values of $\theta$), whereas $\theta$ increases (hence deviations in [\ion{O}{iii}]/H$\beta$ becomes increasingly more relevant) as we move along the SF sequence towards the upper-left region. Whether this has an impact on the results of the ML analysis will be specifically addressed in Section~\ref{sec:nii_relative_importance}.

\section{Machine Learning analysis}
\label{sec:ML}

\subsection{Algorithms, parameters and tasks}
\label{sec:ML_Nii_params}

In the previous section, we have attempted to assess how the position of star-forming galaxies within the [\ion{N}{ii}]-BPT diagram correlates with a variety of physical parameters, by visually inspecting the distribution of such parameters within the diagram and introducing statistics aimed at quantifying the amplitude of variations along and across the SF sequence.
Here, we take a step forward and exploit the statistically sound SDSS database to implement different machine learning algorithms aimed at providing a more robust and quantitative assessment of the drivers of the scatter across the star-forming galaxy sequence in the diagram.
The ultimate goal is to provide a method to robustly identify which physical parameters are statistically more connected with the deviation from the SF locus, adopting a framework which is completely based on observational data and independent from many of the standard prescriptions included in the majority of photoionisation models.

In practice, Artificial Neural Networks (ANN) and Random Forest (RF) of decision trees are trained and tested on our large sample of selected SDSS star-forming galaxies in order to solve both a \textit{classification} and a \textit{regression} problem. The former, aimed at describing which parameters perform better in predicting whether a galaxy is simply located above or below the best-fit curve representative of the SF sequence. The latter, instead, to assess the ability of each variable (and of the whole set) in predicting the exact distance (i.e., the amplitude \textbf{D} of the offset vector described in equation~\ref{eq:nii_distance}) from the SF sequence itself.

The implementation of both ANN and RF algorithms allows us to tackle these two problems from rather different angles.
With the ANN, we aim at exploring the performance of each parameter individually, as well as the maximum potential of the full set as a whole, by means of a model-independent approach free of any underlying assumptions about correlations, linearity and monotonicity within the data.
Unfortunately, when fed with a set of multiple parameters, ANN do not intrinsically provide information about the relative impact that each individual parameter has in contributing to its overall predictive power: one might ask, in fact, whether the full set of parameters is really required to achieve the highest level of accuracy or even a subset could provide comparable results and, ultimately, which parameters specifically contain the informations that maximise the predictivity of the model.
For this reason, we perform the same analysis implementing also RF decision trees, which allows us to disentangle the relative importance of even highly correlated features involved in the prediction algorithm. In other words, by means of the RF analysis we aim at assessing which of the involved (and intercorrelated) parameters are intrinsically the most informative in predicting our target variables when the full set is used in concert.

The set of parameters to be considered in the analysis is taken from the list of observables and properties discussed in Section~\ref{sec:sdss}; in particular, in the last column of Table~\ref{tab:nii_parameters} we mark which parameters are included in the multi parametric ML analysis.
In fact, although each parameter is assessed through the ANN individually (i.e., by feeding the network with the data relative to only one parameter at a time), when evaluating the performances of the algorithms considering multiple parameters altogether it is warranted to perform a careful selection of the quantities to be included in the final set, in order to avoid nuisance parameters which either duplicate the physical information, and/or are trivially correlated with others and with the target labels.
From now on, we refer to the analysis performed on such list of parameters as the `multi-parameter run(s)', and to the list itself as the `multi-parameter set'; accordingly, the RF analysis will also be based upon the subset of parameters included in this list.

First, we decide not to include metallicity at all in the ML analysis. 
In fact, being oxygen abundance mostly derived from the combination of several strong line ratios (including the line ratios which constitute the BPT-axis, see Section~\ref{sec:data} and \citealt[][]{curti_massmetallicity_2020}), such quantity can be trivially recovered from a combination of other emission line-based parameters and the BPT line ratios themselves (which are at the basis of the definition of the distance target label \textbf{D}); hence, in our framework log(O/H) can be treated as a nuisance parameter, not independent from the others, which could bias the performances of both the `multi-parameter' ANN run and the relative feature importance assessment performed by the RF. 
%Indeed, we have tested that including log(O/H) in this form into the analysis, 
However, based on what is shown in Fig.~\ref{fig:Nii_properties} and on the assumptions \textbf{i)} and \textbf{ii)} discussed in Section~\ref{sec:frame}, the contribution from metallicity to setting the offset from the best-fit line appears of little significance, being iso-O/H lines orthogonal to the SF sequence at any point (as also quantified by a $\upDelta_{\perp}$ statistics $\sim 0$ for log(O/H)). Hence, we can add a third assumption to our framework, that is \textbf{iii)} any contribution to the observed offset from the SF sequence from any of the involved parameters is assumed here to occur at fixed metallicity. The validity of such assumption is further discussed later in the text. 

Then, we chose to adopt [\ion{N}{ii}]/[\ion{S}{ii}] instead of [\ion{N}{ii}]/[\ion{O}{ii}] as a primary tracer of the N/O abundance in the fiducial `multi-parameter' analysis of the [\ion{N}{ii}]-BPT diagram.
As stated before, this choice is primarily motivated by the fact that we aim to provide the network with a set of parameters which are as much as possible independent from each other, and whose linear combinations are not trivially connected, from a mathematical point of view, to the position on the \niibpt diagrams itself and to the target labels.
We have tested in fact, that by including both [\ion{N}{ii}]/[\ion{O}{ii}] and [\ion{O}{iii}]/[\ion{O}{ii}] together (even in their $\upDelta$log(p) form) the ANN can reconstruct something very similar to the [\ion{N}{ii}]/[\ion{O}{iii}] ratio (which is closely related mathematically to \textbf{D}), `artificially' boosting its performances in the analysis of the \niibpt. For the same reason, the RF would be strongly biased towards the choice of these two parameters in its relative feature importance computation, well beyond the underlying physical connection of such parameters with the target variable, and hiding potential contributions from other quantities.
Nonetheless, we acknowledge that [\ion{N}{ii}]/[\ion{S}{ii}] intrinsically accounts also for residual dependencies on top of N/O (see discussion in Section~\ref{sec:parameters}), as sulphur may not exactly trace oxygen in case of strong variations in the ionising conditions; hence, such a parameter is less reflective of the `true' nitrogen-over-oxygen abundance than [\ion{N}{ii}]/[\ion{O}{ii}] is.
Therefore, although our fiducial analysis of the `multi-parameter set' is based on [\ion{N}{ii}]/[\ion{S}{ii}] as primary a tracer of N/O, within the text (and more specifically in Appendix~\ref{sec:appendix_A}) we discuss also different combinations of parameters in the set, including [\ion{N}{ii}]/[\ion{O}{ii}] and modifying the list of the other emission lines-based parameters accordingly (for instance, assuming [\ion{Ne}{iii}]/[\ion{O}{ii}] instead of [\ion{O}{iii}]/[\ion{O}{ii}] to trace the ionisation parameter). However, we anticipate and reassure that none of the main conclusions presented in this paper is affected by the choice of different N/O tracers.  

Furthermore, we choose EW(H$\upalpha$) as an independent probe of the sSFR (in order to avoid trivial correlations between sSFR, SFR and M$_{\star}$), and log(\oiii/\oii) as a tracer primarily of the ionisation parameter, because requiring even low-significance (e.g., $2.5\sigma$) detections of the [\ion{Ne}{iii}]$\lambda3869$ emission line would introduce significant sample selection biases (i.e., preferentially removing galaxies in the high-mass, high-metallicity, bottom-right region of the diagram). 
Finally, we also remove H$\upalpha$/H$\upbeta$ from the `multi-parameter set', as the BPT diagrams are, by definition, insensitive to dust extinction (thanks to the small wavelength separation of the emission lines on both x- and y-axis), hence any correlation between such parameter and the location of galaxies in the diagram would necessarily follow from the correlation between the dust content and other physical parameters; moreover, this would further limit the algorithms to perform any trivial algebraic operation between emission line-based parameters.
Before proceeding, each feature in the dataset is properly rescaled by subtracting its average value and normalised by the interquartile (i.e, $25th-75th$ percentile) range.

\subsection{Artificial Neural Networks}
\label{sec:Nii_ANN}

Artificial Neural Networks (ANN) are a set of algorithms with structures that are inspired by the neural networks that constitute the human brain, and whose flexible structure and non-linearity allows to perform a wide variety of tasks \citep{ML_astronomy_2019}.
For the purposes of the present work, a multilayered neural network is designed exploiting the TENSORFLOW\footnote{\url{https://www.tensorflow.org/overview}} package within a PYTHON environment.
The baseline structure of the network is very similar for both the classification and the regression task, however the details (and the relative differences) are described for each of the two cases within the dedicated subsections.
In brief, a typical network consists of an input layer, an output layer, and several hidden layers, where each of these contain neurons that transmit information to the neurons in the succeeding layer. The input data is transmitted from the input layer through the hidden layers, and reaches the output layer, where the target variable is predicted. 
The value of every neuron in the network (except those in the the input layer) is a linear combination of the neurons in the previous layer, followed by the application of a (typically non-linear) activation function.
The weights of the network are model parameters which are optimized during the training stage via back-propagation.

For the purposes of training the network, we randomly select the two-thirds ($\sim 67$ per cent) of the dataset to define a \textit{training sample}, with the remaining one-third ($\sim 33$ per cent) that constitutes the \textit{test sample} over which the performances of the ANN are evaluated (and which the network does not interact with at all during the training stages). 
Given the large available statistics, this choice provides a sufficiently large set to perform an extensive training of the network without sacrificing its ability to generalise; moreover, both sub-samples are selected (and large enough) to be fully representative of the distribution of galaxies in the BPT of the whole parent population. Nonetheless, we stress here that none of the conclusions presented in this paper are affected by a different choice in train-test splitting (e.g, a $50$-$50$ per cent or a $75$-$25$ per cent splitting are also widely adopted approaches).

We perform the analysis by either feeding the network with one parameter at the time, to evaluate their individual connection with the galaxy offset from the SF locus, and with a set of multiple parameters simultaneously (see previous section), in order to explore the maximum predictivity potential of the data.
Because of the increased impact of overfitting in the `multi-parameter' run compared to the individual runs, the structure of the network is slightly different in the former case than in the latter, and its overall complexity is reduced, for both classification and regression analysis, as described more in detail in the following sections.
Each run of the ANN analysis is repeated 30 times, randomly splitting each time the sample into training and testing sub-samples, and taking the average and standard deviation of the resulting distribution in performances as the final score and associated uncertainty, respectively.

\subsubsection{Classification}

We start with a rather simple classification analysis.
The goal is then to determine which parameters are best in predicting whether a galaxy is offset above or below the SF locus in the BPT diagrams. %once the offset is assumed to be orthogonal to the best-fit line. 
In principle, in this case we do not need to assume any particular direction of the `offset vector', but galaxies are just assigned either $1$ or $0$ label according to their position above or below the SF locus (i.e., according to positive/negative values of \textbf{D}), defining a binary classification problem. However, we recall that the $\upDelta$log(p) values of equation~\ref{eq:delta_prop} which are ingested into the network are actually computed by assuming a purely orthogonal deviation.

We design a multilayered network composed of two hidden layers (with 12 and 6 neurons, respectively), with a \textit{rectified linear unit} (ReLu) activation function for the hidden layers and a \textit{sigmoid} function for the one-dimensional (i.e., a binary 0/1 value) output layer. 
The main advantage of using the ReLU function over other activation functions is that it does not activate all the neurons at the same time (i.e., the neurons will only be activated if the output of the linear transformation is larger than zero), whereas the \textit{sigmoid} function is largely used in models where the output layer should return a probability (in this case, the probability of belonging to a given class), since it maps any input values onto the [0, 1] range.
The model is compiled implementing the ADAM solver (with a learning rate $=0.001$) and optimising the standard \textit{binary crossentropy} loss function.
The `accuracy' (i.e., the fraction of galaxies correctly classified) is the metric assumed by the model to assess its performance during the training procedure and when applying its predictions to the test sample.

Such network structure is the result of an extensive direct experimentation with the dataset aimed at maximising the accuracy while keeping overfitting at a minimum. 
As an uncontrolled increase in the complexity of the network can lead to significant overfitting (i.e., the network performing significantly better on the training sample than on the test sample, lowering its ability to generalise its predictions to unseen datasets), we require the difference in accuracy between the performance of the model on the training and test samples to be within a few per cent, and we tune the network hyperparameters accordingly.
As briefly metioned above, in the `multi-parameter' run we decide to reduce the complexity of the network by implementing a single-hidden layer with $10$ neurons only, in order to minimise the impact of overfitting.
% This approach has been already shown to perform quite well in the analysis of large statistical sample of galaxies like the SDSS and MaNGA \citep[see e.g.,][]{Bluck:2019aa}.
For this binary classification problem, the two classes (i.e., \textit{above} and \textit{below} the best-fit line) are also randomly re-sampled in order to be equally represented in both the training and test set (i.e., to have $50$ per cent of galaxies lying \textit{above} and $50$ per cent lying \textit{below} the SF locus in both training and test samples). In any case, we also consider here the area under the true positive rate (TPR)–false-positive rate (FPR) curve (known simply as the area-under-the-curve, `AUC') as an additional metric to evaluate the network performance; one of the advantages of the AUC statistic in fact is that it is insensitive to the fraction of each class provided to the network. 
Furthermore, for the classification problem we focus only on galaxies with values of $|D| > 0.025$, i.e., we remove galaxies which lie so close to the best-fit line to be
potentially misclassified given the typical uncertainties on their measured emission line ratios. However, we note that including also these galaxies only slightly degrades the performances of the network, but does not affect any of the conclusions.

\begin{figure*}

\includegraphics[width=0.7\textwidth]{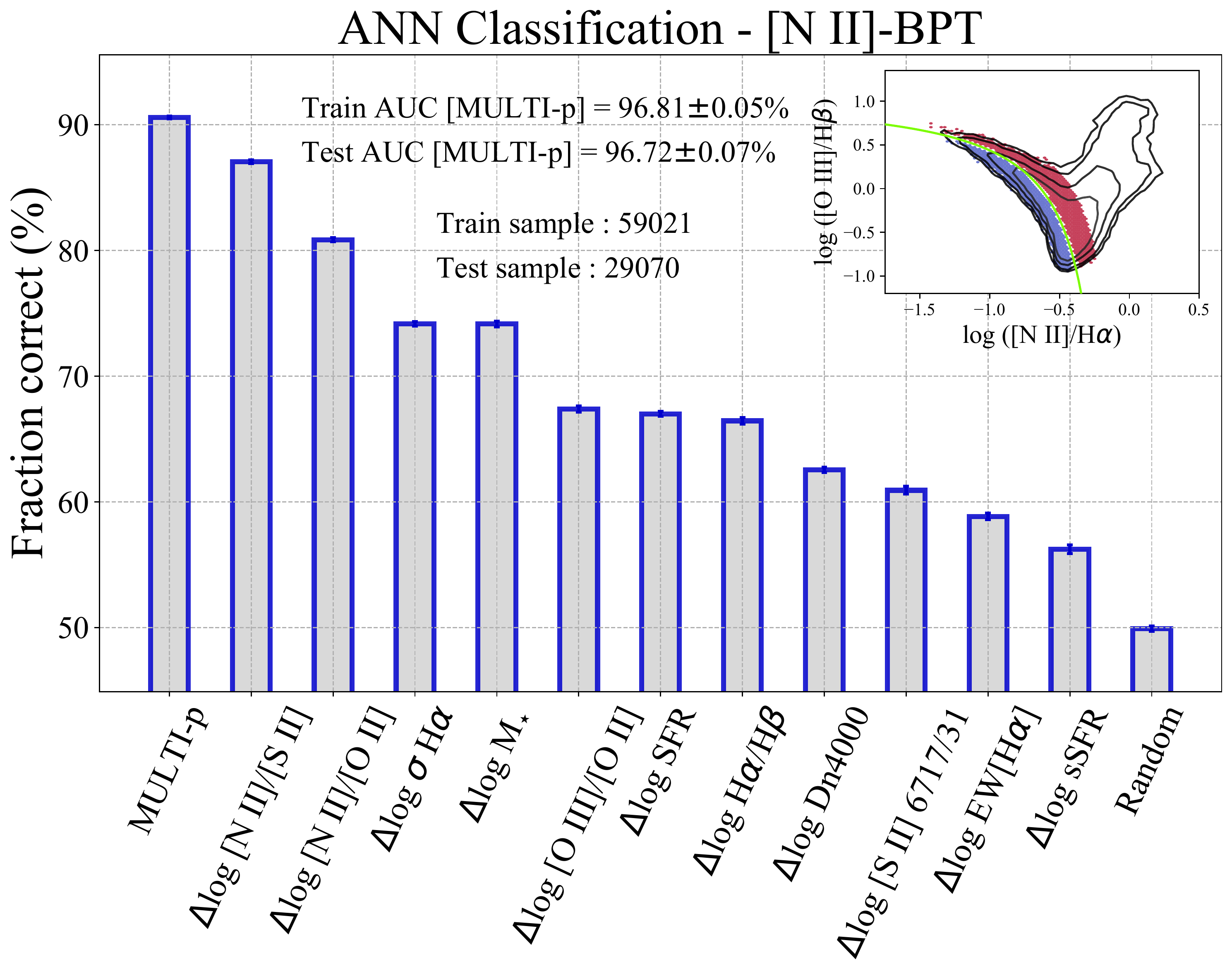}\\
\vspace{0.5 cm}
\includegraphics[width=0.7\textwidth]{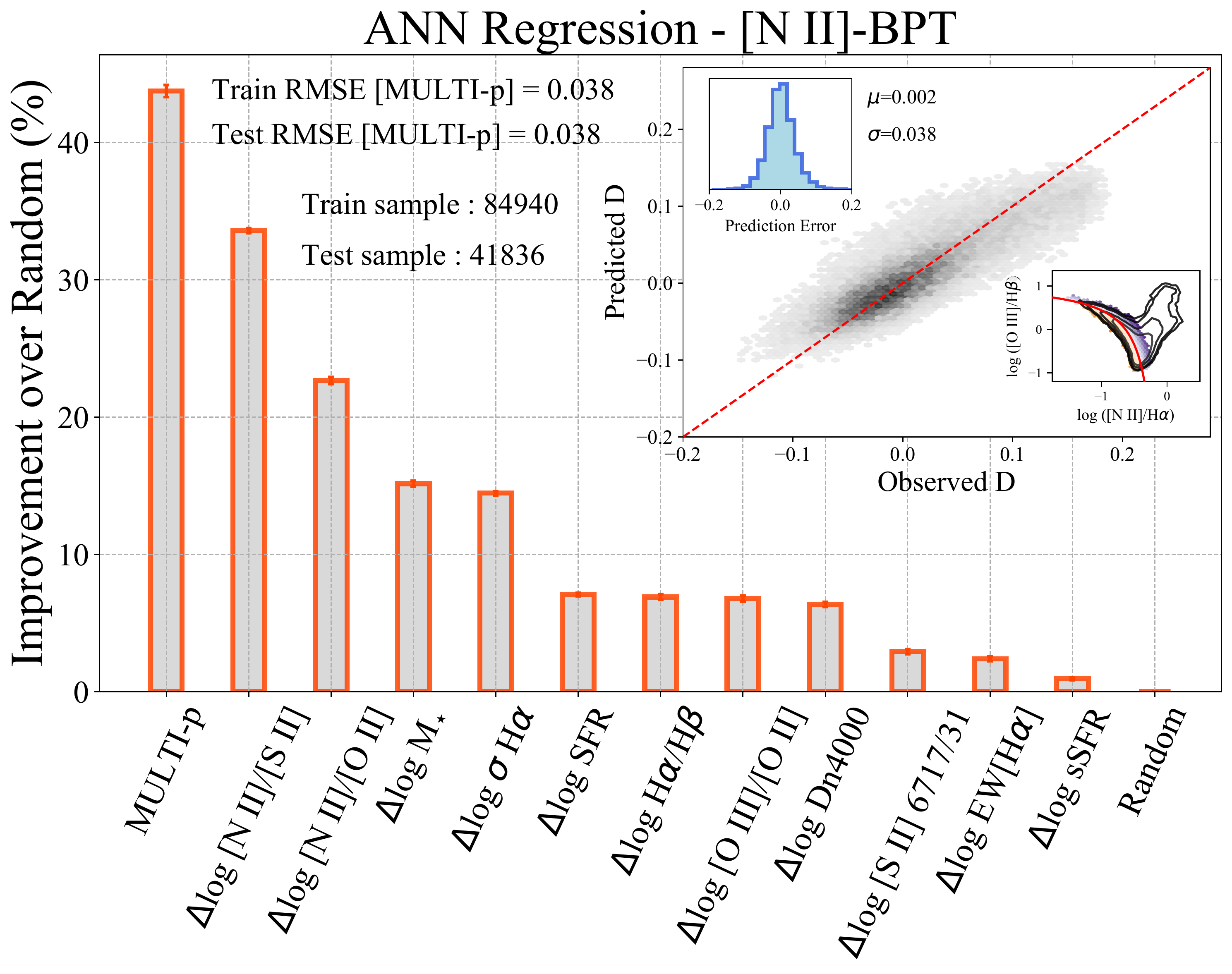}

\caption{\textit{Upper Panel:} Results from the ANN classification analysis aimed at predicting whether galaxies are located above or below the best-fit line of the star-forming galaxies sequence in the [\ion{N}{ii}]-BPT diagram (as schematised in the upper-right inset panel). Histogram bars report the absolute fraction of correctly classified galaxies for both the `multi-parameter' set (see last column of Table~\ref{tab:nii_parameters}), and each parameter taken individually (ordered from most to least predictive); the last bar report, for comparison, the performance of a random variable (which is equivalent to $50$ per cent accuracy in a binary classification problem). \textit{Bottom Panel:} Results from the ANN regression analysis aimed at predicting the exact magnitude \textbf{D} of the offset from the best-fit line of the star-forming galaxies sequence in the [\ion{N}{ii}]-BPT diagram, as defined in Eq.~\ref{eq:nii_distance}. Histogram bars in this case report the `improvement-over-random' (IoR) statistics (ordered from most to least predictive parameter, with the a random variable scoring, by definition, $0\%$ IoR). Within the top-right, inset panel, we compare the \textbf{D} values predicted by the network on the test sample in the `multi-parameter' run to the `true' \textbf{D} target values for the same galaxies. In both panels, we also report the number of galaxies in the training and test sub-samples and the relative AUC and RMSE scores of the `multi-parameter' runs.}
\label{fig:Nii_ANN_classification}

\end{figure*}

In the upper panel of Fig.~\ref{fig:Nii_ANN_classification} we present the results of the binary classification analysis for the set of parameters described in Section~\ref{sec:ML_Nii_params}. The fraction of correctly classified galaxies is shown on the y-axis, and the parameters used to train the network are shown on the x-axis and ordered from the most to the least predictive.
The first bar in the plot refers to the run performed with the `multi-parameter' set, which contains only a sub-set of the full list of parameters, as listed in the last column of Table~\ref{tab:nii_parameters}, and according to what is discussed in Section~\ref{sec:ML_Nii_params}

When all the parameters from the `multi-parameter' set are fed together into the network, the model achieves an impressive classification accuracy of $90.57 \pm 0.11$ per cent (AUC=$96.72 \pm 0.07 $ per cent) on the test sample. 
Therefore, the position of a galaxy with respect to the SF sequence in the [\ion{N}{ii}]-BPT (i.e., whether a galaxy is offset above or below it) can, in principle, be predicted with excellent accuracy by knowing no more than the set of parameters adopted here \footnote{this, however, does not automatically imply that different parameters would not perform equally well, or perhaps even better}.
No significant variation on the performances is obtained from either increasing the network complexity or slightly varying the values of the hyperparameters, further confirming the stability of the result.

In terms of performances of individual parameters (i.e., when the network is fed with only one parameter at the time), \delNO\ achieves the best performance compared to the rest of the set, with an accuracy of $87.07 \pm 0.14$ per cent; adopting \delNOii provides a comparable (though slightly lower) accuracy of $80.85 \pm 0.18$ per cent.
This means that deviations in the N/O abundance from the average value pertaining to galaxies along the SF locus (mainly traced either by [\ion{N}{ii}]/[\ion{O}{ii}] or [\ion{N}{ii}]/[\ion{S}{ii}]) are extremely informative in predicting whether galaxies are offset above or below the SF sequence itself, and perform better than any other individual parameter in our set.
%This is in general not surprising, as we are currently investigating the distribution of galaxies in the [\ion{N}{ii}]-BPT diagram, which contains information about nitrogen abundances embedded in its x-axis. 
%Given the tight connection between metallicity and the position of galaxies along the SF sequence best-fit line as discussed before (see e.g. Fig.~\ref{fig:Nii_properties}), and the existence of a relationship between O/H and N/O in local star-forming SDSS galaxies with non-zero scatter (see e.g, Hayden-Pawson, in prep.), this result suggests that the intrinsic scatter in the O/H vs N/O plane is reflected in the BPT emission lines and account for a large part of the dispersion of galaxies across the SF locus in the diagram.
% Nonetheless, the predictivity of $\upDelta$[\ion{N}{ii}]/[\ion{S}{ii}] alone is roughly $80\%$ of the run involving ''ALL-parameters'', meaning that additional information is required to achieve the best possible result in the classification of star-forming galaxies on such diagram.

Among the other parameters, deviations in M$_{\star}$ and \sigha rank immediately after, although scoring significantly lower accuracies, followed by \delU (primarily tracing variations in U), $\upDelta$log(SFR) and $\upDelta$log(H$\upalpha$/H$\upbeta$) (associated to dust attenuation).
% The good performance of the former parameter is not surprising, because stellar mass is a very good proxy of the state of the chemical evolution in galaxies, tracing the integrated star-formation rate (and hence the metal production) over cosmic time, from which the correlation between M$_{\star}$ and N/O readily follows (refs.)
% The central velocity dispersion of the gas instead is thought to correlate with stellar mass, via relationship with the gas mass and the dynamical state of a galaxy, and SFR, via energy injected into the ISM by supernovae feedback \citep[][]{green_dynamo_2014,krumholz_discs_2018,yu_vel_disp_manga_2019,varidel_turbulence_2020}; however, it also represents a potential tracer of shock-heated gas, and has been already suggested as a feasible parameter to extend and complement the ionisation source classification scheme of the BPT diagrams \citep{dagostino_bpt_2019, law_bpt_2021}. 
% Then follows parameters tracing deviations in the ionisation parameter and the SFR, %once again, we have seen in Section~\ref{sec:frame} how the [\ion{N}{ii}]-BPT diagram intrinsically contains information about the ionisation parameter, being its y-axis well correlated with [\ion{O}{iii}]/[\ion{O}{ii}].
Finally, we note that deviations in sSFR (probed either by $\upDelta$ EW(H$\upalpha$) or directly by the ratio between SFR and M$^{\star}$) and electron density (probed by the [\ion{S}{ii}]$\lambda6717$/$\lambda6731$ ratio) perform only $\sim 10$ percent better than a purely random variable (reported by the last bar and equivalent, in a balanced-sample binary classification problem, to tossing a coin with $50\%$ probability of success), hence proving to be not very informative overall at describing the relative position of galaxies with respect to the SF locus within this diagram.

\subsubsection{Regression}
% \label{sec:ANN_regressio}

We now move to a different part of the analysis, which shares the same goal as the previous one (i.e., describing the connection between relative variations in different physical parameters and the scatter in the BPT diagrams) but set a different target label for the ANN.
In particular, we now want to test the ability of our group of parameters to predict the \textit{magnitude of the offset} (i.e., the amplitude \textbf{D} of the offset vector, taken positive if pointing above the best-fit line) from the sequence of local star-forming galaxies, in a standard regression analysis. Here, following what is discussed in Section~\ref{sec:frame}, and differently from the classification analysis, the offset vector is assumed to be exactly orthogonal to the best-fit curve of the SF sequence, at any given point.
In principle then, there is no reason to assume a priori that the classification and the regression analysis should provide the same results, although the two problems are clearly related to each other.

Similar to the previous case, we create a network with two hidden layers ($12$ and $6$ neurons, respectively) and a \textit{ReLu} activation function. The model is compiled with the ADAM optimiser (with a learning rate $=0.001$) and minimises the \textit{mean squared error} (mse) as the loss function.
Again, for the `multi-parameter' run the complexity of the network is reduced to a single-hidden layer with only $10$ neurons, in order to control the impact of overfitting.
Extensive testing of the network outputs and performances suggests adoption of a \textit{mini-batch gradient descent}\footnote{A gradient descent is an optimization technique used to find the weights of machine learning algorithms. It works by exploiting the error associated to model predictions on the training data to update the parameters in order to reduce the discrepancies at the following steps.
The `mini-batch' gradient descent is a variation of this approach, which splits the training dataset into small batches that are used to calculate the errors and update the model coefficients. Its main advantages over the standard gradient descent are that the model is updated more frequently (which allows for a more robust convergence), 
its computational efficiency is increased, and that it does not require to maintain all the training data in memory at once.} algorithm with a batch size of $128$ and $32$ for the `individual' and `multi-parameter' runs respectively, and to train the network over a total of $100$ epochs.  

In the bottom panel of Fig.~\ref{fig:Nii_ANN_classification}, we report the results of the ANN regression analysis, where the performance of each individual parameter in our set (and of the `multi-parameter` set) is assessed on the basis of that of a purely random variable. Following, e.g., \cite{bluck_are_2019}, we define in fact the `improvement over random' metric as 
\begin{equation}
    \text{IoR}_{i} = \frac{\text{RMSE}_{i} - \text{RMSE}_{\text{rand}} } {0 - \text{RMSE}_{\text{rand}} } \ ,
\end{equation}
where RMSE$_{i,\text{rand}}$ is the root-mean-squared-error of the \textit{i-}th variable and of a purely random variable respectively, whereas zero represents, by definition, the best possible performance in terms of RMSE in a regression problem (i.e., the target variable is predicted with 100 per-cent accuracy).

When trained with the `multi-parameter' set, the network achieves an IoR = $44.81 \pm 0.19 \%$ in predicting the exact distance \textbf{D} from the SF sequence in the test sample.
The values of \textbf{D} predicted for the test sample by the network in the `multi-parameter' run are compared to the \textit{true} target \textbf{D} values as shown in the inset, upper-right panel of Fig.~\ref{fig:Nii_ANN_classification}; we report a median of the errors on the predictions of $\mu = 0.002$ and a standard deviation of $\sigma = 0.038$. 
Similar results are found on the training sample, with no significant overfitting reported.

Individual parameters rank in a very similar order as in the classification problem, with \delNO\ and \delNOii\ (primarily associated with relative deviations in the N/O abundance) being the most predictive quantities (IoR = $33.45 \pm 0.31 \%$ and $22.8 \pm 0.26 \%$, respectively) of the distance from the best-fit line of the SF sequence in the [\ion{N}{ii}]-BPT diagram.
It is interesting to note that, such parameters aside, none of the included quantities scores above $20$ per cent in IoR, with six out of nine parameters marking an improvement below $10$ per cent.
This is somehow expected, and confirms that predicting the exact distance from the SF sequence (in regression) is much more difficult than just classifying a galaxy as offset above or below it; indeed, no individual parameter, except for those primarily tracing variations in N/O, is really capable of providing enough information to predict our target variable \textbf{D} with a very high level of accuracy.
However, when the information from multiple parameters is provided, the predictive power is increased and the network can reproduce the offset of star-forming galaxies in the [\ion{N}{ii}]-BPT with significantly higher accuracy.

\subsection{Random Forest}
\label{sec:Nii_RF}

\begin{figure*}

\includegraphics[width=0.7\textwidth]{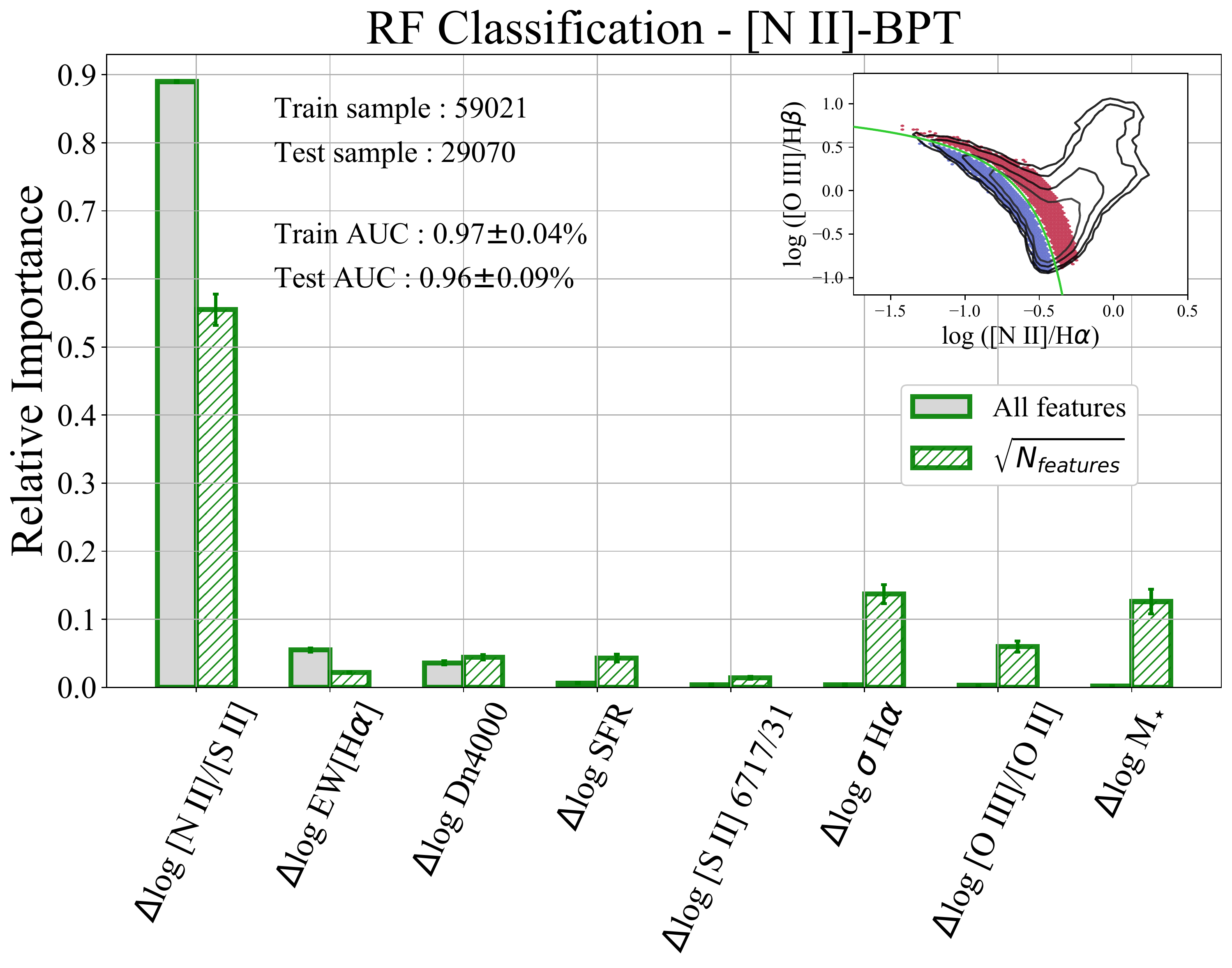} \\
\vspace{0.5 cm}
\includegraphics[width=0.7\textwidth]{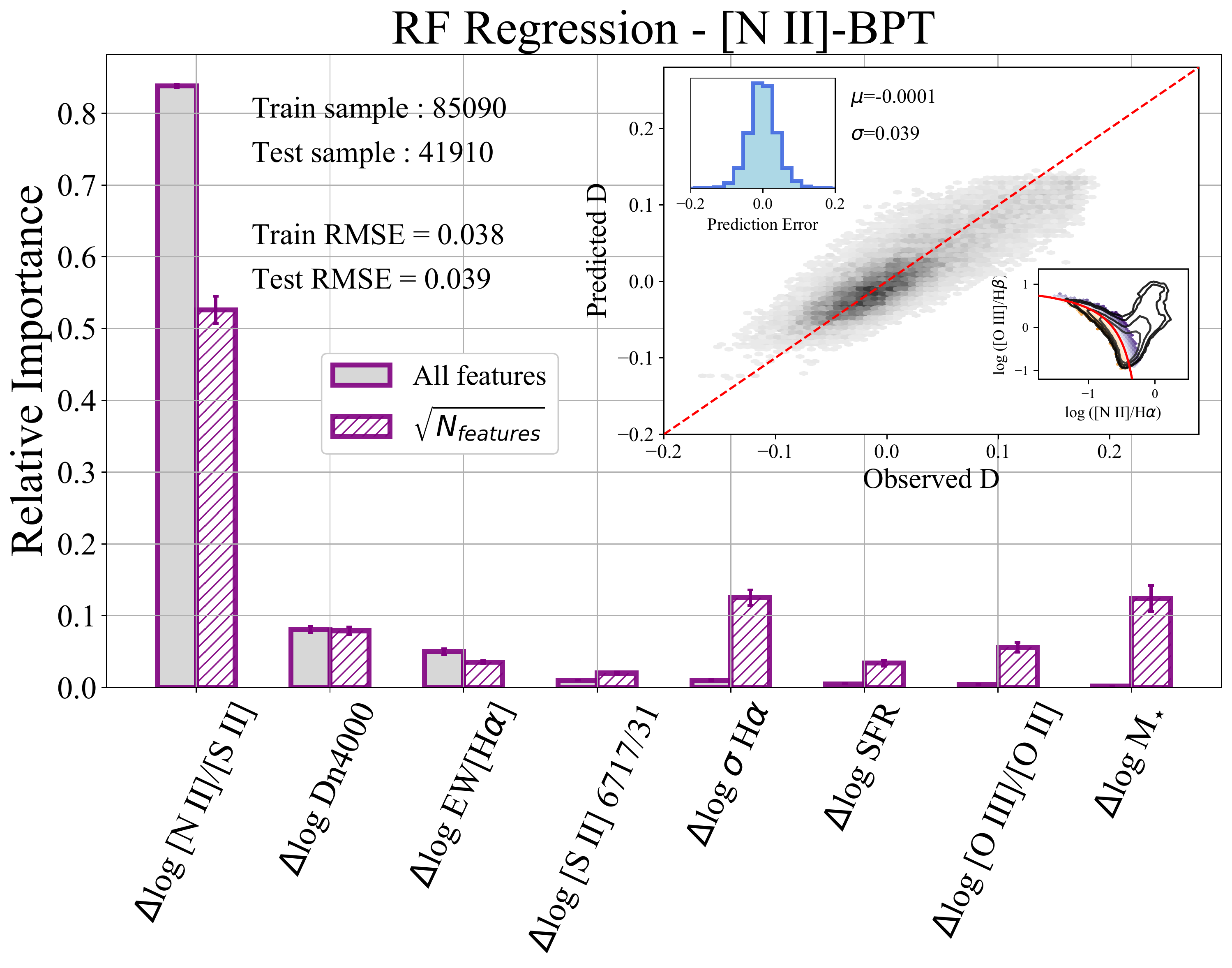}

\caption{Results of the RF classification (\textit{upper panel}) and regression (\textit{bottom panel}) analysis, adopting the `multi-parameter' set under study as listed in the last column of Table~\ref{tab:nii_parameters}. The y-axis reports the relative importance of each of the parameters labelled on the x-axis (with error bars derived from the standard deviation of $30$ independent runs). Colour-filled bars refer to the RF analysis conducted allowing `all features' to be considered at each node of the trees, whereas empty, hatched bars refer to the case where only the square root of the number of features are picked up at each splitting node. The former fully disentangles the inter-dependencies between the various parameters, providing the best possible combination of features to use in concert to maximise the performances of the RF, whereas in the latter a relatively higher importance is retained also by parameters which are (to some extent) correlated with the best ranked ones, being valuable alternatives in case these are not available. In both cases, \delNO, mainly tracing relative variations in the N/O abundance compared to the median SF locus, is robustly identified as the most predictive parameter in either classification and regression tasks. A rather small, though complementary, residual importance is retained by $\upDelta$log(D$_{\text{N}}$(4000)) and $\upDelta$log(\ewha).
}
\label{fig:Nii_RF_class}

\end{figure*}

In this section, we exploit a random forest (RF) of decision trees in order to determine how effective a given parameter is in solving the classification and regression problems addressed before, when considered in direct comparison with the other parameters. In fact, the RF treats multiple parameters as if they were in a competition, selecting the most useful for each decision node. 

In general, a decision tree is a set of consecutive nodes, where each node represents a condition on one feature in the dataset. The conditions are of the form X$_{j}$ > X$_{j,th}$, where X$_{j}$ is the value of the feature at index j and X$_{j,th}$ is a threshold, which is determined during the training stage. The lowest nodes in the tree are usually called `leaves', and carry the final assigned label of a particular path within the tree (e.g., in our classification case, whether a galaxy is labeled as \textit{above} or \textit{below}). 
A RF, then, is simply a collection of decision trees, each of them trained on different bootstrapped, randomly-selected subsets of the original training set (and where, if desired, random subsets of the input features can be selected during the training of each individual tree to construct the conditions in individual nodes).
The final RF prediction is just an aggregate of individual predictions of the trees in the forest, in the form of a majority vote; the main advantage is that, while a single decision tree tends to overfit the training data, the combination of many decision trees in a RF generalizes well to previously unseen datasets.
Furthermore, by quantifying the decrease in impurity provided by each parameter in each fork and within each tree of the RF, the relative importance of the various parameters in the prediction can be established. This competitive approach is especially useful when the parameters considered are highly inter-correlated in a complex and highly non-linear manner. 

We recall here that the following RF analysis is based on the `multi-parameter' subset defined in the last column of Table~\ref{tab:nii_parameters}, whose selection is justified in detail in Section~\ref{sec:ML}. To implement the RF into our analysis, we adopt the RANDOMFORESTCLASSIFIER and RANDOMFORESTREGRESSOR classes from the SCIKIT-LEARN package in PYTHON.

\subsubsection{Classification}

In the binary classification scheme, we set up a forest of $100$ independent estimators, allowing each tree to grow indefinitely but setting a minimum threshold to the number of samples allowed at each leaf-node equal to the number of galaxies in the training sample divided by $250$ (i.e., $\sim350$ samples in our case). 
This choice allows us to control overfitting, which is assessed by requiring the difference in performances between the training and the test sample to be limited to a few percent.
The RF Classifier is set to minimise the \textit{Gini} impurity as the loss-function at each decision node.
The accuracy of the RF classification task is assessed by evaluating the AUC on both the training and the test sample.
We perform $30$ independent runs (randomised at the training-test sample split level), hence evaluating the average and standard deviation of the results over $30\times100 = 3000$ independent estimators.
Consistently with the ANN analysis, only galaxies with |\textbf{D}|$>0.025$ are included in the RF classification analysis.

In the following, we also explore and discuss two different ways for computing the relative feature importance.
In the first case, we leave the RF free to consider the entire set of parameters at each decision split (i.e., what we call the 'All features' case, setting the \textit{max features} hyperparameter of the RF accordingly).
In this way, the algorithm is capable to fully handle the inter-correlations between the different features and find the one (or the group of parameters) which is most intrinsically (and causally) connected with the target variable \citep{bluck_ml_causality_2022}.

In the upper panel of Fig.~\ref{fig:Nii_RF_class}, the results of the RF classification analysis in this first case are shown by the filled bar chart. The parameters are ranked in terms of their relative importance (from the most to the least relevant), which is reported on the y-axis.
The overall performance of the RF model on both the training and test sample is reported in terms of AUC: the RF achieves an $\text{AUC}=96.35 \pm 0.09$ per cent in the binary classification task, a performance comparable to that scored by the ANN when trained with the `multi-parameter' set.
However, although at first sight the ranking in the relative importance of the various parameters resemble that obtained in Fig.~\ref{fig:Nii_ANN_classification} for the ANN analysis (in terms of accuracy of individual features), there are a number of remarkable differences.
In particular, the relative importance of \delNO (mainly tracing deviations in the N/O abundance) is strongly dominant over the other parameters, accounting for more than $80\%$ of the total predictive power, whereas $\upDelta$log(EW[H$\alpha$]) and $\upDelta$log(D$_{\text{N}}$(4000)) (tracing variations in the specific star formation rate and age of stellar populations) are ranked second and third, respectively, retaining together about $10\%$ of the residual importance.
Interestingly, although these parameters were among the least performing, individually, in the ANN analysis, the RF highlights how their information is more complementary to \delNO\ than any other parameter in the set.
On the contrary, the importance of all the remaining parameters is strongly suppressed, revealing how their individual predictive power (as measured by the ANN) was likely due to underlying correlations with one of the best-ranked features. 

In addition, we have also explored the case in which only a fixed number of (randomly selected) features are considered at each node of the RF trees, by setting the \textit{max features} hyperparameter equal to the square root of the total number of parameters in the set (what we label as the `$\sqrt{N_{\text{features}}}$' case).
%and 'forcing' the RF to explore other parameters than only the most important ones.
Although, in this second approach, the correlations between parameters are not fully accounted for in computing the feature importance ranking, this analysis provides an insightful estimate of which parameters perform better in case the most important one(s) is(are) not available. 
The results of this further classification analysis are shown in Fig.~\ref{fig:Nii_RF_class} by the empty, hatched bar chart.
The algorithm is now forced to take into consideration also different features than the most important ones, spreading the final relative importance among a larger number of parameters; in fact, log(M$^{*}$) and $\sigma_{\text{H}\upalpha}$ are now ranked higher than $\upDelta$log(EW[H$\alpha$]) and  $\upDelta$log(D$_{\text{N}}$(4000)). 
Nevertheless, the RF still robustly identifies \delNO\ as the most important parameter, which corroborates the interpretation of variations in the N/O abundance relative to the median SF sequence as the primary physical driver of the scatter in the [\ion{N}{ii}]-BPT. 
A result fully consistent with such interpretation is also recovered when considering \delNOii\ in place of \delNO, and is presented in Appendix~\ref{sec:appendix_A}.

\subsubsection{Regression}
For the regression problem, we design a very similar random forest structure as for the classification task, and only change the loss-function to the \textit{mean squared error}. 
The results of the RF regression analysis are shown, for both the `All features' and `$\sqrt{N_{\text{features}}}$' cases described in the previous section, in the bottom panel of Fig.~\ref{fig:Nii_RF_class}.
Overall, the performance of the RF in predicting the exact distance from the SF sequence is comparable to that achieved by the ANN, with a median and standard deviation of the residuals of $0.0002$ and $0.039$, respectively.
The parameters' ranking closely traces what seen already for the classification problem, with \delNO\ being, by far, the most important parameter and retaining more than $80\%$ of the total predictive power, which increases to $>90\%$ when $\upDelta$log(D$_{\text{N}}$(4000)) and $\upDelta$log(EW[H$\alpha$]) are used in conjunction.
This means that in principle, modulo the assumptions discussed in Section~\ref{sec:frame} and within the residual uncertainties, almost no further information is needed to quantify the magnitude of the offset from the SF locus which a galaxy resides at in the [\ion{N}{ii}]-BPT diagram (we recall here that we are implicitly assuming these variations to occur at fixed metallicity).
Finally, similar to that discussed before, we note that when considering only the square root of the number of features at each node, the relative importance of parameters that are closely connected to \delNO\ (and which can thus perform as good substitutes of such parameter) increases to a level that matches that of the second and third parameters in the ranking.

Summarising, from the joint ANN and RF analysis we infer that variations in the N/O abundance (associated here to \delNO) with respect to the average of galaxies along the SF locus are the primary drivers of the deviation of star forming galaxies from their median loci in the [\ion{N}{ii}]-BPT diagram, once the offset is considered orthogonal at any point to the best-fit line and at fixed oxygen abundance.
This result is further confirmed in case \delNOii\ is included (in place of \delNO) in the RF analysis, although in that case the relative importance of the other parameters is impacted. In fact, we stress again here that because the RF disentangles the relative importance of a set of features used \textit{in conjunction} with each other, changing even one parameter only within the set can have an impact on the relative importance retained by all of the remaining variables too (see also the discussion in section~\ref{sec:discuss_nii} and Appendix~\ref{sec:appendix_A}).
% We discuss the implications For more details, we refer to Appendix~\ref{sec:appendix_A} and to the discussion of section~\ref{sec:discuss_nii}. 

\subsection{Does the relative parameter importance change across the diagram ?}
\label{sec:nii_relative_importance}

\begin{figure}

\includegraphics[width=0.97\columnwidth]{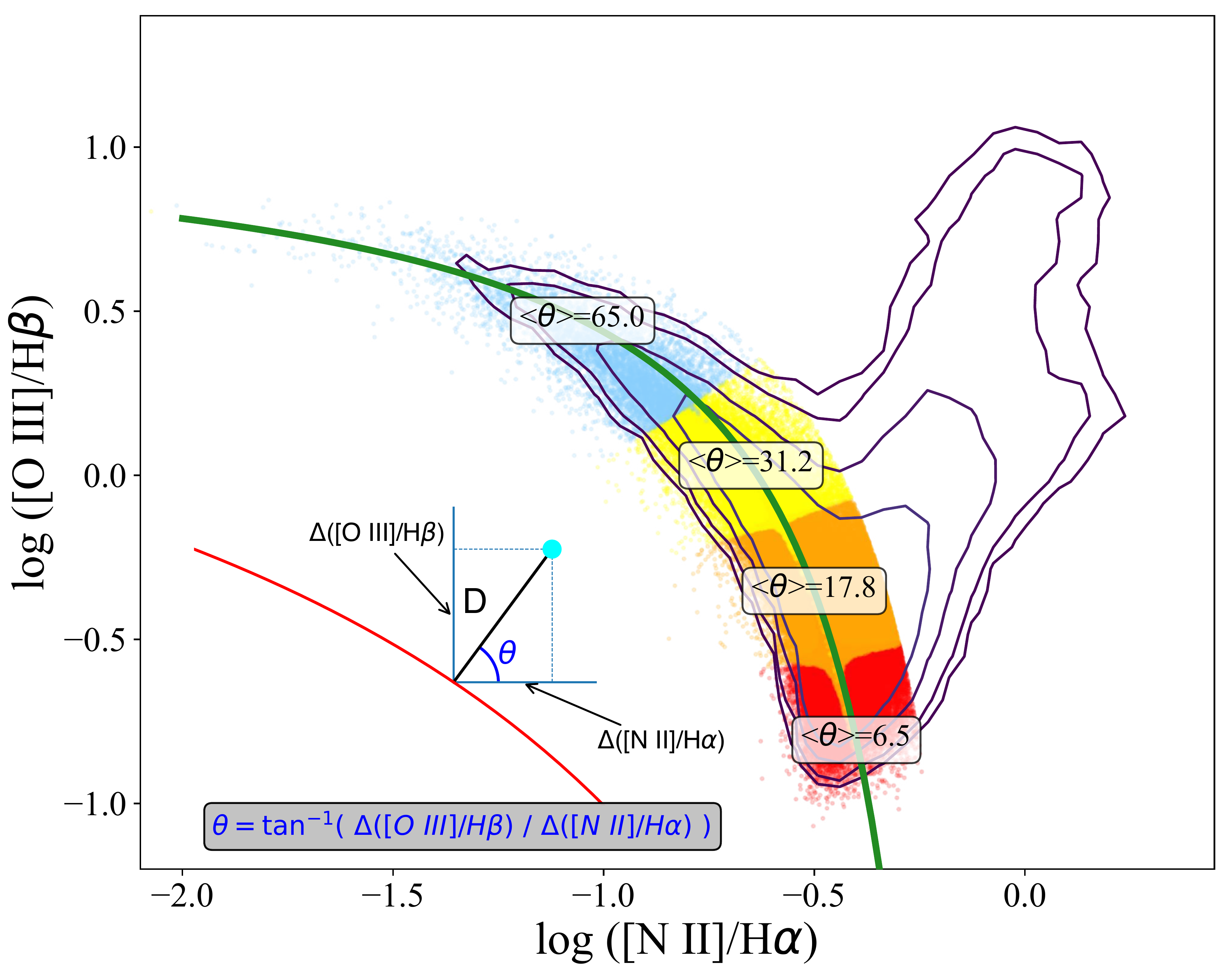}

\caption{The distribution of star-forming galaxies in the [\ion{N}{ii}]-BPT diagram is divided in four sectors,  defined by different intervals in the <$\theta$> angle as formed by the `offset vector' of each galaxy with the horizontal axis. In this way, we aim to study whether and how the results from the ML analysis presented in Section~\ref{sec:Nii_ANN} and ~\ref{sec:Nii_RF} vary as a function of the position of galaxies along the SF sequence.}
\label{fig:Nii_sectors_binned}

\end{figure}

In the previous sections, we have analysed the connection between the scatter of galaxies in the [\ion{N}{ii}]-BPT and different physical parameters, and found \delNO\ as the most predictive parameter of both the direction (classification task) and amplitude (regression task) of the offset vector from the best-fit curve of the SF sequence.
However, the distribution of star-forming galaxies in the diagram is not homogeneous, with the highest density of galaxies concentrated in the high-metallicity, bottom-right region, where the offset vector is primarily directed along \niiha.
As shown already in Fig.~\ref{fig:Nii_bpt_metrics} in fact, the relative strength of the two components of an orthogonal `offset vector' changes as we move along the sequence of star-forming galaxies in the diagram.
This effect can be parametrised in terms of the arctangent of the angle (positive counterclockwise) formed by the `offset vector' with the x-axis, and indicated with $\theta$ in equation~\ref{eq:theta_angle}: moving from the bottom-right to the upper-left region of the sequence $\theta$ increases, and so it does the relative strength of the offset along \oiiihb\ compared to that along \niiha.
Therefore, it is worth asking if either the individual absolute performance and/or the relative importance of the various parameters involved in the ML analysis changes, as a function of the location considered along the star-forming sequence.

For this reason, in this Section we repeat the analysis presented in Section~\ref{sec:Nii_ANN} and ~\ref{sec:Nii_RF} by splitting the [\ion{N}{ii}]-BPT diagram in four \textit{sectors}, defined on the basis of the different inclinations of the offset vector with respect to the horizontal axis, i.e., of different intervals spanned by the $\theta$ angle. 
The choice of the number and `width' of these sectors is empirical, and driven by the aim, on the one hand, to obtain a segregation of the diagram as homogeneous as possible (i.e., to avoid having sectors spanning too different ranges in $\theta$), while on the other, to have a minimum reasonable number of galaxies within each sector in order to perform a meaningful statistical analysis. 

\begin{figure*}

 \includegraphics[width=0.98\columnwidth]{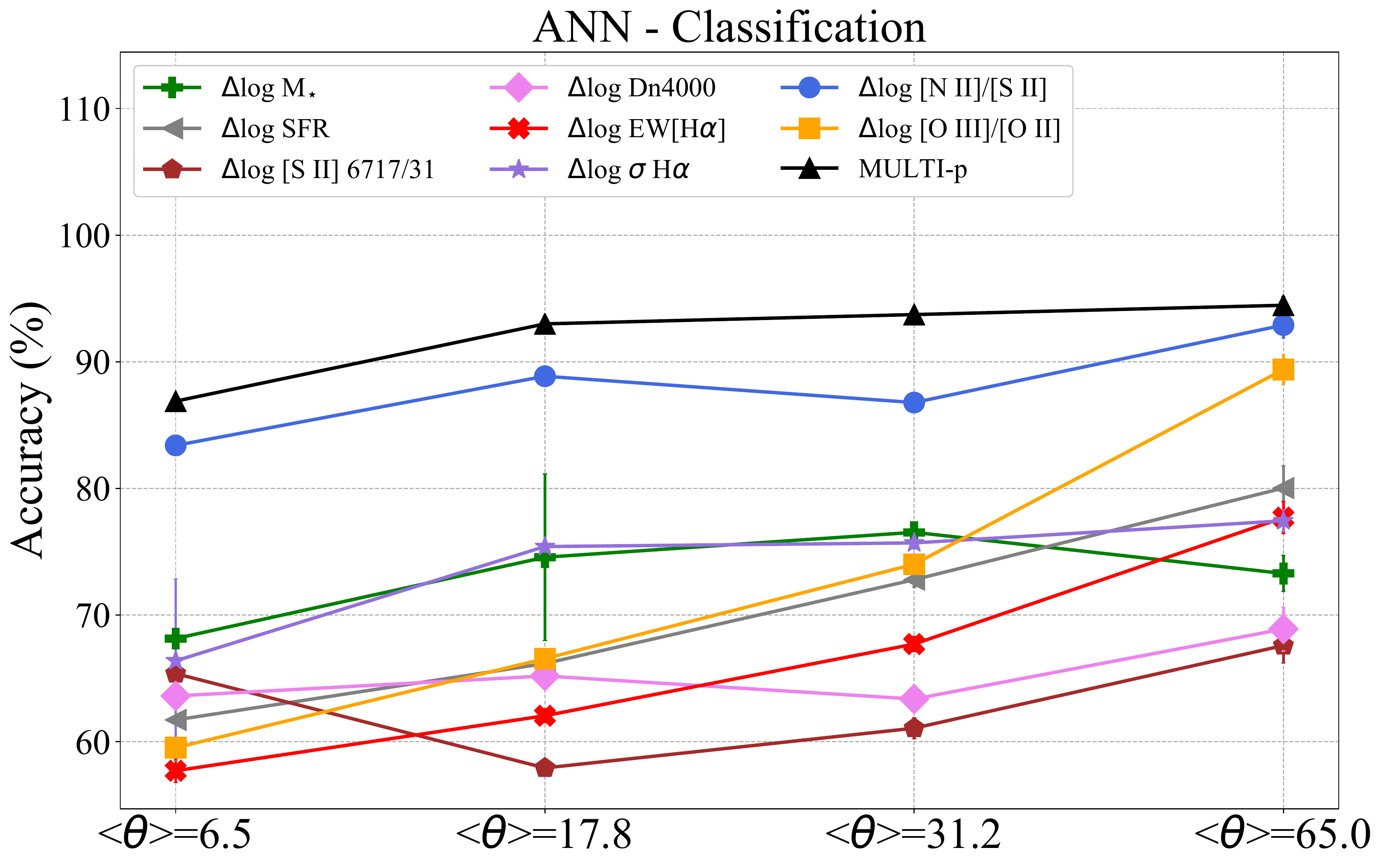}
 \hspace{0.2cm}
 \includegraphics[width=0.98\columnwidth]{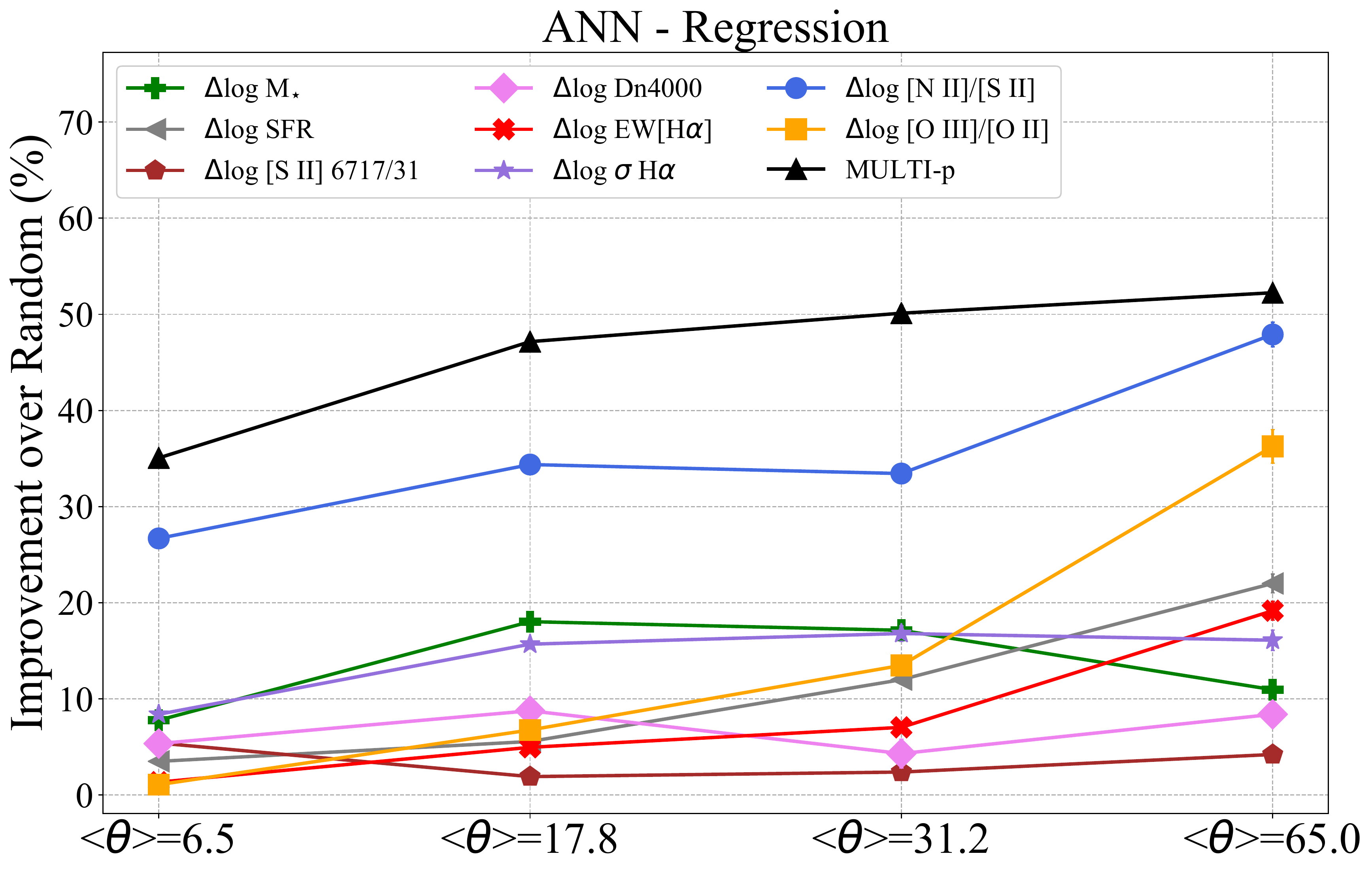}
 
\vspace{0.2cm}
\includegraphics[width=0.98\columnwidth]{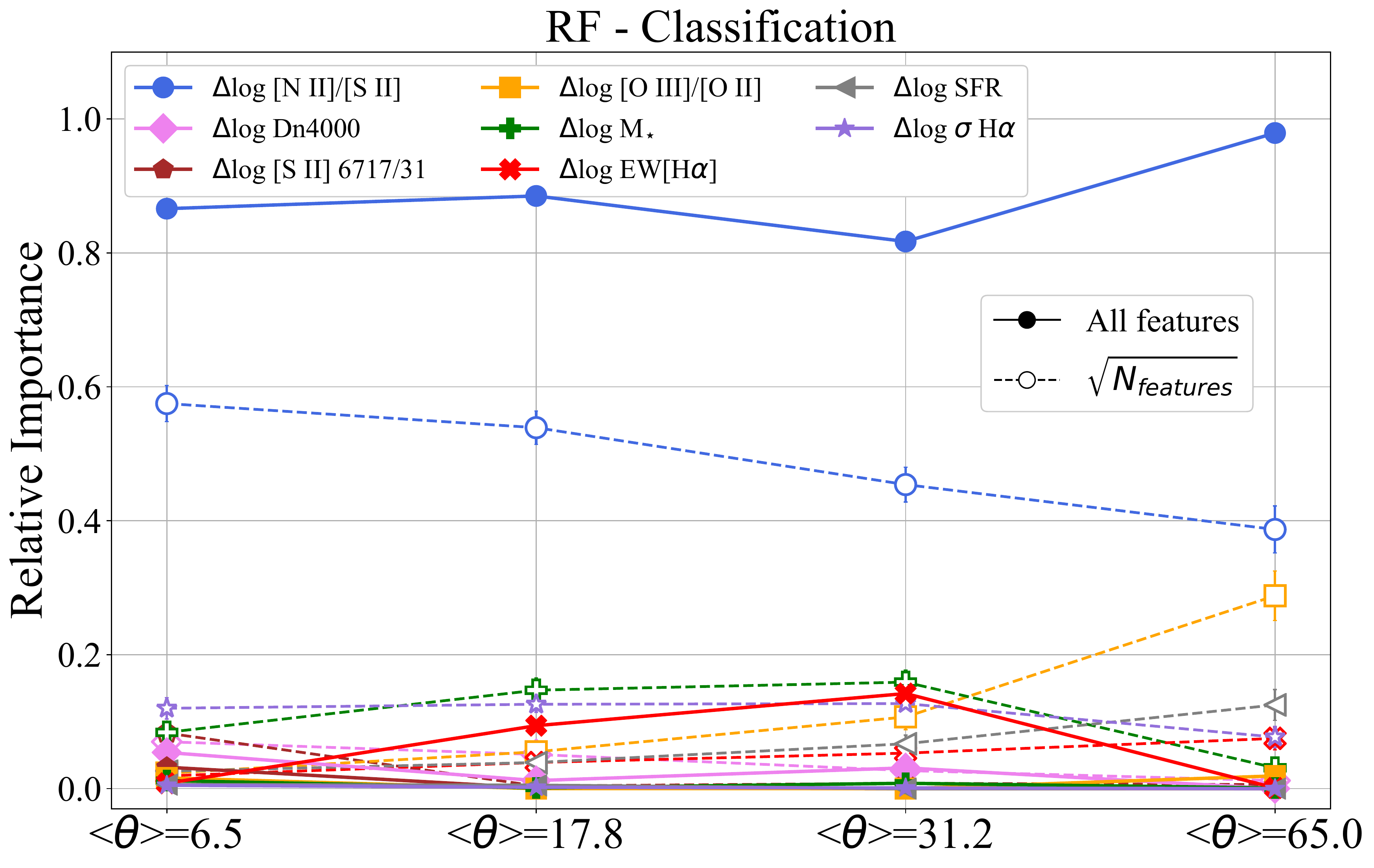}
\hspace{0.2cm}
 \includegraphics[width=0.98\columnwidth]{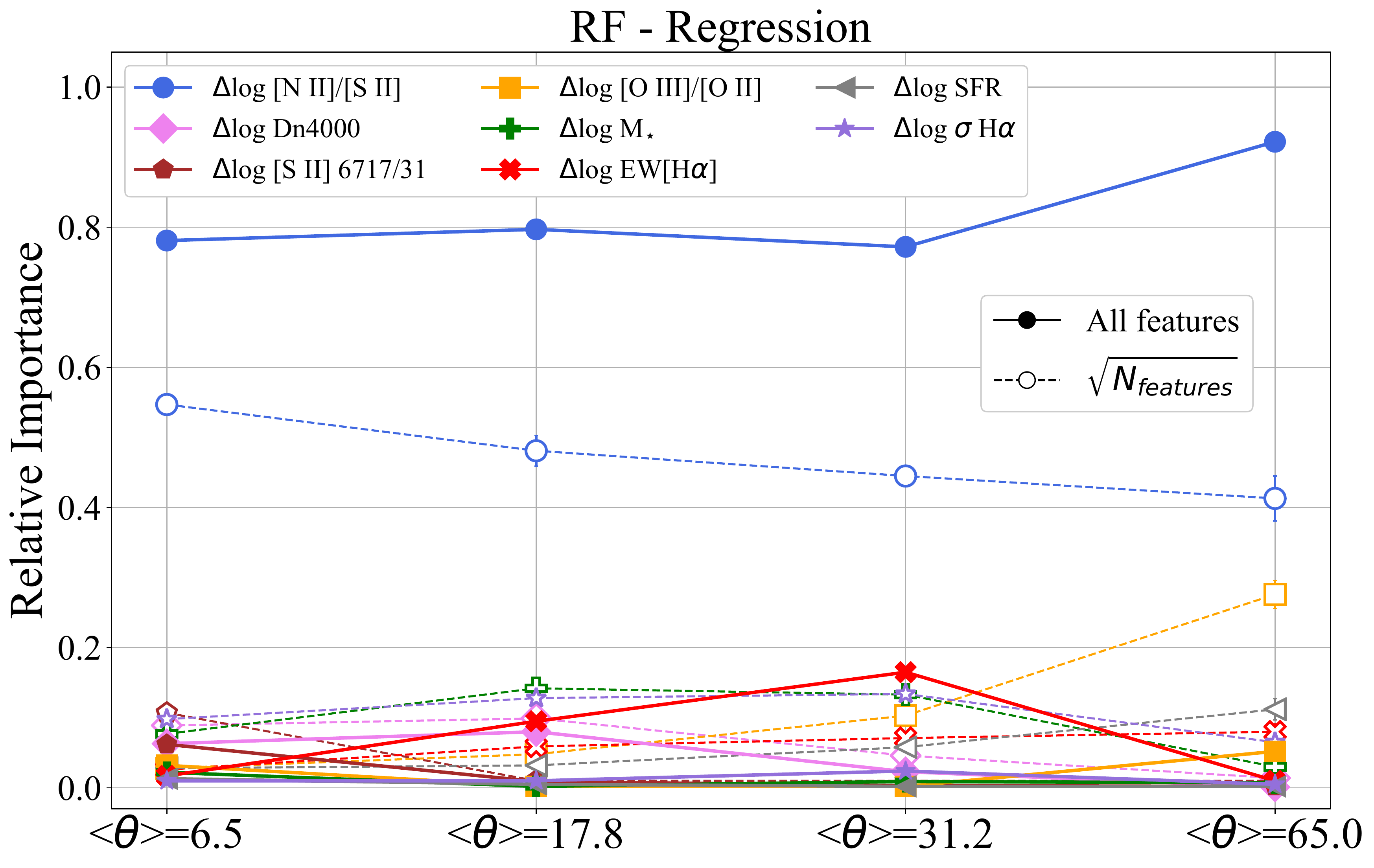}
 
\caption{\textit{Upper panels:} Results for the ANN classification (left) and regression (right) analysis for each of the four sectors in which the [\ion{N}{ii}]-BPT diagram has been divided (see Fig.~\ref{fig:Nii_sectors_binned}). The different tracks follow the accuracy and IoR as a function of the average angle $<\theta>$ of the `offset vector' in each region, and for each of the parameters of interest (colour-coded as in the legend). When trained with the `multi-parameter' set, the network achieves excellent accuracy and IoR across the entire diagram, with \delNO\ achieving the best individual performance in all sectors. \ \textit{Bottom panels:} Same as the upper panels, but for the RF analysis. The relative importance of the various features is now reported on the y-axis and plotted as a function of the average $<\theta>$ of each sector: straight lines are representative of the RF analysis with all features allowed at each node, whereas dashed lines are for the $\sqrt{\text{N}_{\text{features}}}$ case. For both regression and classification problems, \delNO\ is found as the most relevant parameter across the entire diagram.}
\label{fig:Nii_theta_comparison}

\end{figure*}

The partition of the \niibpt diagram in four sectors is graphically represented in Fig.~\ref{fig:Nii_sectors_binned}. %and their main properties are reported in Table XXX.
Because of the different numbers of galaxies within each sector (with the bottom-right ones, i.e. at low $<\theta>$ values, being much more populated than the others), the input values for the hyper-parameters of both ANN and RF models are tuned to adapt to the varying sampling, especially in the upper-left sector <$\theta \sim 65\degree$> where the total number of sources falls below $10,000$.
For instance, for the ANN analysis of individual parameters in such sector the sample is more unevenly split ($80$-$20$ per-cent) in training and test galaxies (in order to feed the model with still a sufficiently large number of galaxies for training), the \textit{batch size} of the stochastic gradient descent algorithm is set to one-third of the training sample size, and the model is trained for $300$ Epochs instead of $100$. For the `multi-parameter' run instead, the batch size is fixed to $8$. We have tested that tweaking the hyper-parameters of the network in this way allows us to maintain a reasonable balance between performances and overfitting, the latter becoming of increasing concern especially in small datasets.
% Similarly, in the RF analysis of sectors four and five the minimum number of samples allowed at a leaf node is decreased to $100$. 

The results of the ANN analysis for the four different sectors are reported and compared in the upper panels of Fig.~\ref{fig:Nii_theta_comparison}, where the classification \textit{Accuracy} and the \textit{IoR} in regression are plotted, for both individual features and the `multi-parameter' set, as a function of the average $<\theta>$ of each of the regions in which the diagram has been divided into.
In each sector and for each parameters set, $30$ independent ANN runs are performed and the average (and standard deviation) of the performances are evaluated.

As a first remarkable result, we find the performances of the network to be quite stable across the entire diagram, scoring $\gtrsim 90$ per-cent accuracy in classification and $\gtrsim 40$ per-cent IoR in regression in all sectors.
In terms of performances of the individual parameters, those associated to the chemo-dynamical state of the galaxy (M$_{\star}$, N/O, \sigha) score the largest accuracy and IoR in the first three sectors, whereas the performances of parameters associated with star formation and ionisation conditions (like [\ion{O}{iii}]/[\ion{O}{ii}], SFR, EW[H$\upalpha$]) increases as moving towards the upper-left part of the diagram, becoming almost dominant in the top-left sector.
Nonetheless, \delNO\ maintains a stable level of performance across the entire diagram, scoring the highest accuracy and IoR everywhere, whereas overall very weak dependency exists, for instance, between the scatter of galaxies in the [\ion{N}{ii}]-BPT and variations in electron density (traced by $\upDelta$[\ion{S}{ii}]$\lambda 6717/31$) in all sectors but the first one, where this parameter matches the individual performances of M$_{\star}$ and \sigha.

The Random Forest analysis of the four independent sectors is reported instead in the bottom panels of Fig.~\ref{fig:Nii_theta_comparison}, where the relative importance of the various features is plotted as a function of the average <$\theta$> spanned by each region.
Deviations in the \ion{N}{ii}/\ion{S}{ii} parameter (tracing primarily variations in the N/O abundance) are by far the most relevant quantity to consider (for both the classification and the regression tasks) throughout all the sectors of the diagram, especially when all features are considered at each node (solid lines), and hence the RF truly exposes the parameter which is intrinsically most connected to the target label.
Interestingly, variations in EW(H$\upalpha$) (i.e., tracing variations in the sSFR)
gain a significant $\sim 20 \%$ relative importance in the central sectors. 
In case only $\sqrt{N_{\text{features}}}$ are considered at each splitting-node instead (dashed lines), stellar mass, $\sigma_{\text{H}\alpha}$ and density are found as the most useful alternative parameters to \delNO\ in the first sector, while EW[H$\upalpha$], SFR and \delU\ overcome them in the two uppermost regions. 

% Summarising, \delNO is robustly identified as the most predictive and most relevant parameter in both classification and regression tasks across the entire diagram, regardless of the average inclination of the offset vector (and thus, regardless of the strength of its two components along \niiha and \oiiihb).

\subsection{Discussion}
\label{sec:discuss_nii}

The analysis presented in the previous sections unambiguously suggests that relative variations in parameters mainly tracing the nitrogen-over-oxygen abundance are the primary physical driver of the deviation from the median locus of star-forming galaxies in the \niibpt.
In fact, \delNO\ (or, equivalently, \delNOii) is robustly identified as the most predictive (individual) parameter and the most relevant feature (among the `multi-parameter' set) in both classification and regression tasks, for either the global analysis of the sample and within separated regions across the diagram, regardless of the average inclination of the offset vector (and thus, regardless of the strength of its two components along \niiha\ and \oiiihb).

\begin{figure}
    \centering
    \includegraphics[width=0.97\columnwidth]{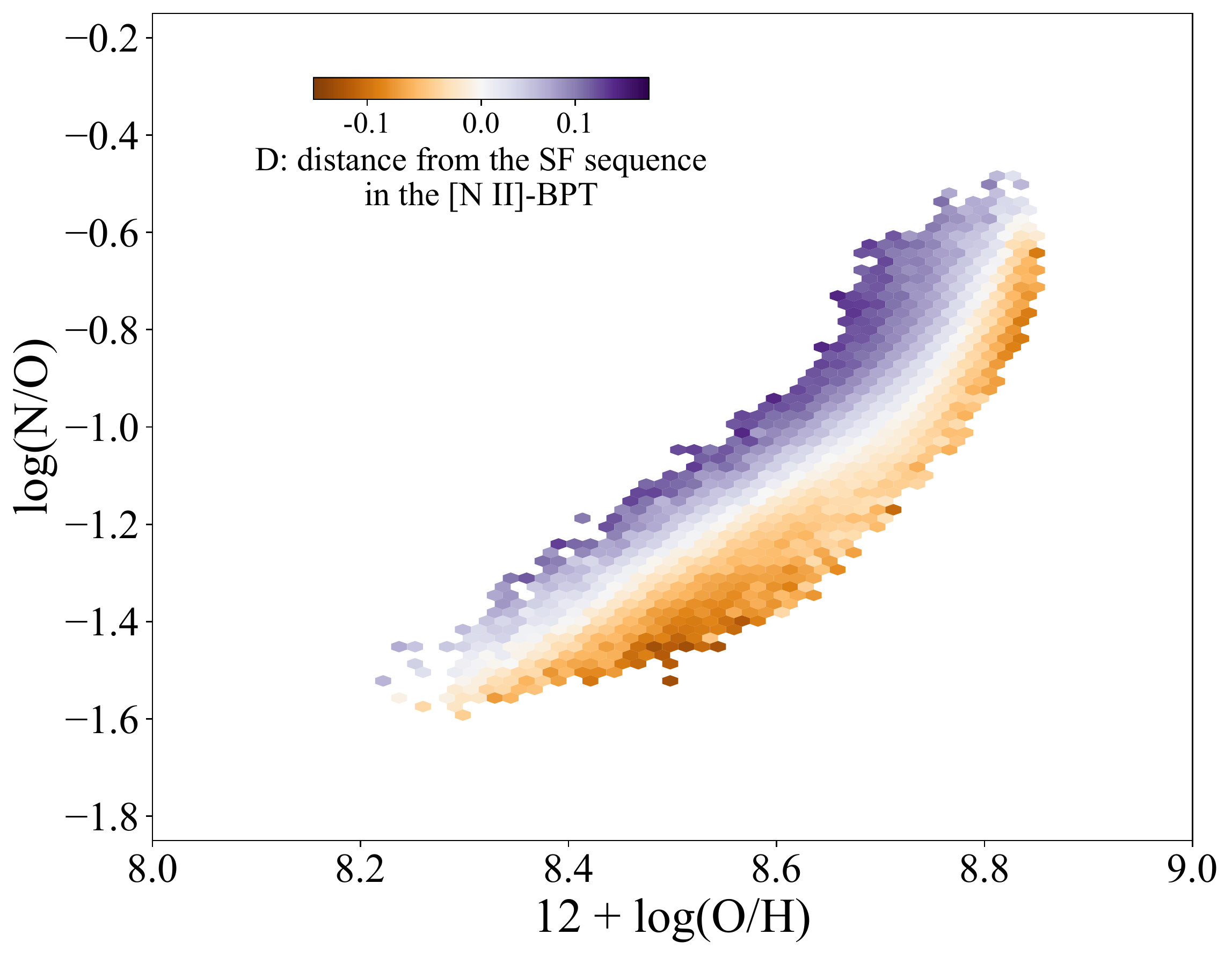}
    \caption{The relationship between N/O and O/H for local SDSS star-forming galaxies is colour-coded by the magnitude \textbf{D} of the offset vector from the SF sequence of the \niibpt diagram. A clear segregation in \textbf{D} is seen in N/O, at fixed metallicity.}
    \label{fig:NO_OH_dist}
\end{figure}

If we recall the tight, monotonic dependence of the position of galaxies along the SF sequence in the diagram with metallicity (as outlined in Section~\ref{sec:nii_params_sequence}), we can interpret our global results of Fig.\ref{fig:Nii_ANN_classification} and~\ref{fig:Nii_RF_class} as a manifestation of the existence of an O/H vs N/O relation for SDSS star-forming galaxies, whose intrinsic scatter is reflected and, to some extent, translated into the observed distribution of emission line ratios within the \niibpt.
%Low- and intermediate-mass (LIM) stars, with mass in the range 4<m<8M⊙, are dominant stellar nucleosynthesis sources of N, when experiencing the AGB phase
A tight relationship between O/H and N/O abundances is indeed observed in both \Hii regions and local galaxies, especially at M$^{\star}\gtrsim 10^{9.5}$M$_{\odot}$ \citep[][]{vila_costas_nitrogen-ratio_1993,van_zee_spectroscopy_1998,perez-montero_impact_2009, pilyugin_abundance_2012, andrews_mass-metallicity_2013, hayden_pawson_NO_klever_2021_arxiv}, and it is set by the predominant nucleosynthetic origin of nitrogen from CNO burning of pre-existing stellar carbon and oxygen in low- and intermediate-mass stars experiencing the AGB phase (i.e., the `secondary' nitrogen production mechanism, \citealt[][]{kobayashi_isotopes_MW_2011,ventura_agb_yields_2013, vincenzo_nitrogen_2016}); alternatively, \cite{vincenzo_extragalactic_2018} reproduced the observed N/O–O/H relation introducing failed supernovae (SNe) in massive stars within their cosmological simulations.
Recently, such relationship between O/H and N/O has been suggested as even tighter than the one between M$^{\star}$ and N/O \citep{hayden_pawson_NO_klever_2021_arxiv}, in contrast to what claimed by previous studies \citep[e.g.,][]{andrews_mass-metallicity_2013, masters_tight_2016}.
In light of our results, this would confirm that deviations in N/O at fixed O/H are more likely to be related to the offset from the SF sequence in the \niibpt\ than relative variations in M$^{\star}$, although the two are clearly physically correlated.
The connection between the two diagrams is also readily evident if we look at the distribution of our galaxy sample in the N/O vs O/H diagram, as shown in Fig.~\ref{fig:NO_OH_dist} (where \nii/\oii\ is converted to N/O following the Te-based calibrations presented in \cite{hayden_pawson_NO_klever_2021_arxiv}); here, each hexagonal bin is colour-coded by the average distance \textbf{D} of galaxies from the best-fit line of the \niibpt, almost perfectly tracing the scatter around the median N/O vs O/H relation.

In general, and especially for galaxies located in the bottom-right, high-metallicity region of the diagram (the majority of the sample, with $\sim 70$ per-cent of them characterised by $\theta<22$\degree), variations in the parameters which mainly trace N/O are associated to galaxies of different stellar masses, and can be thus interpreted as age-related effects: galaxies with higher M$^{\star}$, in fact, are more chemically mature than their lower mass counterparts (i.e., those located along iso-O/H lines and with negative \textbf{D} values), in the sense that they had more time to enrich the ISM with nitrogen produced by low- and intermediate-mass stars on longer timescales.
Hence, to a positive $\upDelta$log(M$^{\star}$) corresponds a positive $\upDelta$log(N/O) (and viceversa, with relatively lower mass galaxies at fixed O/H which still have nitrogen partly locked in stars), producing the offset in the \niibpt. Indeed, $\upDelta$log(M$_{\star}$) here acts as a good proxy for \delNO\ in the ML analysis, reaching high scores in the ANN runs whilst scoring almost zero importance in the RF, but subtracting nonetheless $\sim10$ per-cent of relative importance from \delNO\ in the $\sqrt{\text{N}_{\text{features}}}$ case.
Another important aspect that could affect the N/O abundance of the gas-phase at fixed metallicity (and which however is not directly constrained here observationally) is a variation in the dust-to-metal ratio, given the different dust depletion properties of oxygen and nitrogen \citep{gutkin_modelling_2016, hirschmann_synthetic_2017}.

Interestingly, a small but significant ($\sim$13 per-cent) part of the global feature importance is retained by $\upDelta$log(
D$_{\text{N}}$(4000)) ($\sim$8 per-cent) and $\upDelta$log(EW(H$\upalpha$)) (another $\sim$5 per-cent). 
We interpret this as an additional contribution to the offset which is still associated to galaxy ageing, but that it is complementary to the information already provided by \delNO.
In particular, we associate it to a differential impact of older stellar populations (e.g., hot, post-AGB stars, which dominate ionising photon production after $\sim0.1$Gyr), boosting intermediate- and low-ionisation emission lines from a more diffuse ionised gas (DIG) and setting the relative distance of these galaxies from the `composite' and `LI(N)ER' area of the diagnostic diagram  \citep[e.g.,][]{zhang_sdss-iv_2017, byler_lier_pagb_2019}.
Such an effect is accounted for in particular by $\upDelta$log(
D$_{\text{N}}$(4000)) in the `first' sector of the \niibpt, whereas it is more prominently seen in $\upDelta$log(EW(H$\upalpha$)) in the `central' ones: in both cases, galaxies offset above the sequence are characterised by signatures of relatively older stellar populations (higher D$_{\text{N}}$(4000) and lower \ewha) compared to on-sequence galaxies, and viceversa (see also Fig.~\ref{fig:nii_delta_prop}).

Moving upwards along the SF sequence, parameters related to star formation and ionisation state of the gas score progressively higher accuracies in the ANN, although variations in N/O tracers are still identified as the primary driver of the scatter, with $\sim92$ per-cent of relative importance scored in the RF.
Although we acknowledge that the small number of galaxies in this sector ($< 10,000$) might impact the ability of the RF of correctly retrieving the exact relative importance of each of the parameters, nonetheless the overwhelming success of \delNO\ in a region where the offset vector has a strong component also along \oiiihb\ prompts us some reflections.
In particular, one alternative interpretation involve breaking our initial assumption that the offset occur along iso-metallicity lines (i.e., at fixed O/H): as can be seen in Fig.~\ref{fig:nii_delta_prop} in fact, variations in metallicity across the SF sequence, although very mild (i.e., of the order of $<0.05$ dex in $\upDelta$ log(O/H)), are present in such region of the diagram, and are opposite to variations in N/O tracers. 
Therefore, the connection between the offset from the median sequence and relative variations in N/O tracers here could just, at least partially, reflect metallicity variations. 
A decrement in metallicity coupled with an increase in N/O could be explained, in fact, by invoking the presence of differential outflows (i.e., preferentially removing oxygen from the ISM) from relatively younger, low metallicity galaxies with prominent star-formation \citep[e.g.,][]{vincenzo_nitrogen_2016,magrini_gaia_eso_NO_2018}. Interestingly then, in the $\sqrt{\text{N}_{\text{features}}}$ realisation of the random forest large part of the \delNO\ importance is taken by both \delU\ and \delsfr.  
The former variable mainly traces the ionisation parameter, which has a strong dependence on the stellar metallicity (hence, indirectly on the gas abundances), whereas the latter could be tracing indeed the differential impact of star-formation driven outflows in this galaxy population.
Moreover, the \oiii/\oii\ ratio itself is known to have an intrinsic, although secondary, dependence on metallicity \citep[][]{kewley_using_2002}.
Further analysis based on large samples of galaxies with independent and `direct' metallicity estimates in such region of the \niibpt\ could certainly help to either confirm or deny such interpretation.
Another possibility invokes variations in the hardness of the ionising radiation (for which we do no have direct and independent observational constrains) at fixed metallicity and N/O, impacting the ionisation parameter and the relative strength of the N$^{+}$ and S$^{+}$ emission lines, hence the [\ion{N}{ii}]/[\ion{S}{ii}] ratio (beside [\ion{O}{iii}]/[\ion{O}{ii}]), as predicted by various photoionisation models \citep[e.g.,][]{byler_nebular_2017}.

Finally, we note that variations in \sigha, although performing overall well in the ANN analysis, picks basically no relative feature importance at all in the RF anywhere across the diagram. 
Its performance, indeed, closely follows the trend seen for \mstar, and this suggests that any information carried by \sigha, likely tracing the dynamical mass of the system \citep[][]{green_dynamo_2014,krumholz_discs_2018}, is already embedded into the \mstar\ parameter (and/or other parameters, like SFR, \citealt{yu_vel_disp_manga_2019,varidel_turbulence_2020}) within the population of star-forming galaxies.
% The central velocity dispersion of the gas instead is thought to correlate with stellar mass, via relationship with the gas mass and the dynamical state of a galaxy, and SFR, via energy injected into the ISM by supernovae feedback \citep[][]{green_dynamo_2014,krumholz_discs_2018,yu_vel_disp_manga_2019,varidel_turbulence_2020}; however, it also represents a potential tracer of shock-heated gas, and has been already suggested as a feasible parameter to extend and complement the ionisation source classification scheme of the BPT diagrams \citep{dagostino_bpt_2019, law_bpt_2021}. 
However, we also note that, if \delNOii\ is adopted instead of \delNO\ to trace variations in the N/O abundance, then \sigha\ retains a non negligible amount of (complementary) relative importance in the RF analysis. We interpret this as a signature of a correlation between \nii/\sii\ and \sigha, at fixed \nii/\oii; these results are discussed more in detail in Appendix~\ref{sec:appendix_A}.
As a final remark, we also note that the relatively small dynamical range of \sigha\ across the star-forming galaxy populations (whose gas emission lines originates from kinematically `cold' \Hii regions with typical velocity dispersions of $\sim30$km s$^{-1}$), coupled with the intrinsic spectral resolution of the SDSS-II spectrograph of $\sim70$km s$^{-1}$, also make any inference based on central \sigha\ measurement more challenging to physically interpret. Exploiting the improved calibration of the instrumental response for the MaNGA spectrograph \citep{law_manga_LSF_2021}, together with the information provided on spatially resolved scales, in a forthcoming paper we aim at revisiting the significance of our ML results on gas velocity dispersion measurements in star-forming galaxies.

\section{The [\ion{S}{ii}]-BPT diagram}
\label{sec:S2_ML}

\subsection{Parameters and metrics}

We now shift our focus on the [\ion{S}{ii}]-BPT diagram, for which we perform an analysis similar to that previously described for the [\ion{N}{ii}]-BPT. 
Compared to the [\ion{N}{ii}]-BPT, in the \siibpt\ 
two separate branches can be seen departing from the star-forming galaxy abundance sequence, one connected to the Seyfert region, while the other one connected to the region where LI(N)ERs are located. 
%The two branches are not observed on the BPT diagram, because the [Nii]/Hα ratio does not provide sufficient separation between the two branches, which overlap.
Moreover, the distribution of star-forming galaxies in the  diagram appears less tight than in the [\ion{N}{ii}]-BPT, with a larger scatter, especially in the \siiha\ line ratio at fixed \oiiihb. 
It can be also be seen, for instance, that galaxies which are located on (or very close to) the best-fit line in the [\ion{N}{ii}]-BPT, i.e. which have by construction \textbf{D}$\sim0$ according to equation~\ref{eq:nii_distance}, are instead more scattered across the best-fit line of the SF sequence in the [\ion{S}{ii}]-BPT (with a standard deviation of $0.07$ dex in \siiha\ at fixed \oiiihb).
This rather simple observation already suggests that the scatter in this diagram might be primarily associated with different physical mechanisms than in the \niibpt.

\begin{figure*}
    \centering
    \includegraphics[width=0.95\textwidth]{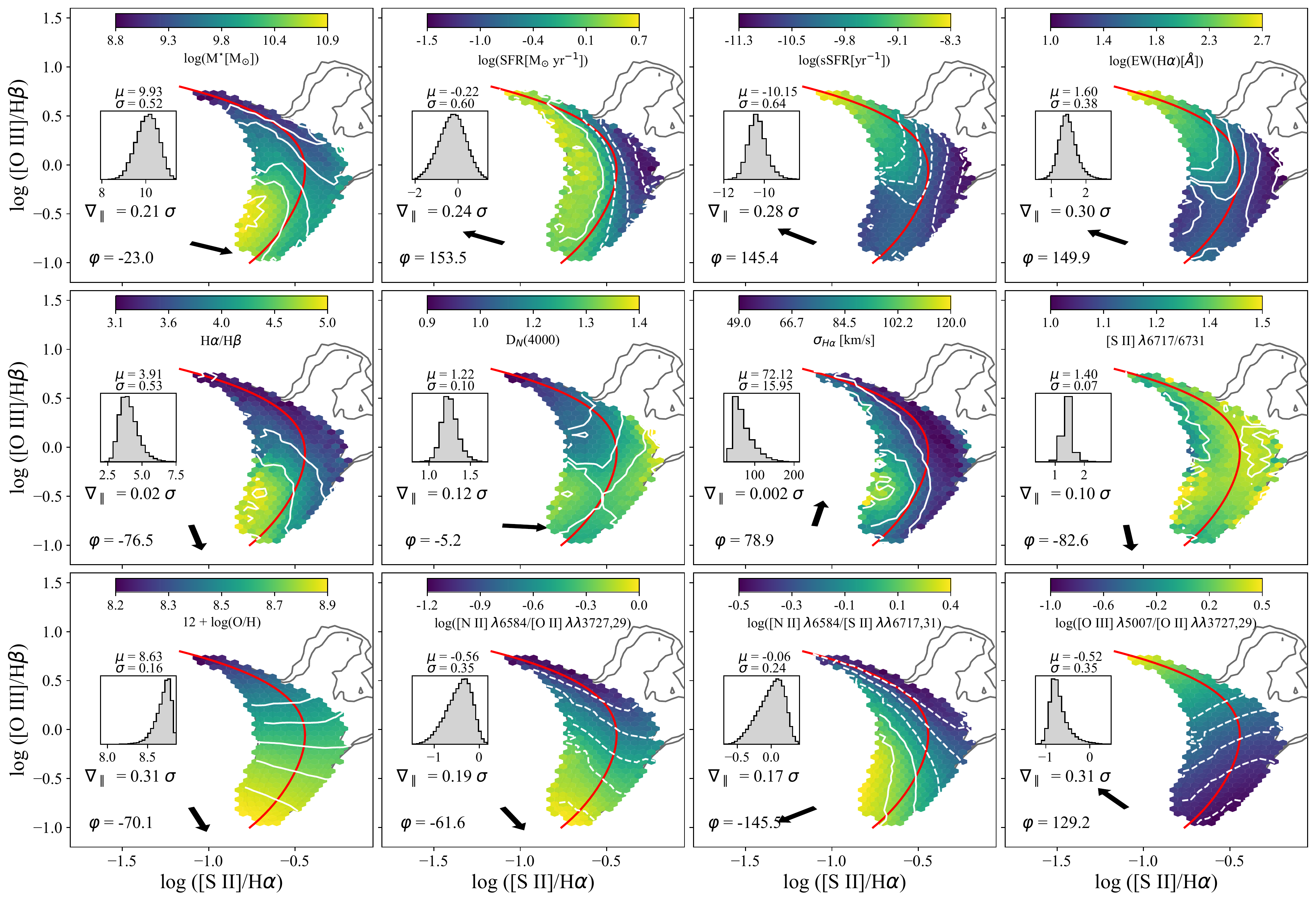}
    \caption{This figure is organised as in Fig.~\ref{fig:Nii_properties}, but for the [\ion{S}{ii}]-BPT diagram.}
    \label{fig:S2_bpt_properties}
\end{figure*}

\begin{figure*}
    \centering
    \includegraphics[width=0.95\textwidth]{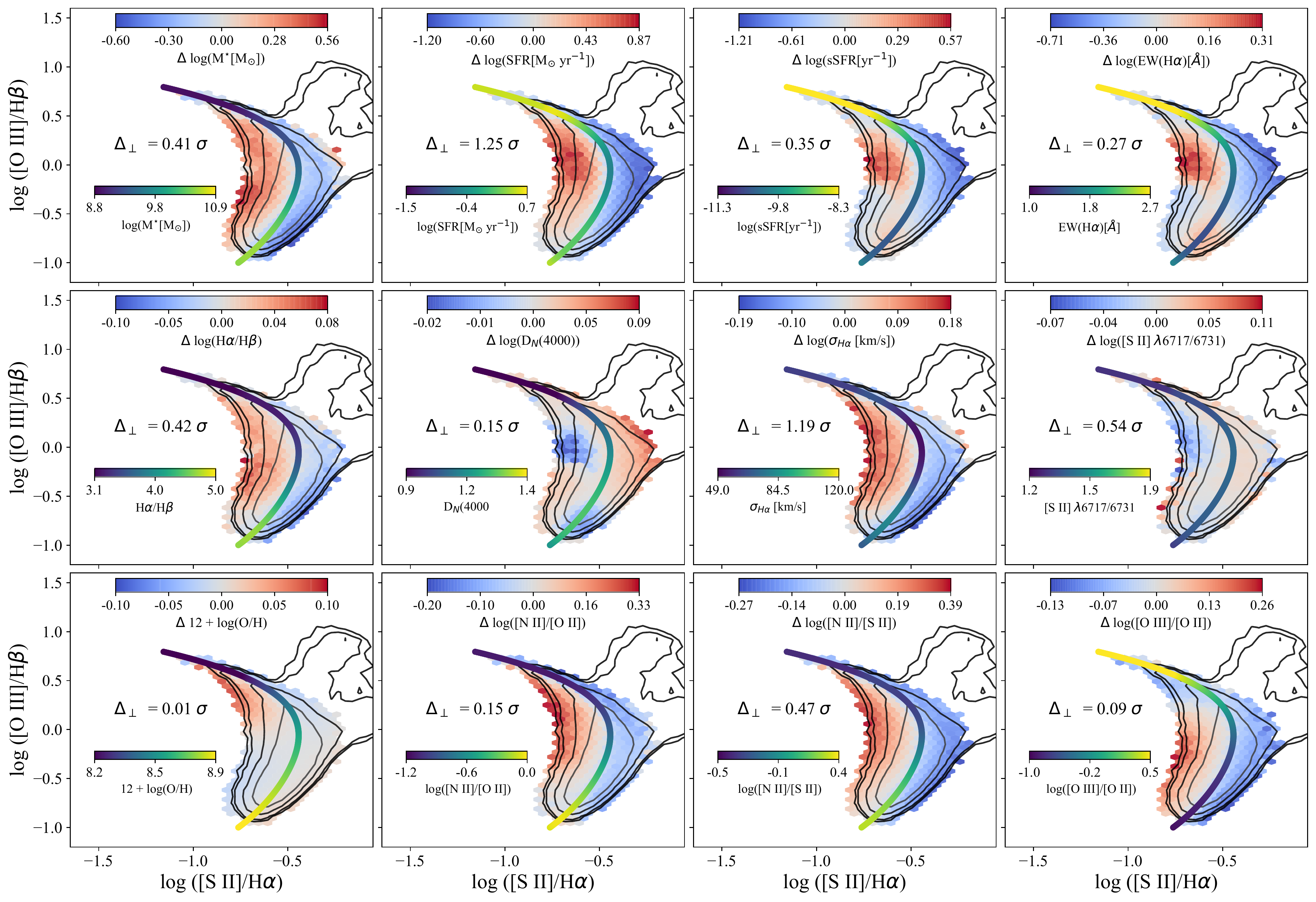}
    \caption{This figure is organised as in Fig.~\ref{fig:nii_delta_prop}, but for the [\ion{S}{ii}]-BPT diagram.}
    \label{fig:S2_bpt_delta_properties}
\end{figure*}

\begin{figure*}
    \centering
    \includegraphics[width=0.85\textwidth]{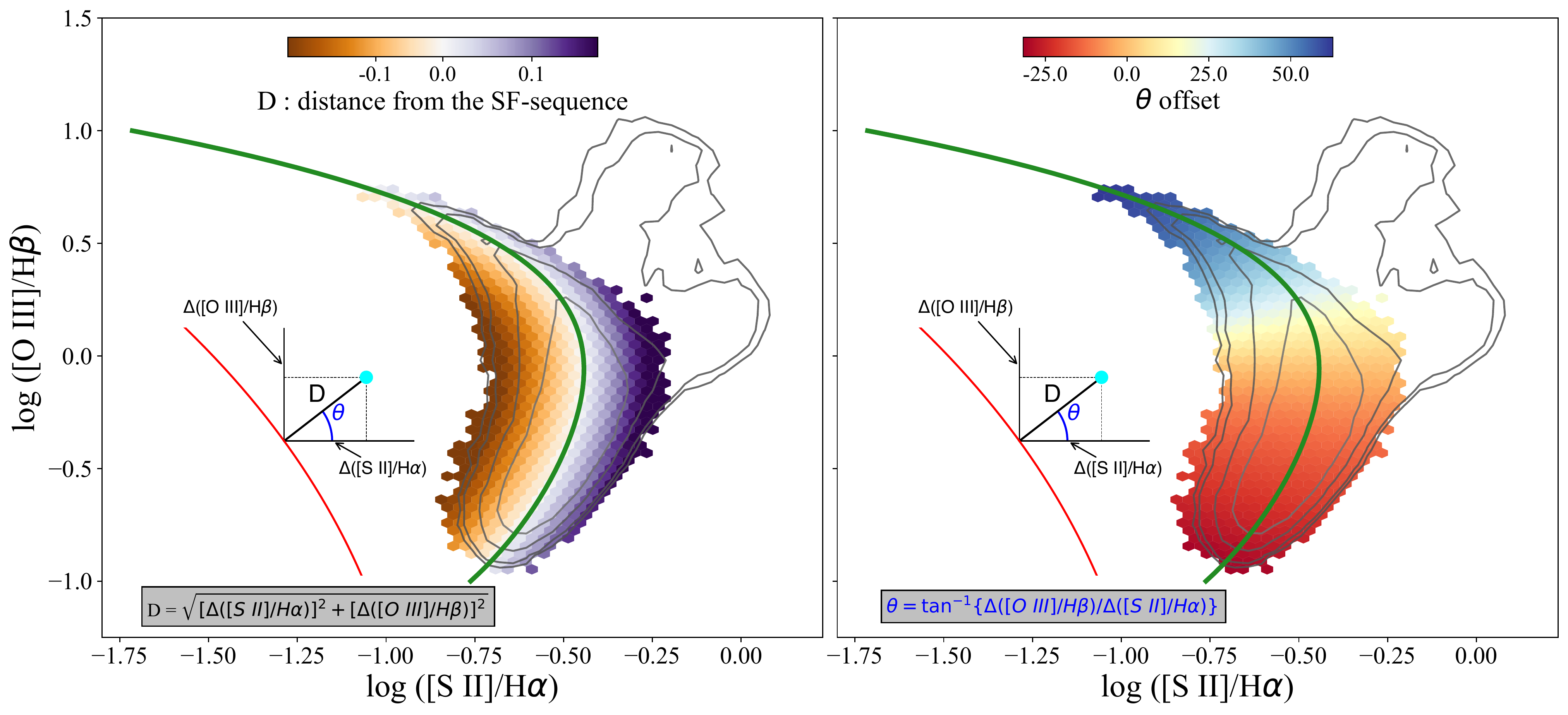}
    \caption{This figure is organised as in Fig.~\ref{fig:Nii_bpt_metrics}, but for the [\ion{S}{ii}]-BPT diagram.}
    \label{fig:S2_bpt_metrics}
\end{figure*}

The distribution of our set of parameters among the star-forming galaxy population within the [\ion{S}{ii}]-BPT diagram is shown in Fig.~\ref{fig:S2_bpt_properties}. Compared to what seen for the [\ion{N}{ii}]-BPT in Fig.~\ref{fig:Nii_properties}, there are a few remarkable differences.
Firstly, the best-fit line of the SF sequence is not monotonic, but it is double-valued in \siiha, presenting two distinct branches at high and low \oiiihb, with a turnover point around log(\oiiihb) $=0$.
We perform a fit to the median \oiiihb\ values in small bins of \siiha\ and provide a fourth-order polynomial representation of the best-fit line of the SF sequence in this diagram, as expressed by the following:
%	poly_coeff = np.array([-0.20545192, -0.41137273, -0.58825772, -0.06522904, -0.44463318])
%\sum_{k=1}^N k^2
\begin{equation}
\displaystyle    \text{log}([\ion{S}{ii}]/H\alpha) = \sum_{n=0}^4 p_{n} \cdot \text{log}([\ion{O}{iiii}]/H\beta)^n \,
%\displaystyle \sum_{i=0}^n a_i 
\end{equation}
where the coefficients p$_{n}$ are = [-0.20545, -0.41137, -0.58826, -0.06523, -0.44463] (from 0-th to 4-th order, respectively).

In terms of galaxy properties (and similar to the [\ion{N}{ii}]-BPT) the [\ion{S}{ii}]-BPT diagram is characterised by a strong sequence in metallicity and ionisation parameter, which can be both clearly visualised in Fig.~\ref{fig:S2_bpt_properties} and quantified by a $\nabla_{\parallel}$ statistics equal to $0.31 \sigma$; moreover, we note that, once again, lines of constant metallicity are almost perfectly orthogonal to the best-fit curve everywhere along the sequence.
% However, because of the double-valued behaviour of both \oiiihb\ and \siiha\ with oxygen abundance \citep[see e.g.,][]{curti_massmetallicity_2020}, 
%but are almost perfectly parallel to the x-axis, suggesting a weak dependence on \siiha\ at fixed \oiiihb\ (as also shown by the direction of the gradient vector, $\varphi=-69.5\degree$).
% Given the strong correlation between metallicity, ionisation parameter and N/O abundances, such quantities are seen to form clear sequences across the diagram too.
Interestingly, lines of constant star formation rate are almost parallel to the best-fit line of the SF sequence across the entire diagram, whereas significant variations are seen to occur when crossing the line.
This suggests a potential strong correlation between the offset from the SF locus and variations in the SFR, which can be also visualised in Fig.~\ref{fig:S2_bpt_delta_properties} (and quantified by a $\upDelta_{\perp} = 1.25 \sigma$), where, in a similar fashion as for Fig.~\ref{fig:nii_delta_prop}, we plot the logarithmic deviation in each parameter from the average value measured on the closest point along the best-fit line of the sequence (we refer to Section~\ref{sec:metrics} for details about how the $\upDelta$log(p) metric is computed). 
% The subsequent machine learning analysis will help assessing to what extent such trend in SFR is intrinsically connected with the observed offset in the diagram or whether it just follows from correlation with other parameters. 
In Table~\ref{tab:sii_parameters}, we summarise properties and statistics associated to each of the parameters of interest for the [\ion{S}{ii}]-BPT.

\begin{table*}
    \centering
    \begin{tabular}{lccccc}
        \hline
        \hline

Parameter & Physical property & $\nabla_{\parallel}$ & $\varphi$ & $\upDelta_{\perp}$ & multi-parameter set \\
\hline

log(M$^{\star}$[M$_{\odot}$]) & Stellar mass & $0.21\sigma$ & $-23.0\degree$ & $0.41\sigma$ & \cmark \\
log(SFR[M$_{\odot}$ yr$^{-1}$]) & Star formation rate & $0.24\sigma$ & $153.5\degree$ & $1.25\sigma$ & \cmark \\
log(sSFR[yr$^{-1}$]) & Specific SFR & $0.28\sigma$ & $145.4\degree$ & $0.35\sigma$ & \xmark \\
log(EW[H$\alpha$]) & Specific SFR & $0.3\sigma$ & $149.9\degree$ & $0.27\sigma$ & \cmark \\
H$\alpha$/H$\beta$ & Dust extinction & $0.02\sigma$ & $-76.5\degree$ & $0.42\sigma$ & \xmark \\
D$_{\text{N}}$(4000) & Age of stellar populations & $0.12\sigma$ & $-5.2\degree$ & $0.15\sigma$ & \cmark \\
$\sigma_{H\alpha}$ [km/s] & Gas velocity dispersion & $0.002\sigma$ & $78.9\degree$ & $1.19\sigma$ & \cmark \\
$[\ion{S}{ii}] \lambda 6717/6731$ & Gas density & $0.11\sigma$ & $-82.8\degree$ & $0.54\sigma$ & \cmark \\
12 + log(O/H) & Oxygen abundance & $0.31\sigma$ & $-70.1\degree$ & $0.01\sigma$ & \xmark \\
log([\ion{N}{ii}] $\lambda 6584$/[\ion{O}{ii}] $\lambda\lambda 3727,29$) & N/O abundance & $0.19\sigma$ & $-61.6\degree$ & $0.15\sigma$ & \cmark \\
log([\ion{N}{ii}] $\lambda 6584$/[\ion{S}{ii}] $\lambda\lambda 6717,31$) & N/O abundance & $0.17\sigma$ & $214.5\degree$ & $0.47\sigma$ & \xmark \\
log([\ion{O}{iii}] $\lambda 5007$/[\ion{O}{ii}] $\lambda\lambda 3727,29$) & Ionisation parameter (U) & $0.31\sigma$ & $129.2\degree$ & $0.09\sigma$ & \cmark \\
% log([\ion{Ne}{iii}] $\lambda$3869/[\ion{O}{ii}] $\lambda\lambda$3727,29) & Ionisation parameter (U) & $0.86\sigma$ & $85.1\degree$ & $0.13\sigma$ & \cmark \\

         \hline
    \end{tabular}
     
    \caption{List of parameters considered in the analysis of the [\ion{S}{ii}]-BPT diagram. The statistics defined in Section~\ref{sec:frame}, and the list of parameters included in the `multi-parameter' set, are also reported.}
    \label{tab:sii_parameters}
\end{table*}

\subsection{Machine Learning Analysis}
For the the machine learning analysis, we replicate the framework described in Section~\ref{sec:ML}, with just a few differences as described below.
In particular, the set of parameters included in the ML analysis for the [\ion{S}{ii}]-BPT is detailed in Table~\ref{tab:sii_parameters}.
Similarly to the [\ion{N}{ii}]-BPT case, only a limited number of parameters is selected for the purposes of assessing the performances of a 'multi-parameter set', whereas all the parameters are evaluated in the individual ANN runs, with the exception of metallicity, because its derivation involves exactly the same line ratios of the BPT-axis.
As already noted however, metallicity iso-contours appear orthogonal to the best-fit curve of the SF sequence everywhere across the diagram (as clearly shown by Fig.~\ref{fig:S2_bpt_properties}). 
Therefore, the considerations made in Section~\ref{sec:ML} for the [\ion{N}{ii}]-BPT remain valid for the [\ion{S}{ii}]-BPT as well, and removing metallicity from the ML analysis is not expected to bias the final results significantly, as any contribution to an orthogonal offset from variations in log(O/H) can be assumed, in this framework, negligible a-priori.

We further note instead that, in contrast to what was done for the [\ion{N}{ii}]-BPT, here we select the more `direct' N/O abundance tracer given by the [\ion{N}{ii}]/[\ion{O}{ii}] ratio: in this way, not only do we rely on a more physically motivated tracer for N/O, but any trivial correlation between the [\ion{S}{ii}]/H$\upalpha$-axis and [\ion{N}{ii}]/[\ion{S}{ii}] is also removed.
The other parameters are selected according to the same criteria outlined in Section~\ref{sec:ML} for the [\ion{N}{ii}]-BPT.

Finally, the distance \textbf{D} and angle \textbf{$\theta$} metrics for the [\ion{S}{ii}]-BPT are computed in the same way as in equations~\ref{eq:nii_distance} and ~\ref{eq:theta_angle}, as illustrated in Fig.~\ref{fig:S2_bpt_metrics}. Here we note that, because of the double-branched nature of the SF sequence and the orthogonality of the offset vector, the $\theta$ angle can assume also negative values.

\subsubsection{Artificial Neural Networks}

The results of the ANN classification analysis are presented in Fig.~\ref{fig:Sii_ANN_class}. As done previously, only galaxies with |\textbf{D}|>0.025 are included in the classification analysis, to reduce the noise introduced by the potential misclassification of sources located extremely close to the best-fit line of the SF sequence. However, we note that including all galaxies slightly reduces the performances of the network, but does not impact at all the ranking of the parameters nor affect the interpretation of the results.

Overall, the network achieves a $\sim 87\%$ accuracy ($\sim 94\%$ AUC) in the binary classification task when fed with the `multi-parameter' set, only slightly worse than the performance achieved in the [\ion{N}{ii}]-BPT. % although it does not match the outstanding performance achieved in the [\ion{N}{ii}]-BPT.
In terms of individual parameters, the distribution of feature performances is quite different from that found for the [\ion{N}{ii}]-BPT: here, in fact, $\upDelta$log(SFR) is the most predictive variable, followed by deviations in stellar mass, whereas the most predictive feature in the [\ion{N}{ii}]-BPT (i.e., relative variations in N/O, here traced by \delNOii) scores an accuracy of only $\sim63\%$ in galaxy classification.
%once the global metallicity (which determines the position along the best-fit sequence) is fixed.

Moving to the regression problem, i.e., trying to reproduce the minimum distance \textbf{D} of each point from the best-fit line of the star-forming sequence, the network achieves an overall RMSE of $0.045$ and a $\sim 37\%$ IoR on the test sample in the `multi-parameter' run, with the comparison between the predicted and the observed target distance \textbf{D} shown in the inset, upper-right panel of Fig.~\ref{fig:Sii_ANN_class}.
Again, relative variations in star-formation rate score the best performance among the individual parameter runs, followed by $\upDelta$log(M$_{\star}$) and \delU. In general, the ranking of individual parameters in the regression task is fully consistent to what found when solving the classification problem.
% These results confirm what visually suggested already by Figure~\ref{fig:S2_bpt_delta_properties}, that is on global scale the scatter  

% \begin{figure*}

% \centering

% % \includegraphics[width=0.4\textwidth]{Sii_SDSS_NN_regression__9var_128406_elem.png}
% \includegraphics[width=0.85\textwidth]{Sii_improvements_ANN__9var_128406_elem.png}

% % \includegraphics[width=0.48\textwidth]{SDSS_offset_angle.png}

% \caption{...}
% \label{fig:Sii_ANN_regression}

% \end{figure*}

\subsubsection{Random Forest}

\begin{figure*}

\includegraphics[width=0.7\textwidth]{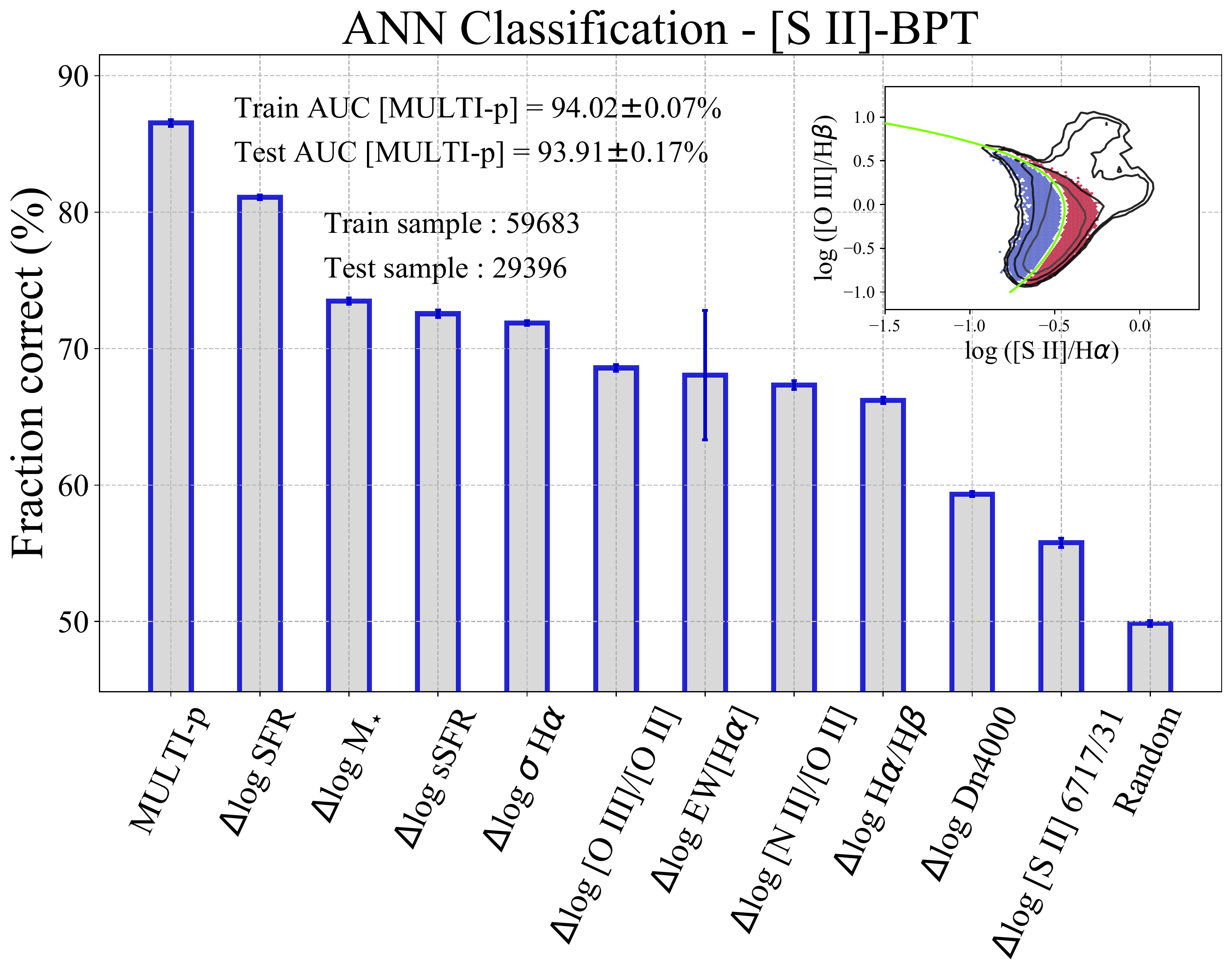}\\
\vspace{0.5cm}
\includegraphics[width=0.7\textwidth]{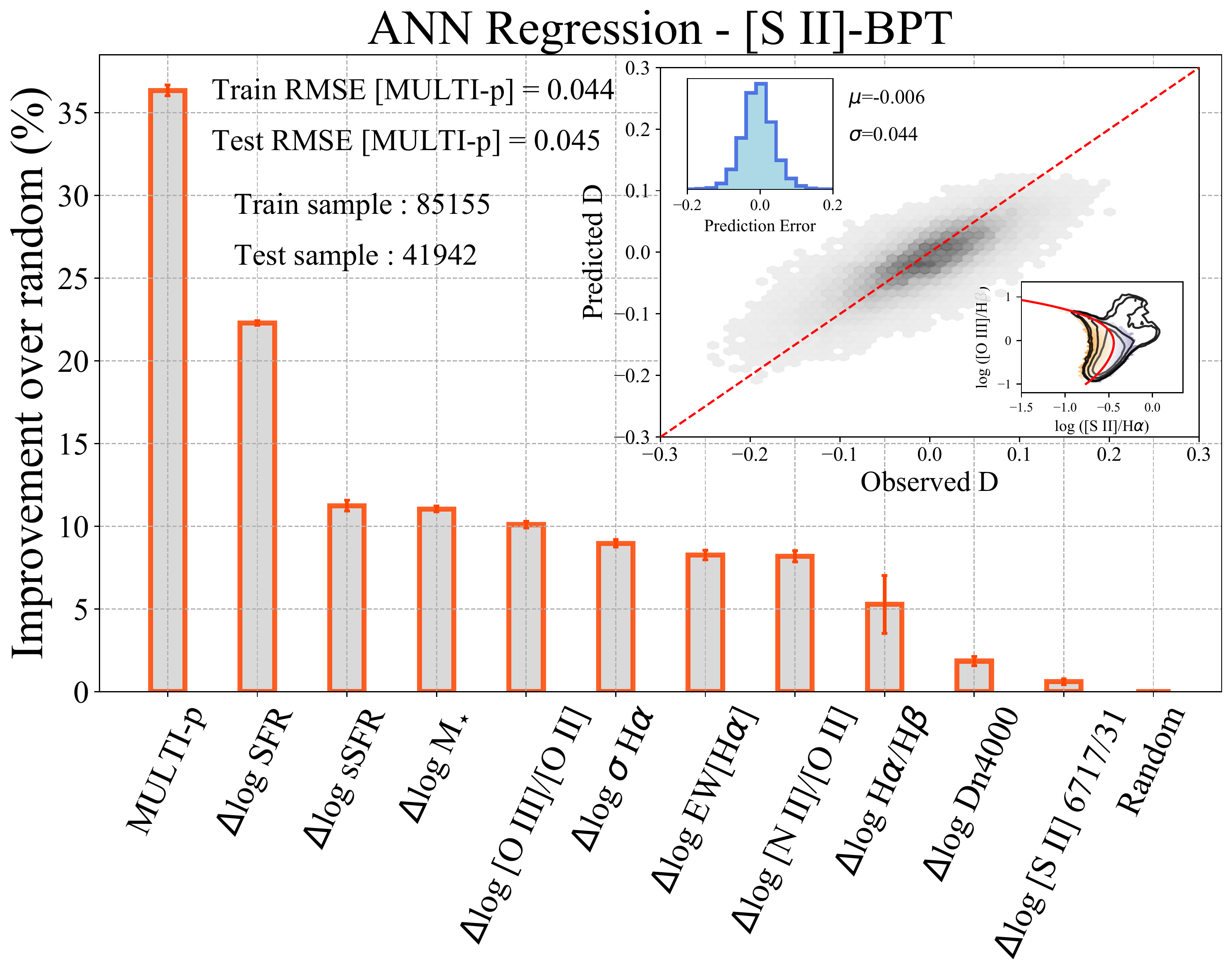}

\caption{Results of the ANN classification (upper panel) and regression (bottom panel) analysis on star-forming galaxies in the [\ion{S}{ii}]-BPT diagram. The panels are structured as in Fig.~\ref{fig:Nii_ANN_classification}. Overall, the network achieves an $\sim87\%$ accuracy in classification and a $\sim 37\%$ IoR in regression when trained with the `multi-parameter' set. For both problems, $\upDelta$log(SFR) (i.e., relative variation in the star-formation rate) is the parameter that individually performs best.}
\label{fig:Sii_ANN_class}

\end{figure*}

\begin{figure*}

\centering
\includegraphics[width=0.7\textwidth]{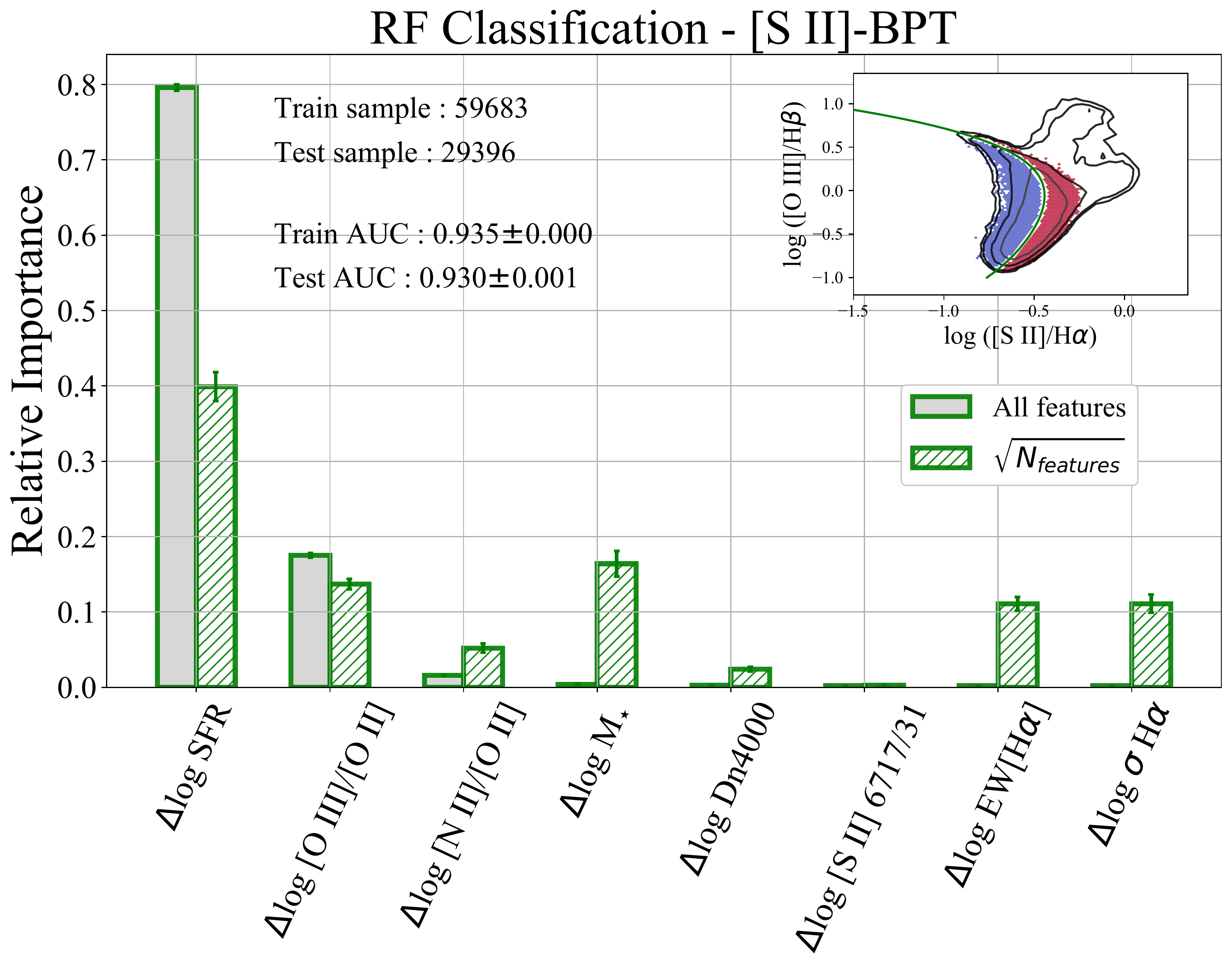}\\
\vspace{0.5 cm}
\includegraphics[width=0.7\textwidth]{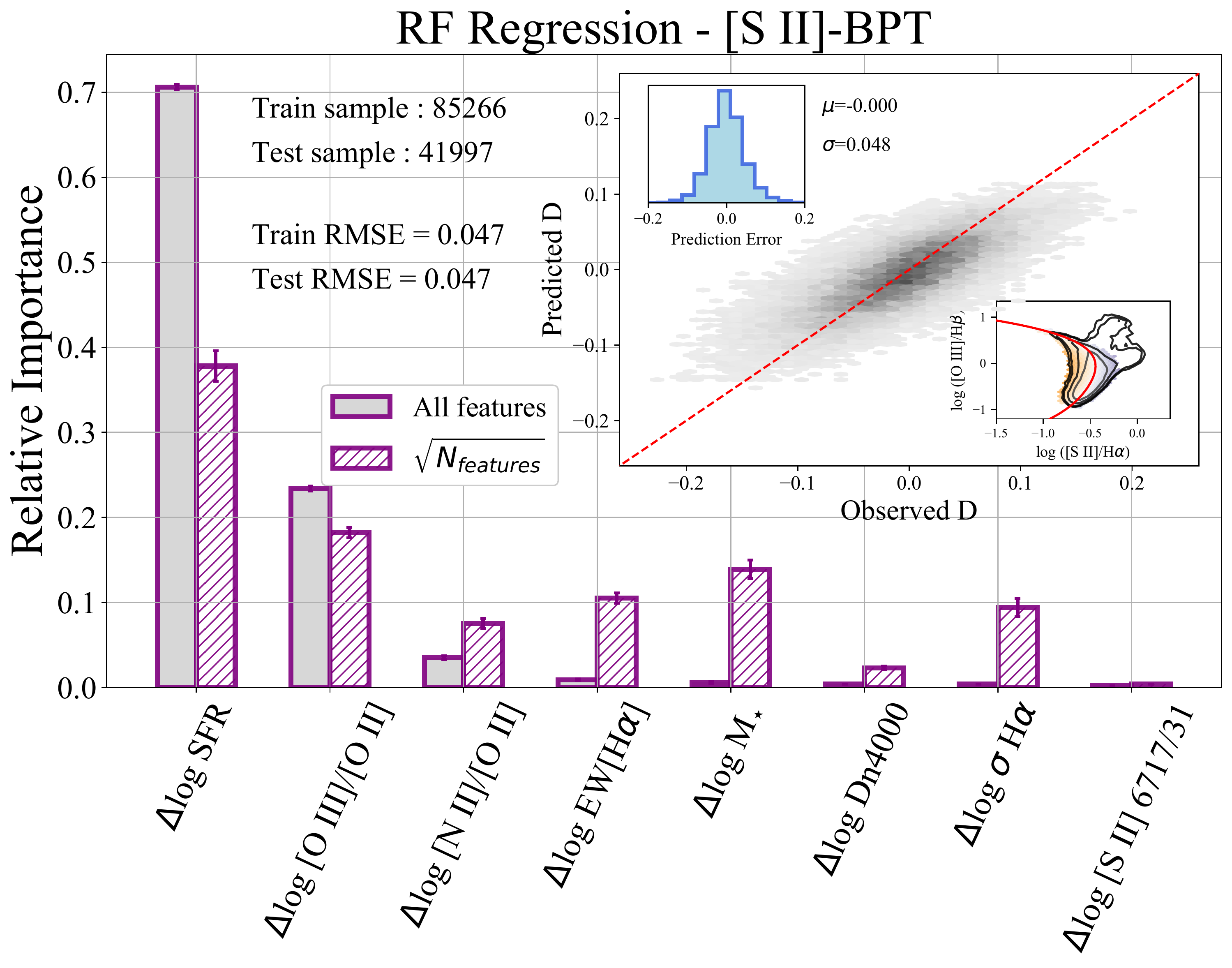}

\caption{Results of the RF classification (upper panel) and regression (bottom panel) analysis on star-forming galaxies in the [\ion{S}{ii}]-BPT diagram. The panels are structured as in Fig.~\ref{fig:Nii_RF_class}. $\upDelta$log(SFR) is, by far, the most relevant parameter for predicting the offset from the best-fit of the median SF sequence in the [\ion{S}{ii}]-BPT diagram, and combined with \delU\ account for more than $90\%$ of the total predictive power of the RF. When the trees are randomised in the selection of $\sqrt{N_{\text{features}}}$ features at each fork, deviations in M$_{\star}$, EW[H$\upalpha$] and \sigha are the parameters retaining the larger part of the residual importance.
}
\label{fig:Sii_RF_class}

\end{figure*}

% \begin{figure*}

% % \includegraphics[width=0.4\textwidth]{Sii_SDSS_RF_regr_comparison_9var_128406_elem.png}
% \includegraphics[width=0.85\textwidth]{Sii_SDSS_RF_regr_comparison_9var_128406_elem.png}

% \caption{...}
% \label{fig:Sii_RF_regression}

% \end{figure*}

\begin{figure}
    \centering
    \includegraphics[width=0.95\columnwidth]{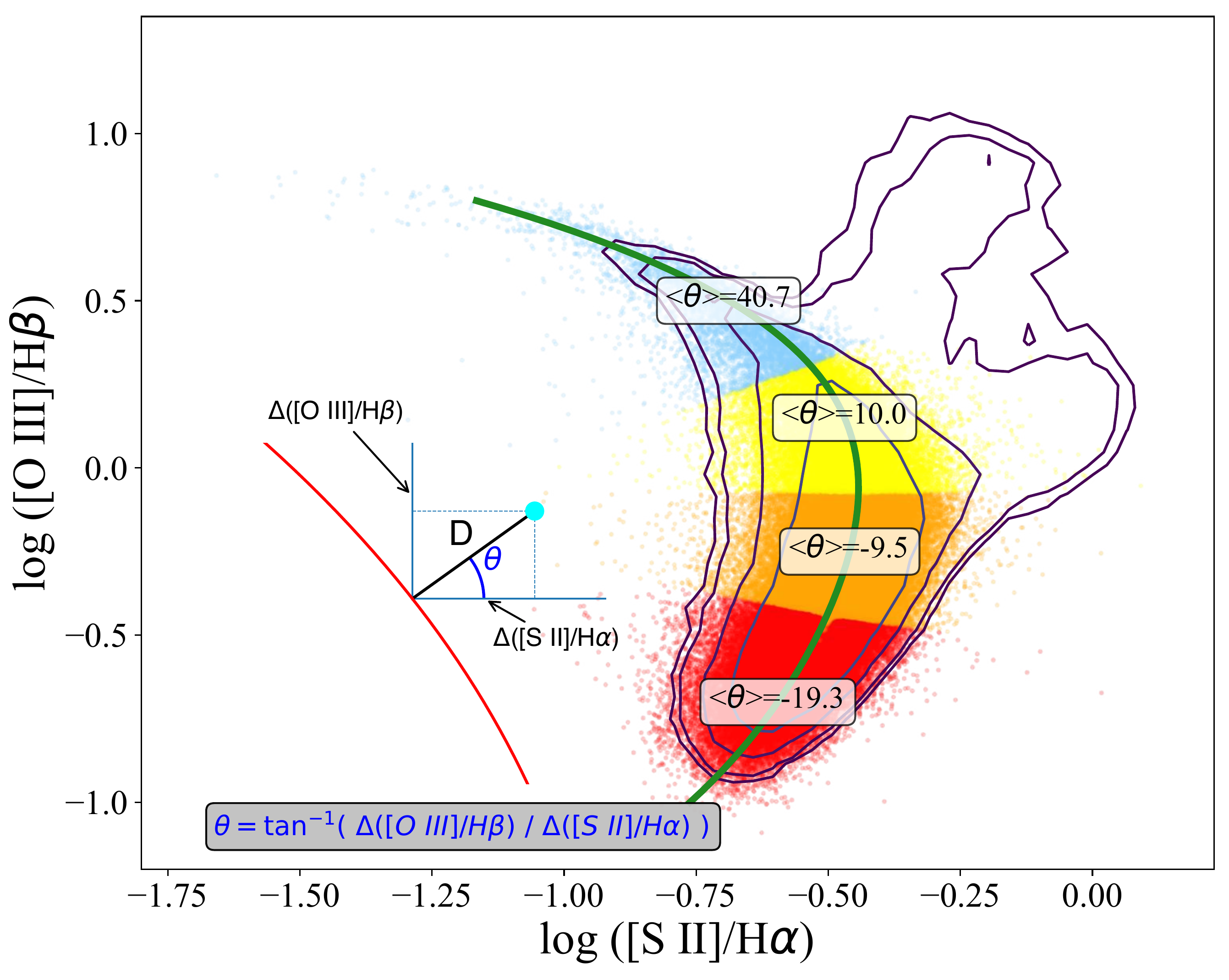}
    \caption{The four sectors in which the \siibpt\ is divided in order to assess the variation of parameters' performance as a function of the position of galaxies along the sequence.}
    \label{fig:Sii_sectors_binned}
\end{figure}

\begin{figure*}

 \includegraphics[width=0.95\columnwidth]{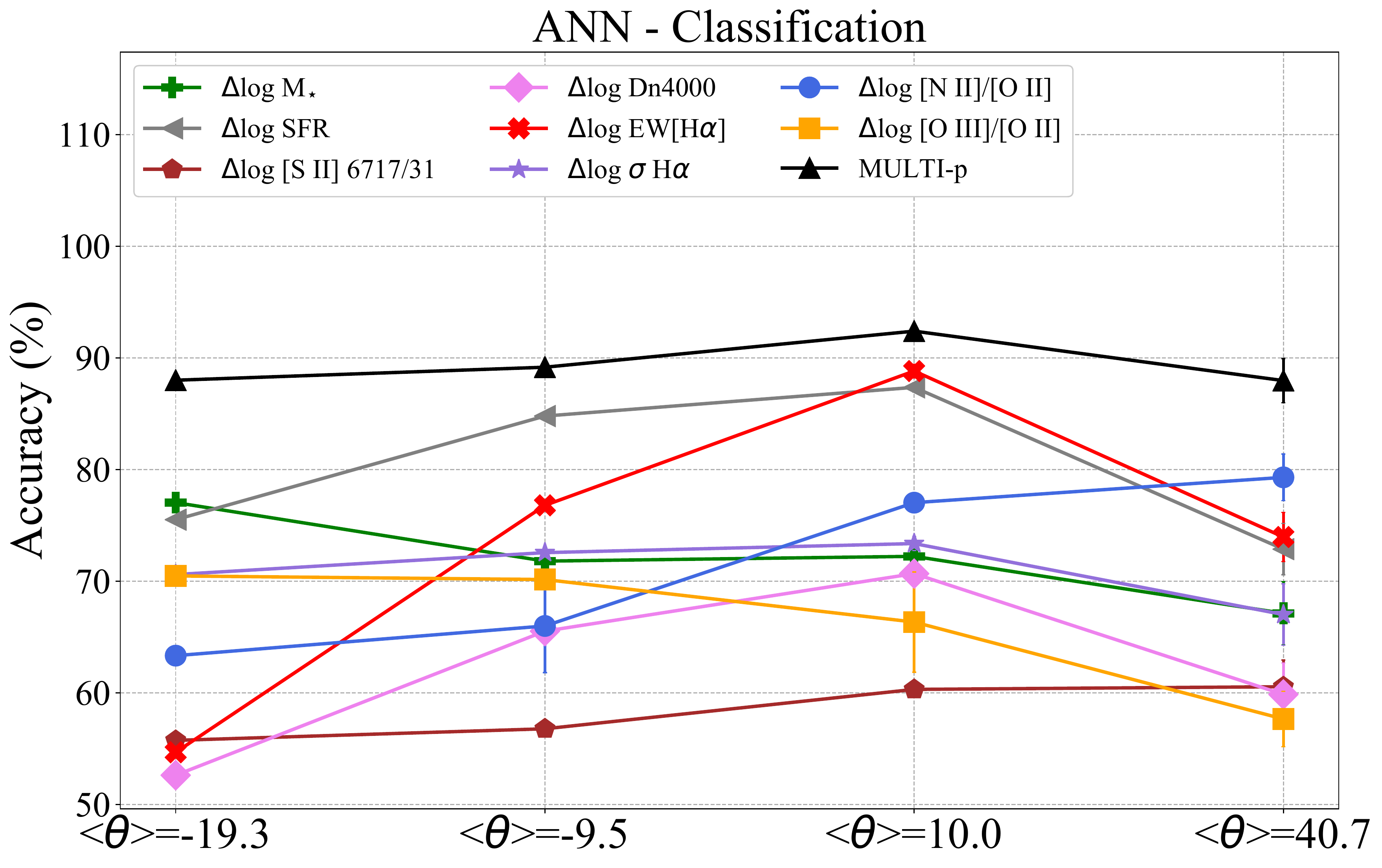}
  \hspace{0.2cm}
 \includegraphics[width=0.95\columnwidth]{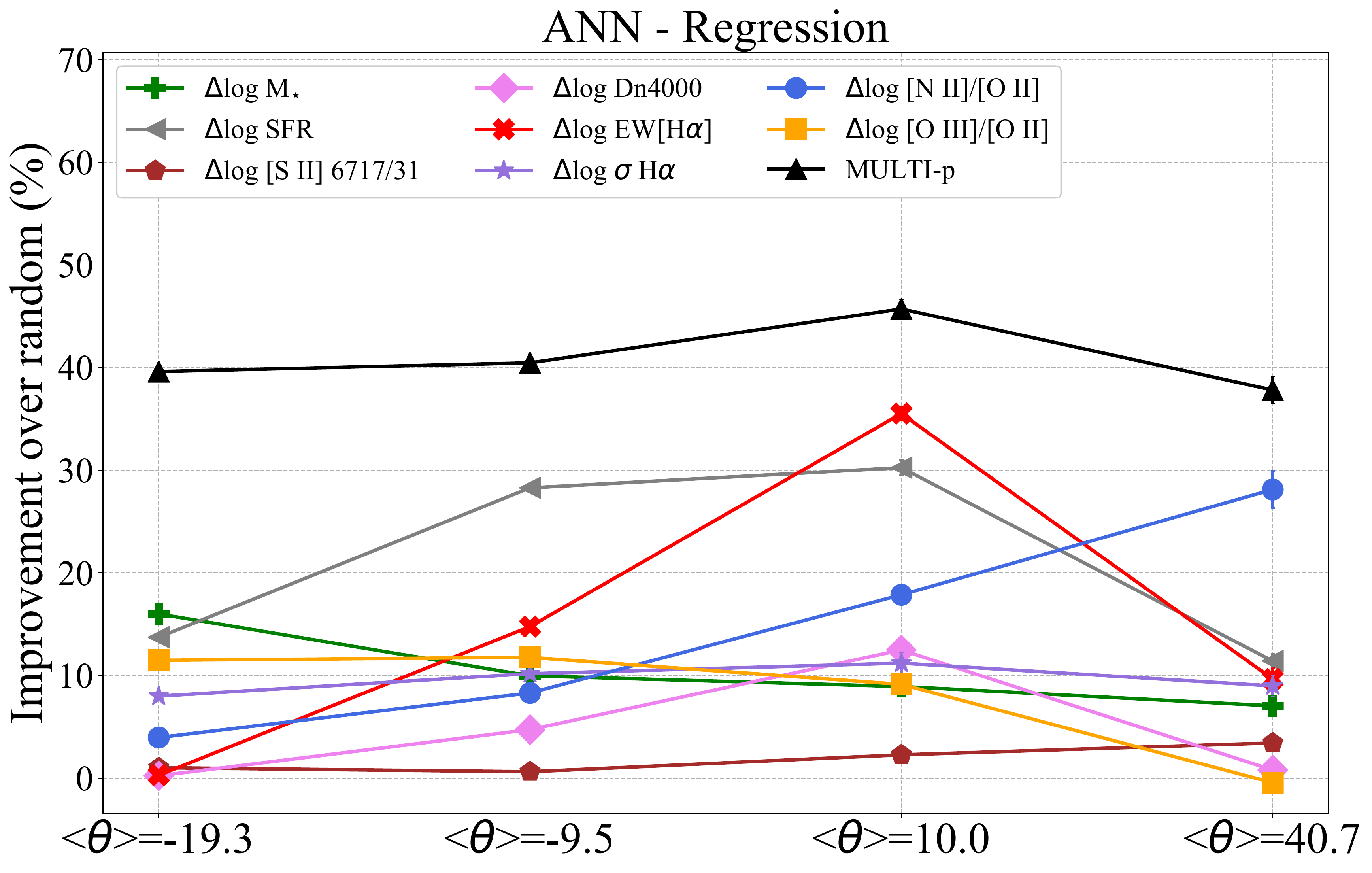}
 \vspace{0.2cm}
\includegraphics[width=0.95\columnwidth]{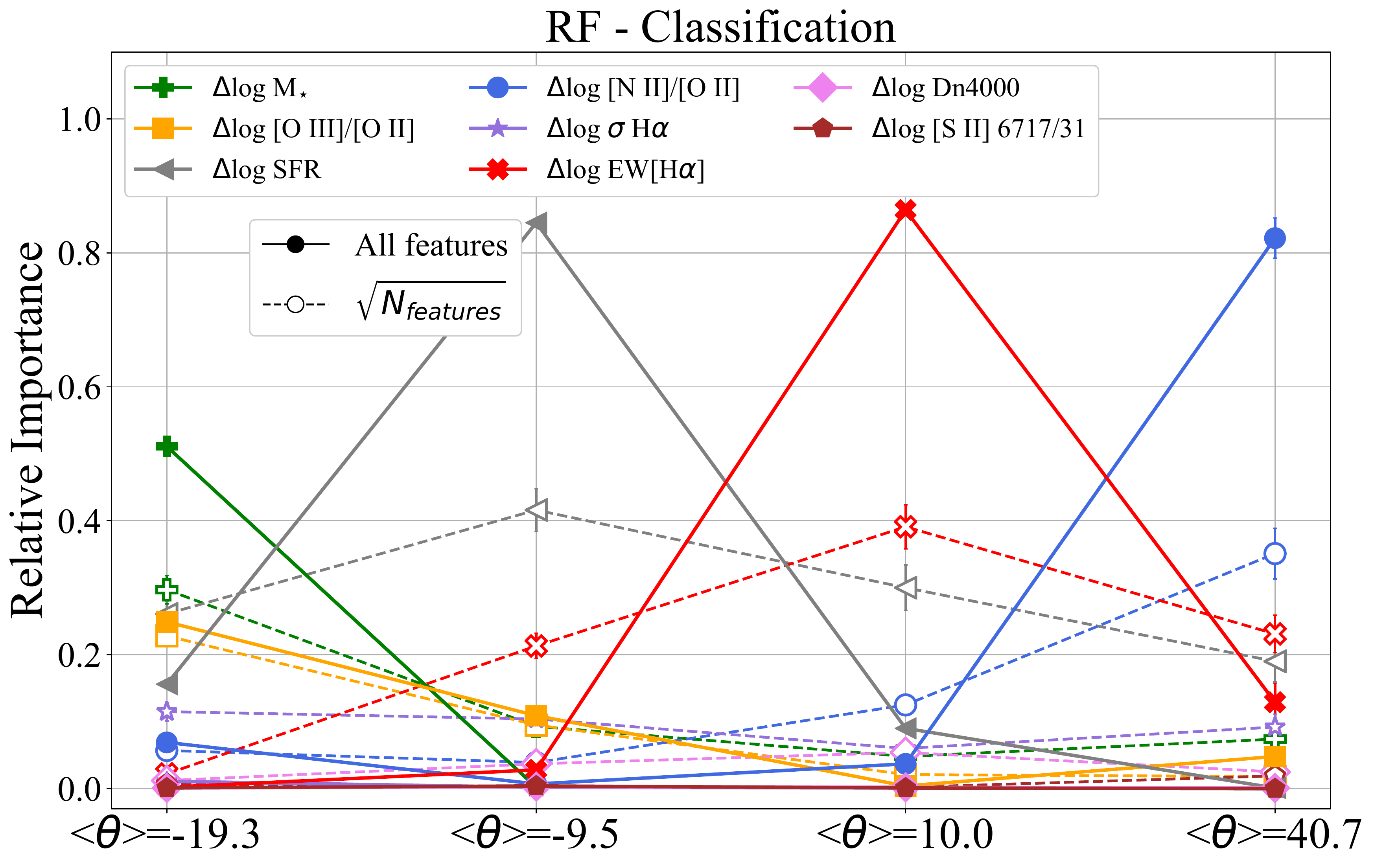}
\hspace{0.2cm}
\includegraphics[width=0.95\columnwidth]{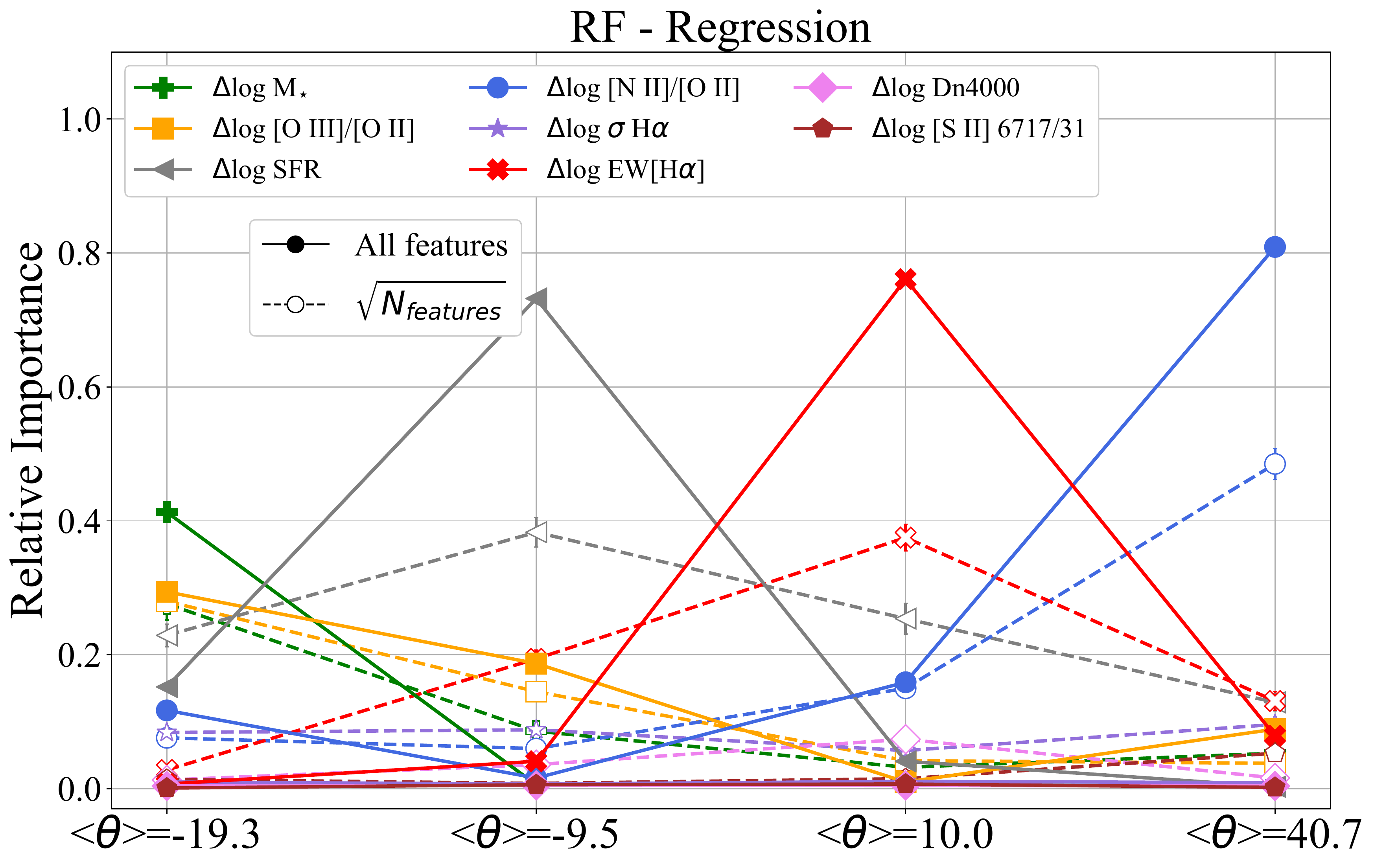}

\caption{Same as in Fig.~\ref{fig:Nii_theta_comparison}, for the four sectors the [\ion{S}{ii}]-BPT diagram has been divided into. The accuracy of the ANN is stable across the entire diagram in both classification ($\sim 90\%$) and regression ($\gtrsim 40\%$ IoR).
Parameters connected with star formation ($\upDelta$log(SFR), $\upDelta$log(EW[H$\upalpha$])) dominate the relative contribution to the observed scatter in the central regions, where the offset vector is mainly directed along the [\ion{S}{ii}]/H$\upalpha$-axis (i.e., low <$\theta$> values), whereas at the edges of the SF sequence, parameters related to the chemodynamical properties of galaxies (e.g., $\upDelta$log(M$_{\star}$), \delNOii) increase their impact on the predicted offset.
}
\label{fig:Sii_theta_comparison}

\end{figure*}

The outcome of the random forest analysis on the [\ion{S}{ii}]-BPT diagram is presented in Fig.~\ref{fig:Sii_RF_class}, for both classification (upper panel) and regression (bottom panel).
Deviations in SFR clearly rank as the most important parameter in classifying galaxies within the diagram, whereas \delU\ is ranked as the second most important variable to be used in conjunction with $\upDelta$log(SFR).
Because of the way the RF computes the relative importance of the parameters (fully accounting for their mutual correlations), once again the relative importance of some variables appears here suppressed compared to their absolute performance shown in Fig.~\ref{fig:Sii_ANN_class}, demonstrating that part (or all) of their individual predictive power followed purely from correlation with other parameters.
In the regression task, the RF achieves similar accuracy as the ANN, and the ranking of relative importance closely follows that of the classification task, with $\upDelta$(SFR) and \delU\ dominating over the other variables.

When the RF is forced to randomly select only ($\sqrt{N_{\text{features}}}$) features at each fork, deviations in M$_{\star}$, EW[H$\upalpha$] and \sigha\ retain part of the residual importance at the expenses of the two main parameters (as shown by the empty, hatched bars in Fig.~\ref{fig:Sii_RF_class}).
Nonetheless, $\upDelta$log(SFR) is still strongly identified as the most predictive parameter in the set. Whether such a trend is maintained along the full SF sequence is the subject of the analysis of the following section.

\subsubsection{[\ion{S}{ii}]-BPT sectors}

In a similar fashion to what done for the [\ion{N}{ii}]-BPT, we here divide the [\ion{S}{ii}]-BPT into four sectors, in order to assess how the predictivity and relative importance of each parameter change when considering galaxies in different specific regions across the diagram, as parametrised by the inclination of their offset-vector with respect to the horizontal axis.
The results are shown in Fig.~\ref{fig:Sii_theta_comparison} for ANN (upper panels) and RF (bottom panels) respectively, where we compare the performance and relative importance of each parameter in the classification (left panels) and regression (right panels) tasks as a function of <$\theta$>, the median angle (positive counterclockwise from the horizontal axis) formed by the offset vectors of galaxies pertaining to a given sector.
Overall, the `multi-parameter' run in the ANN maintains a constant performance level in both tasks across the entire diagram, with an accuracy close to $90$ per cent in classification and an IoR $\gtrsim 40\%$ in regression. The ranking of individual parameters is also roughly constant with increasing <$\theta$>, although $\upDelta$log(EW[H$\upalpha$]) does see the sharpest increases in the central regions whilst declining, similar to $\upDelta$SFR and \delU, in the last sector. 
In contrast, deviations in the N/O abundance (here mainly traced by [\ion{N}{ii}]/[\ion{O}{ii}]) see a steady but constant increase of its performance moving from the `bottom` to the `top` region of the diagram.
The dependence of the scatter on relative variations in gas density (traced by the [\ion{S}{ii}] doublet) remains instead almost negligible across the entire diagram.

For what concerns the RF analysis, we see that in the bottom part of the diagram (i.e., negative values of <$\theta$>) deviations in stellar mass and ionisation parameter dominate the relative contribution to the RF predictivity, while parameters associated with star formation like SFR and EW[H$\alpha$] gain more importance in the central regions (intermediate <$\theta$> values), where the offset occurs preferentially along \siiha. Interestingly, in the last sector (i.e., the one including galaxies lying at the top-left of the sequence) the scatter is dominated by deviations in \delNOii, which hold $\sim80\%$ of the total information, whereas the other parameters are strongly suppressed.

\subsection{Discussion}
Overall, both ANN and RF analysis suggest that the scatter in the [\ion{S}{ii}]-BPT diagram is primarily sensitive to parameters associated to the recent star formation activity in galaxies. 
Assuming that the offset occurs at fixed metallicity (following Fig.~\ref{fig:S2_bpt_properties} and \ref{fig:S2_bpt_delta_properties}, and the considerations of Section~\ref{sec:metrics}), one possible explanation invokes the size of the\Hii regions, and the relative fraction occupied by the S$^{+}$ ions, in galaxies with different levels of star formation.
H$\rm{II}$ regions are in fact stratified, with higher
ionization species like S$^{++}$ much more common closer to the ionizing source while lower ionization species like S$^{+}$ relatively more abundant in the outer parts \citep[see e.g.,][]{levesque_EL_models_2010, xiao_hii_modelling_2018, mannucci_dig_2021}.
Since sulphur has a lower ionisation potential than both oxygen and nitrogen, the S$^{+}$ zone is typically much more extended than the O$^{+}$ or N$^{+}$ ones within the same \Hii region, and this has an impact on the ratios between sulphur lines and hydrogen recombination lines (like \siiha). It is possible then, that galaxies with a relatively larger ongoing star formation (compared to median galaxies on the SF sequence) are characterised by a large number of nearby \Hii regions which eventually merge, reducing the effective emitting size of the S$^{+}$ region \citep[as also suggested by][]{masters_tight_2016}.
We note that, because almost $60\%$ of the total sample of star-forming galaxies are characterised by an offset vector pointing between $\theta=-20\degree$ and $\theta=40\degree$ (hence directed predominantly along \siiha), this would likely explain why the contribution from variations in SFR dominates the overall behaviour within the diagram, as exposed by the global ANN and RF results of Fig.~\ref{fig:Sii_ANN_class} and \ref{fig:Sii_RF_class}.
Interestingly, the sharp rise of the importance of $\upDelta$log(EW[H$\upalpha$]) (hence, variations in sSFR) at the expenses of $\upDelta$log(SFR) in the third sector provides us with additional information on the status of these galaxies, which not only are characterised by higher/lower levels of star-formation compared to `on-sequence' galaxies, but they are also forming stars at higher/lower pace than in their past history.

The random forest analysis also suggests that coupling \delsfr\ with variations in the \oiii/\oii\ ratio (sensitive primarily to U), which picks around $20$ per-cent relative importance (hence providing complementary information to that hold by \delsfr\ only) maximises the predictivity of the algorithm. 
An increase in the ionisation parameter in fact (beside that already associated with an increase in SFR) could on the one hand provoke a suppression of the \siiha\ line ratio at fixed \oiiihb, as the abundance of doubly ionised sulphur increases at the expense of S$^{+}$, while on the other it could boost the \oiiihb\ ratio itself at fixed O/H.

Interestingly, in the lowest part of the sequence, as probed by the $<\theta>=-19\degree$ sector, a relative variation in stellar mass is the most predictive quantity of the observed offset from the median sequence; nonetheless, star-formation rate and U tracers still contribute significantly to the total predictivity.
This result is probably driven by the presence of a group of high-mass galaxies located at around log(\oiiihb)$=-0.5$, log(\siiha)$=-0.75$ (see Fig.~\ref{fig:S2_bpt_properties}); these sources likely represent a sub-population of relatively older, high-mass, chemically mature galaxies, whose central \sigha\ (of the order of $\sim80-100$km/s) is more typical of bulge structures or other largely pressure-supported systems rather than thin disk structures. In this sense, the large relative importance picked by \mstar\ is likely driven by the information brought by $\sigma_{\star}$ (which is not represented in our parameter set), plus part of the importance `borrowed' from SFR, with the relative importance of \delsfr\ and $\upDelta$log(\mstar) indeed almost matching in the RF run with $\sqrt{\text{N}_{\text{features}}}$ allowed at each node.

Finally, in the uppermost region of the diagram the ML analysis identifies \delNOii\ (mainly tracing variations in N/O) as the most informative parameter for predicting the scatter in the \siibpt.
Being the \siibpt\ free of nitrogen lines however, any variation in the N/O abundance cannot have a direct impact on the BPT-line ratios by itself, but should be considered as a reflection of some other underlying physical effect.
As already discussed in Section~\ref{sec:discuss_nii}, one possibility invoke to break the initial assumption of offsets occurring along iso-metallicity lines in this region of the diagram.
If this is the case, the connection between the offset from the median sequence and relative variations in [\ion{N}{ii}]/[\ion{O}{ii}] in the \siibpt\ could just effectively trace metallicity variations.
However, in contrast to what is seen for the \niibpt, these galaxies are observed to deviate in stellar mass, O/H and N/O according to the standard mass-metallicity-N/O relation (i.e., to an increase in M$^{\star}$ correspond an increase in both O/H and N/O, and viceversa).
Hence, this suggest that the high relative importance kept by \delNOii\ in this sector might just be the reflection of the average relationship between these quantities.
% Nonetheless, we note that the uppermost sector is comprised by only $\sim 14,000$ objects, hence the results provided by the ML are likely to be less robust than for other regions of the diagram, and their interpretation even less immediate.
% Further analysis based on large samples of galaxies with independent and direct metallicity estimates in such region of the \siibpt
% could certainly help to either confirm or deny such scenario.

\section{Summary and Conclusions}
\label{sec:summary}

In this paper, we have presented a novel approach to study the distribution of galaxy properties in the BPT diagnostic diagrams, attempting to link variations in such properties (via different observational tracers) to the variations in the line ratios space observed within the diagrams, following a purely empirical, data-based approach.
In particular, artificial neural networks (ANN) and random forest (RF) of decision trees have been trained and tested over a large sample of SDSS galaxies, in order to assess which physical parameters are the most intrinsically connected with the observed offset of local star-forming galaxies from their median sequence in both the [\ion{N}{ii}]- and [\ion{S}{ii}]-BPT diagrams.
Relative variations in a set of physical parameters (in the form of the $\upDelta$ log(p) metric, equation~\ref{eq:delta_prop}) are linked to the deviation from the sequence itself, with an offset assumed orthogonal to the best-fit line of the sequence at any given point. 

The performances of our set of parameters (individually and as a whole), as well as their relative importance, are evaluated in solving both a classification (i.e., predicting whether a galaxy is offset above or below the median sequence) and a regression (i.e., predicting the exact magnitude of the offset) problem.
The key results of the paper are summarised below. 

\begin{itemize}
     
    \item The distribution of star-forming galaxies in both the [\ion{N}{ii}] and the [\ion{S}{ii}]-BPT diagrams primarily traces a sequence in the gas-phase metallicity.
    A significant gradient in log(O/H) along the best-fit curve of the SF sequence is in fact observed in both diagrams ($\nabla_{\parallel}$ = 0.31$\sigma$), coupled with zero (or very mild) variations assessed orthogonal to it ($\upDelta{\perp} \sim $0, see also iso-contours of O/H in Fig.~\ref{fig:Nii_properties} and ~\ref{fig:S2_bpt_properties}).
    Hence, in our framework we assume the gas-phase metallicity to be the main parameter that set the position of galaxies \textit{along} the SF sequence, whereas contributions from relative variations in O/H are not considered to significantly impact the deviations from it, when described by a purely orthogonal `offset vector'.
    
    % \item We define the $\upDelta$ log(p) metric to quantify the (logarithmic) variation in a parameter \textit{p} for a given galaxy, as compared to the average values of \textit{p} in galaxies which lie on the closest point along the median SF sequence. This defines our set of parameters,
    % which we analyse through both ANN and RF in order to explore their connection with the observed deviation from the median SF sequence in both the \niibpt\ and \siibpt\ diagrams.

    \item When trained with multiple parameters, the ANN is capable of classifying whether a galaxy is offset above or below the best-fit of the median SF sequence with $>90$ per cent accuracy (AUC$=0.96$) in the [\ion{N}{ii}]-BPT, and to predict the magnitude of the offset from the sequence itself with a RMSE$=0.038$ on the test sample.
    When assessing the correlation of individual parameters in the ANN, \delNO\ and \delNOii\ (mainly tracing relative variations in the N/O abundance compared to `on-sequence' galaxies) are robustly assessed as the most accurate features in both classification and regression tasks for the \niibpt diagram (Fig.~\ref{fig:Nii_ANN_classification}). 
    
    \item From the RF analysis, we find that \delNO\ is, by far, the most relevant parameter for predicting the offset from the SF locus in the \niibpt, gathering more than $80\%$ of the total importance among the whole `multi-parameter' set (Fig.~\ref{fig:Nii_RF_class}). Therefore, this suggests that any offset from the median sequence of star-forming galaxies in this diagram is primarily associated to relative variations in their N/O abundance. 

    \item The impact of the individual parameters on the offset of galaxies from the best-fit curve of the SF median loci in the \niibpt\ changes as a function of the position along the sequence. In the bottom-right region of the diagram, the offset-vector is almost horizontal, and the deviation from the SF sequence is nicely predicted by properties related to the chemo-dynamical evolution of galaxies (e.g., \delNO, $\upDelta\sigma_{\text{H}\alpha}$ and $\upDelta\text{M}^{\star}$), whereas moving along the SF locus, parameters associated to ongoing star-formation in galaxies (e.g., $\upDelta$SFR, \delU, $\upDelta$EW[H$\upalpha$]) increase their predictivity (Fig.~\ref{fig:Nii_theta_comparison}, upper panels).
    Nonetheless, \delNO\ remains the most relevant parameter of the set for predicting the offset from the SF sequence throughout the entire diagram, regardless of the relative amplitude of the components of the offset vector.
    
    \item If we assume the offset to occur at fixed metallicity, these results can be interpreted as a manifestation of the relationship between N/O and O/H (mainly driven by the `secondary' nucleosynthetic production of nitrogen in high mass galaxies), whose median behaviour and scatter is to a large extent reflected in the distribution of galaxies within the \niibpt\ (Fig.~\ref{fig:NO_OH_dist}).
    When only $\sqrt{\text{N}_{\text{features}}}$ are considered at each fork of the RF, feature importance is partially shifted from the most important variable (i.e., \delNO) to variables which are strongly correlated with it, like \mstar, which acts then as a good proxy for N/O, reflecting the different level of chemical maturity in these galaxies. %\textbf{Differential depletion of oxygen and nitrogen onto dust grains can also impact the N/O abundance of the gas-phase at fixed metallicity.}
    
    \item Parameters associated with the age of the stellar populations (D$_{\text{N}}$(4000)) and specific star formation rate (EW(H$\upalpha$)) provide complementary information to that brought by \delNO, which is required to maximise the predictivity of the RF (Fig.~\ref{fig:Nii_RF_class}). These are likely associated with the differential contribution of younger/older stellar populations, which impact the strength of intermediate- and low-ionisation emission lines originating from both inside and the warm, diffuse ionised gas outside the \Hii regions.
    
    \item In the [\ion{S}{ii}]-BPT diagram, the overall scatter of galaxies around the best-fit SF sequence is primarily associated with relative variations in SFR (Fig.~\ref{fig:Sii_ANN_class}).
    Among the other parameters, \delU (primarily sensitive to variations in the ionisation parameter) retain the most complementary information to $\upDelta$SFR that maximises the accuracy of both classification and regression tasks (Fig.~\ref{fig:Sii_RF_class}). 
   
    \item In particular, parameters associated with recent star formation activity (\delsfr\ and $\upDelta$log(EW[H$\upalpha$])) dominate the relative contribution to the offset in the central part of the diagram, where the majority of galaxies reside (and where the orthogonal offset vector is primarily directed along \siiha).
    We primarily interpret this in terms of the relative change in the extension of the 'effective' emitting zone of the low-ionisation S$^{+}$ ions in \Hii regions within galaxies with different levels of ongoing star formation.
    Further contribution from variations in the ionisation parameter can either impact the \oiiihb\ ratio and affect the relative S$^{++}$/S$^{+}$ ionic abundance.
     
    \item At the edges of the sequence, where the best-fit line of the SF locus in the \siibpt\ diagram bends, parameters tracing chemo-dynamical properties of galaxies (e.g., \delNOii, $\upDelta$log(M$_{\star}$) increase their scores in the ANN, and gain a higher amount of relative importance in the RF (Fig.~\ref{fig:Sii_theta_comparison}). This is likely driven by a sub-population of high-mass, bulge dominated galaxies in the bottom-left region, whereas might partially trace residual metallicity variations, which are mainly accounted for by \delNOii, in the upper-left part.

\end{itemize}

In conclusion, we have shown how the distribution of star-forming galaxies in the BPT diagnostic diagrams can be well described by a framework in which the offset from the median location of sources along the SF sequence can be ascribed to relative variations in different physical conditions, once the position along the sequence has been set by the knowledge of their gas-phase oxygen abundance.
Exploiting a variety of machine learning techniques, we have identified relative variations in N/O abundance tracers to primarily govern the scatter in the \niibpt\ diagram, whereas relative variations in parameters associated with star-formation are most relevant to predict the behaviour of galaxies in the \siibpt\ diagram.
This framework could be refined and tested in the future on different and multi-dimensional diagnostic diagrams (accounting for higher order effects on top of those considered in the present work), as well as on high redshift galaxy samples, to provide new and complementary insights for models and improve our understanding of the evolution of the physical conditions of galaxies across cosmic time as inferred from their observed spectral properties.

%%%%%%%%
%%%%%%%%
\section*{ACKNOWLEDGEMENTS}
We gratefully acknowledge the anonymous referee for their insightful comments that helped improving the paper.

MC, CH-P and RM acknowledge support by the Science and Technology Facilities Council (STFC) and from European Research Council (ERC) Advanced Grant 695671 `QUENCH'. 
RM also acknowledges funding from a research professorship from the Royal Society.
We are also very grateful for several insightful and stimulating discussions on this work, and on machine learning techniques in general, especially to Asa Bluck, Joanna Piotrowska, and Sim Brownson.

Funding for the SDSS and SDSS-II has been provided by the Alfred P. Sloan Foundation, the Participating Institutions, the National Science Foundation, the US Department of Energy, the National Aeronautics and Space Administration, the Japanese Monbukagakusho, the Max Planck Society and the Higher Education Funding Council for England.

\section*{Data Availability}
The SDSS data used in this work are publicly available at:
\url{https://www.sdss.org/dr12/spectro/galaxy_mpajhu/}.
%All other data contributing to this article will be shared on reasonable request to the corresponding author.

%%%%%%%%
%%%%%%%%
\bibliographystyle{mnras}
\bibliography{main} 

%%%%%%%%
%%%%%%%%
\appendix

\section{Tests on the Random Forest}
\label{sec:appendix_A}
\subsection{Different sets of parameters}

% \begin{itemize}
%      \item using [Ne III]/[O II] instead of [O III]/[O II] to trace ionisation parameter 
%      \item parameters which are not trivially connected at all with the BPT lines (\delNOii and/or \delNO share the same [N II] emission line fluxes with the BPT-axis, hence with the target variable)
%     \item randomised shuffling of [O II] fluxes to create an hybrid [Nii]/[Oii] ratio.

%  \end{itemize}
 
Throughout the paper, we have analysed the connection between the scatter of local star-forming galaxies across the median sequence in the BPT diagrams and a number of physical quantities, traced by either direct or indirect spectro-phototmetric observables.
As discussed already in Section~\ref{sec:parameters}, some of these parameters can be traced by means of different ratios of emission lines: for instance, if we consider the SDSS galaxy sample at the basis of this work, the ionisation parameter can be primarily traced either by the \oiii/\oii\ or the [\ion{Ne}{iii}]$\lambda3869$/[\ion{O}{ii}]$\lambda3727,29$ ratio, whereas the N/O abundance is mainly traced by either \nii/\sii and \nii/\oii\ (although the latter is a more `direct' probe than the former). 
The selection of our fiducial set of parameters to be included in the ML analysis is discussed and motivated in Section~\ref{sec:ML}.
However, in this appendix we want to test how much the results and conclusions presented in the main body of the paper are robust to the choice of a different set of parameters, either by changing some of the original tracers, and/or by removing one or more variables.
In particular, we assess which impact this might have on the RF analysis in terms of the estimated relative feature importance.

In first instance, we start by considering \nii/\oii\ to trace N/O instead of \nii/\sii. As discussed in Section~\ref{sec:ML}, including \ion{N}{ii}/\ion{S}{ii} in the fiducial RF analysis was required to avoid inducing any trivial correlation between \delNOii, \delU, and \textbf{D} in the ML analysis of the \niibpt, which would have provided biased relative importance for these two features in the prediction of our target label (which is based on a combination of the \oiiihb\ and \niiha\ line ratios).
In order to include \nii/\oii\ and keep at the same time the other parameters in the set independent, we can follow two different approaches, i.e., \textbf{i)} change the main ionisation parameter tracer, from [\ion{O}{iii}]/[\ion{O}{ii}] to [\ion{Ne}{iii}]/[\ion{O}{ii}] or \textbf{ii)} remove [\ion{O}{iii}]/[\ion{O}{ii}] at all (hence, any variable primarily related to the ionisation parameter) from the analysis.

Although case \textbf{i)} might sound the most obvious and physically motivated choice, we note that, given the low signal-to-noise of the [\ion{Ne}{iii}]$\lambda3869$ emission line in many individual SDSS spectra, requiring significant detections in such line inevitably impact the final selected sample, introducing a bias which is directly correlated with the position of galaxies in the BPT diagram itself (i.e., galaxies would be preferentially removed from the bottom-right, metal-rich region of the diagram).
To mitigate this effect, for the purposes of the present test we require only a $2.5\sigma$ detection in the [\ion{Ne}{iii}]$\lambda3869$ emission line (on top of the S/N requirements outlined in Section~\ref{sec:data}), which is enough to provide a $\gtrsim 3\sigma$ significance in the [\ion{Ne}{iii}]/[\ion{O}{ii}] ratio; this brings the final selected galaxy sample to $22,840$.

\begin{figure*}
    \centering
    \includegraphics[width=0.48\textwidth]{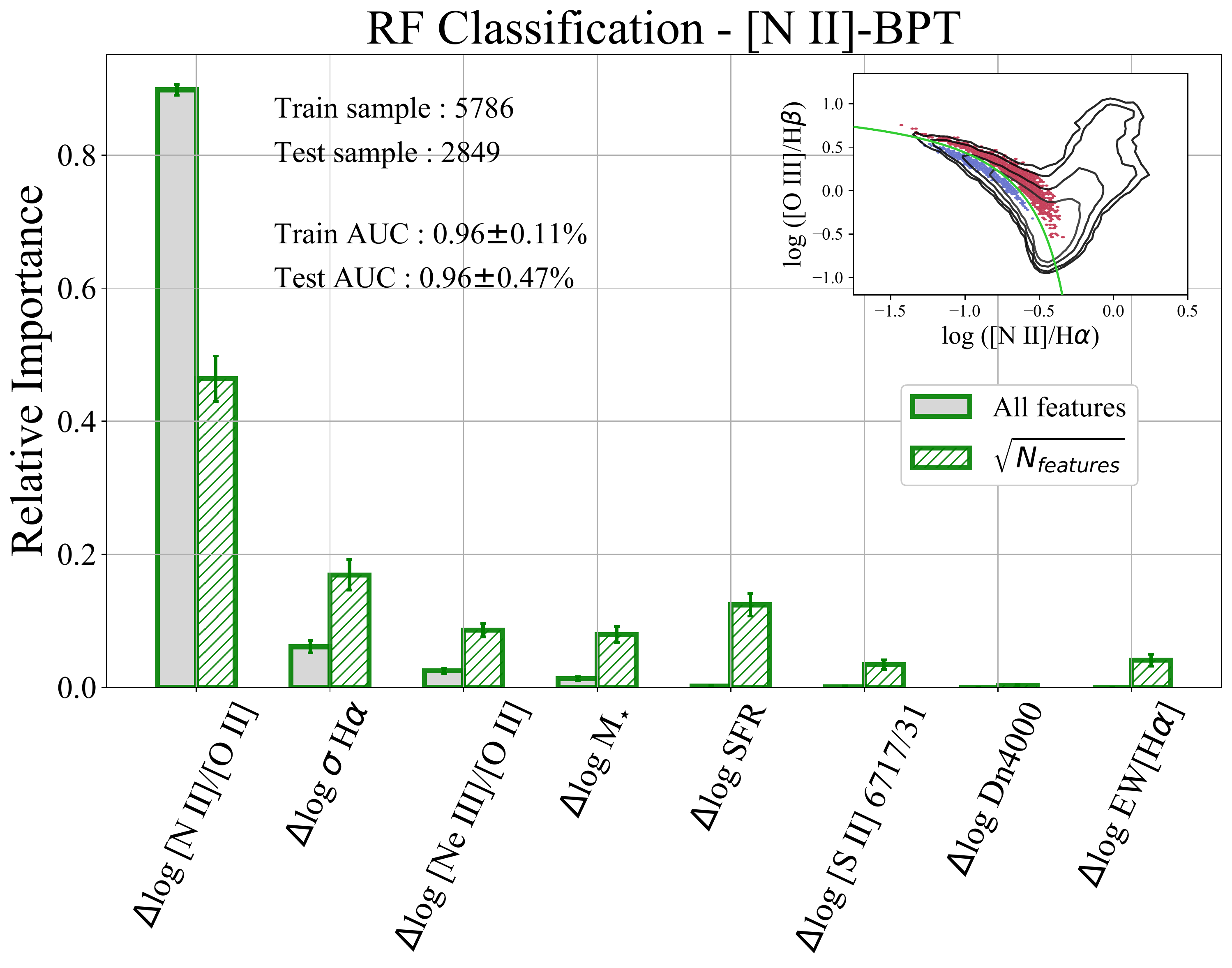}
    \includegraphics[width=0.48\textwidth]{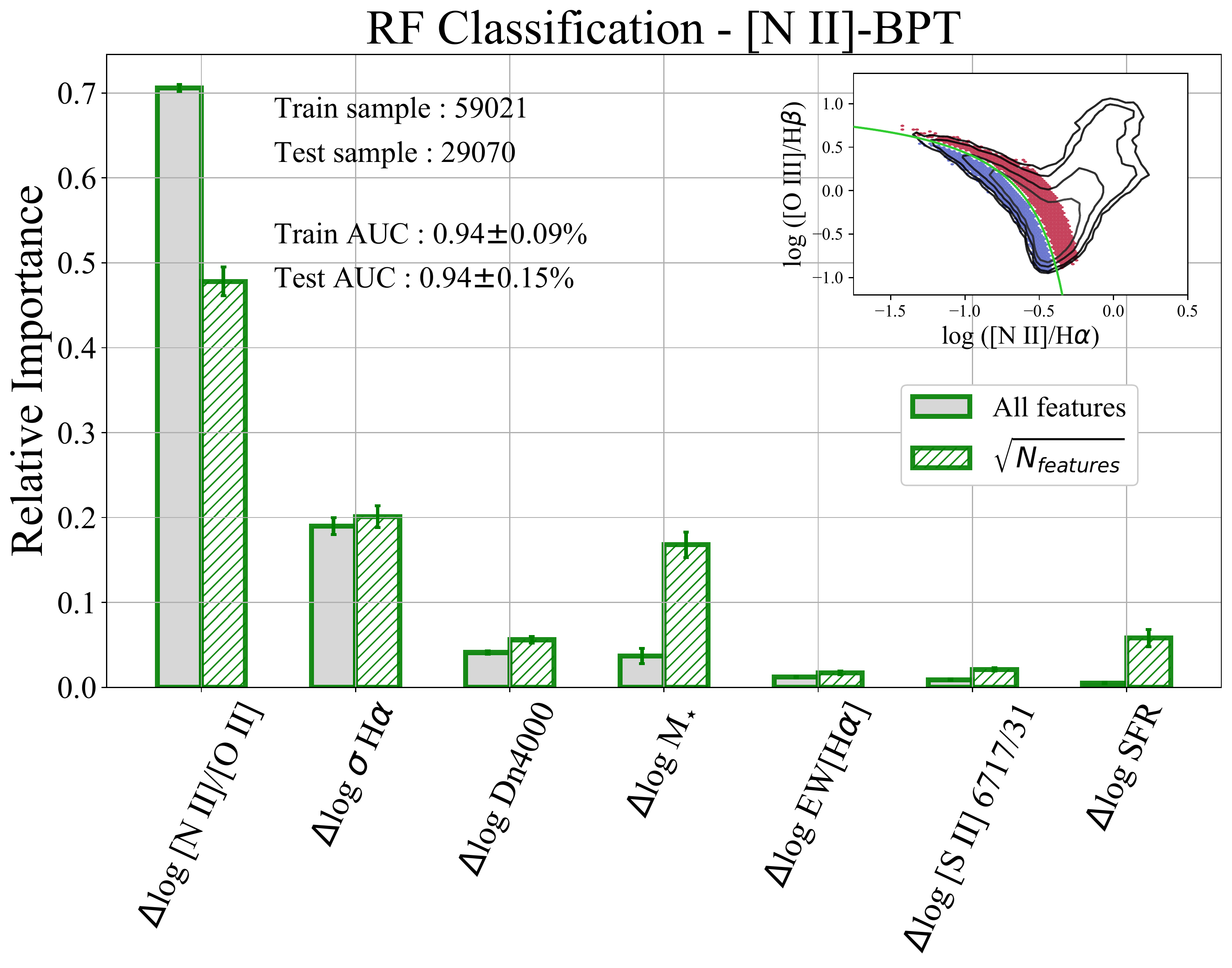}
    \caption{This figure replicates Fig.~\ref{fig:Nii_RF_class}, but with a difference choice of the involved parameters. In particular, [\ion{N}{ii}]/[\ion{O}{ii}] is adopted instead of [\ion{N}{ii}]/[\ion{S}{ii}] as a tracer of the N/O abundance: in the \textit{left panel}, [\ion{Ne}{iii}]/[\ion{O}{ii}] is also included to independently trace the ionisation parameter, causing a strong reduction of the number of selected galaxies, whereas in the \textit{right panel} no primary tracer of U is adopted at all, and the full star-forming sample is hence considered. In both cases, the RF picks \delNOii\ (tracing, again, primarily deviations in N/O) as the most relevant feature in the classification task, confirming the results presented in the main body of the paper. For sake of brevity, we do not show here the regression analysis, which nonetheless leads to equivalent conclusions.}
    \label{fig:appendix_ne3o2}
\end{figure*}

The output from the RF classification analysis for such dataset are shown in the left panel of Fig.~\ref{fig:appendix_ne3o2} (which replicates the structure of Fig.~\ref{fig:Nii_RF_class}): \delNOii\ is ranked as the most important parameter in the set, confirming the results obtained with \delNO, whereas $\upDelta$log([\ion{Ne}{iii}]/[\ion{O}{ii}]) retains a similar level of relative importance (in combination with \delNOii) as that originally scored by \delU\ in combination with \delNO\ as shown in Fig.~\ref{fig:Nii_RF_class}.

In case \textbf{ii)}, we decide instead to remove completely any parameter primarily associated with the ionisation parameter (whose importance is, as we have seen, overall minimal when variations in parameters mainly tracing N/O are already accounted for) and perform the RF analysis on the original full sample of galaxies by including \nii/\oii\ to trace N/O.
We stress here again that any change in the composition of the set of parameters to be included into the RF might affect the overall distribution of relative importance among the different quantities.
Nonetheless, in this way we can not only assess the performance of \nii/\oii\ in the RF exploiting our full statistical power, but also test to what extent removing the other emission line-based parameter from the set would impact its final score (i.e., whether a large part of the relative importance of \delNOii\ is just driven by trivial correlations with other parameters based on emission line ratios).
The results of this second test are shown, for the RF classification task, in the right panel of Fig.~\ref{fig:appendix_ne3o2}.
Again, \delNOii\ is assessed as the most relevant parameter from the RF, whereas the small residual importance associated with the ionisation parameter tracers is now mostly accounted for by $\upDelta$log(\sigha) and $\upDelta$log(M$_{\star}$).
Both these tests confirm that relative variations in the N/O abundance are most predictive for characterising the position of a galaxy with respect to the median SF sequence in the \niibpt, regardless of the choice of the N/O tracer and of the inclusion of different emission lines-based features.
For the sake of brevity, we do not discuss the RF regression analysis here, which we have verified to give fully comparable results for both case \textbf{i)} and \textbf{ii)} discussed above.

We also further comment on the amount of relative importance retained by $\upDelta$log(\sigha) in the RF analysis, when \delNOii\ is adopted instead of \delNO\ to trace primarily variations in N/O. 
In fact, where \nii/\oii\ traces more directly N/O by virtue of the closest ionisation potential of N$^{+}$ and O$^{+}$, \nii/\sii\ presents further, secondary dependence on different parameters on top of that on N/O (we also refer to the discussion in Section~\ref{sec:parameters}).
A partial correlation analysis reveals, in fact, that a strong correlation exists between [\ion{N}{ii}]/[\ion{S}{ii}] and SFR, M$^{\star}$ and \sigha, at fixed [\ion{N}{ii}]/[\ion{O}{ii}], with partial Spearman ranks equal to $0.9583$, $0.9178$ and $0.9418$, respectively. 
It is plausible, then, that decoupling the N/O tracer from such dependencies is reflected into the RF `seeing' the relative importance of these parameters, and in particular $\upDelta$log(\sigha), increasing with respect to the others. We also note in fact that central \sigha, tracing predominantly the dynamical mass of galaxies, is one of the parameters showing the strongest 'absolute' variation across the SF sequence compared to the variation along it (see Fig.~\ref{fig:Nii_properties} and ~\ref{fig:nii_delta_prop}), with a $\nabla_{\parallel}$ = 0.002$\sigma$ and $\upDelta_{\perp}$ = 0.74$\sigma$. Therefore, relative variations in \sigha\ are well connected to the deviations from the SF sequence in the \niibpt, and this is now exposed also by the RF, whereas previously this information was already partially embedded in the adoption of \delNO. 
The performance of $\upDelta$log(\sigha) is likely driven by the highest masses galaxies at the edge of the star-formation \cite{kauffmann_dependence_2003} diving line; indeed, the relative importance carried by $\upDelta$log(\sigha) and $\upDelta$log(\mstar) is almost equivalent in the RF analysis with $\sqrt{N_{\text{features}}}$ selected at each node.

\begin{figure*}
    \centering
    \includegraphics[width=0.48\textwidth]{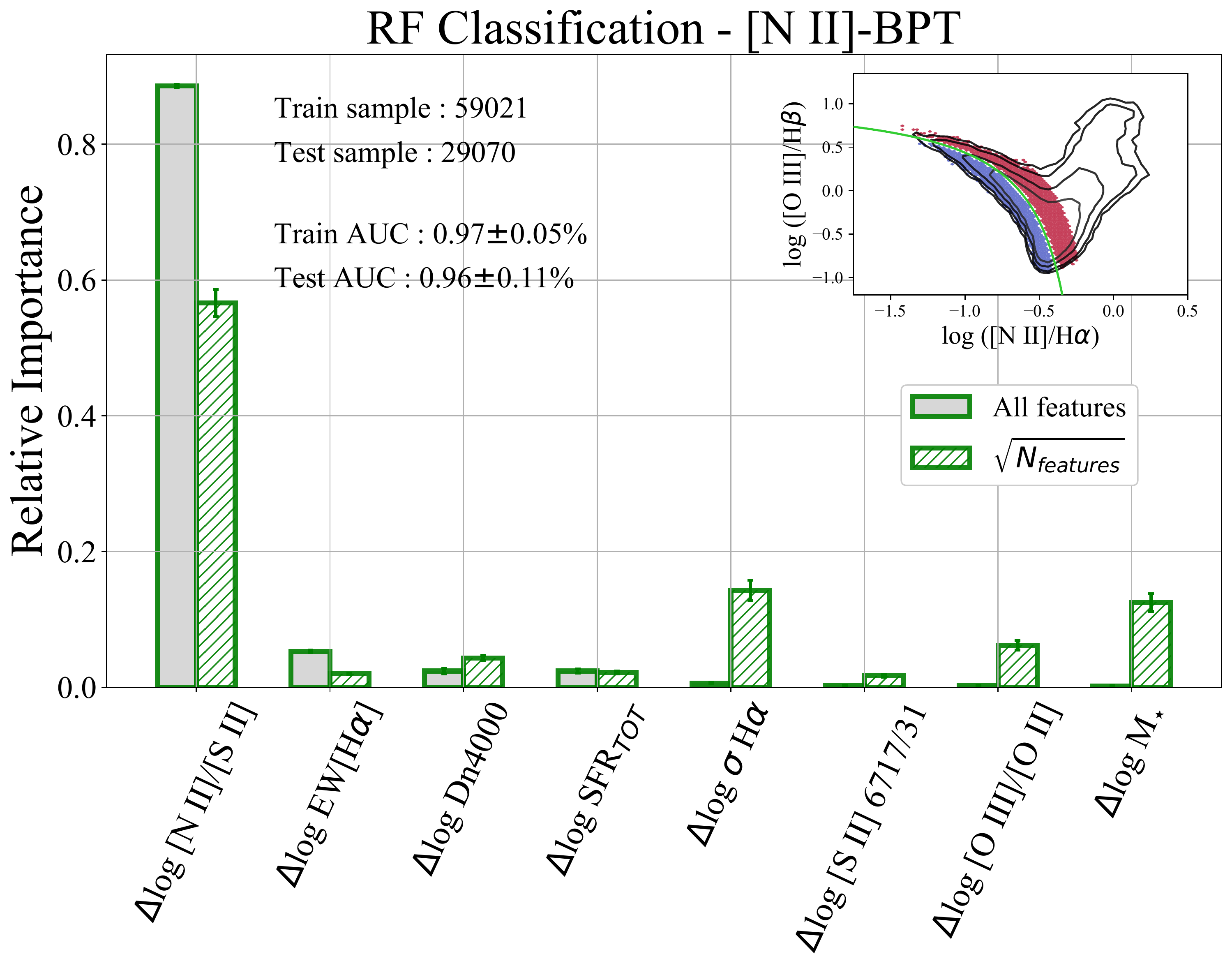}
    \includegraphics[width=0.48\textwidth]{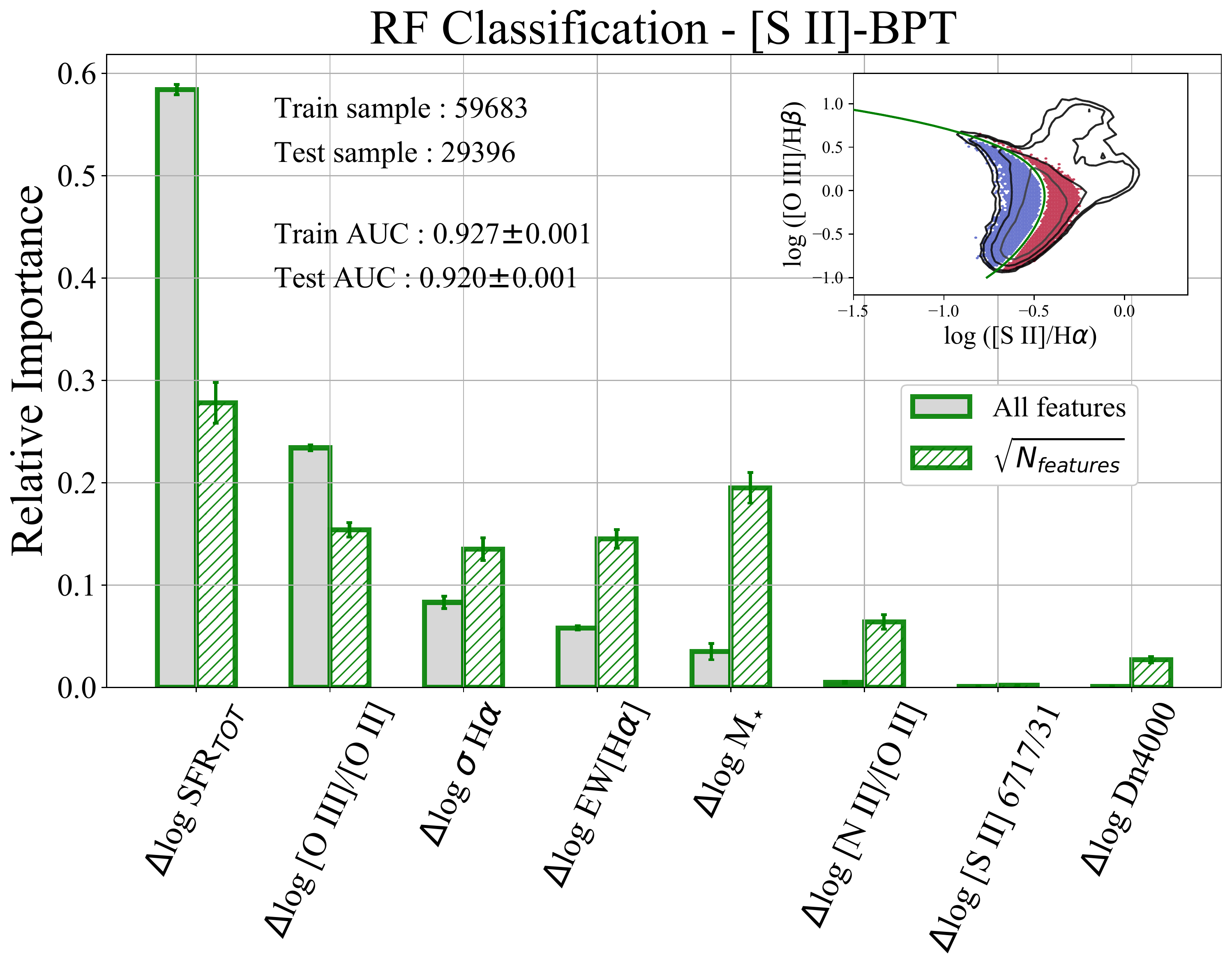}
    
    \caption{Random forest classification analysis for the \niibpt\ (left panel) and \siibpt\ (right panel), respectively, including the total SFR in the parameter set, as computed by applying the aperture corrections of \citealt{brinchmann_physical_2004,salim_uv_2007} to the SFR derived from the H$\upalpha$ flux inside the SDSS fibre.}
    \label{fig:appendix_sfr_tot}
\end{figure*}

Finally, we test the impact of assuming the total SFR (i.e., applying the aperture corrections provided in \citealt{brinchmann_physical_2004, salim_uv_2007} to the SFR derived within the fibre) on the outcomes of the ML analysis.
In particular, in Fig.~\ref{fig:appendix_sfr_tot} we report the results of the classification tasks for the \niibpt\ (right-hand panel) and \siibpt\ (left-hand panel). 
In the \niibpt, \delsfr\ now scores only a slightly higher importance compared to the fiducial analysis presented in Fig.~\ref{fig:Nii_RF_class}, comparable to that of $\upDelta$log(D$_{\text{N}}$4000), but the overall interpretation and the \delNO\ performances are not significantly affected.

The \siibpt\ deserves more attention, as \delsfr\ is originally identified as the most important parameter in regulating the deviations from the median sequence of star-forming galaxies, a result now confirmed also by the analysis involving the total SFR.
On the one hand, this demonstrates that the inferred correlation does not just trivially follow by adopting the H$\upalpha$ flux inside the fibre in both the feature and the target label, as \delsfr\ now also contains information on galaxy-wide scales provided by the photometry-based aperture corrections. 
On the other, it is interesting (and perhaps not surprising) to see that part of the total relative importance of \delsfr\ in the `All features' case is now redistributed to quantities which are both correlated with the SFR, but still computed within the fibre, like $\upDelta$log(\ewha) and $\upDelta$log(\sigha), which now score a relative importance similar to those achieved in the `$\sqrt{N_{\text{features}}}$' case of the fiducial analysis.

\subsection{Assessing trivial correlations between the target labels and emission line-based parameters}

One potentially critical point of the hereby presented analysis concern the level of trivial correlation which exists between our target label \textbf{D} in the \niibpt, which is based on a combination of \oiiihb\ and \niiha, and some of the involved parameters (in particular \delNO\, \delNOii, and \delU), which share one of their emission lines with the target label itself (i.e, \nii\ and \oiii, respectively).
In this section we aim to test if, and to what extent, the final outputs of the ML algorithms (especially the very good performances of \delNO\ and \delNOii) are just trivially recovered from the covariance between the target label and these emission lines-based parameters.

In order to perform this test, we keep the flux of the \nii\ emission line fixed for all galaxies in our sample, while randomly shuffling the \sii\ line fluxes among the full selected star-forming galaxies.
In such way, we create a `hybrid' variable, which we refer to as [\ion{N}{ii}]/$\widetilde{[\ion{S}{ii}]}$, which share the emission line at the numerator with the target label of the ML, but it does not retain any longer a clear, physical interpretation as a N/O tracer. 
Therefore, if the predictivity of the fiducial \delNO\ parameter resided only (or mostly) in sharing the \nii\ line flux with \textbf{D}, the RF should still pick $\upDelta$log([\ion{N}{ii}]/$\widetilde{[\ion{S}{ii}]}$) as the (or one of the) most relevant variable among the full set of parameters; on the contrary, if the relative importance of $\upDelta$log([\ion{N}{ii}]/$\widetilde{[\ion{S}{ii}]}$) is strongly suppressed, that would suggests that the information resides in the full line ratio (hence, primarily in the variations of the actual N/O abundance) rather than just being driven by trivial mathematical covariance.

\begin{figure*}
    \centering
    \includegraphics[width=0.48\textwidth]{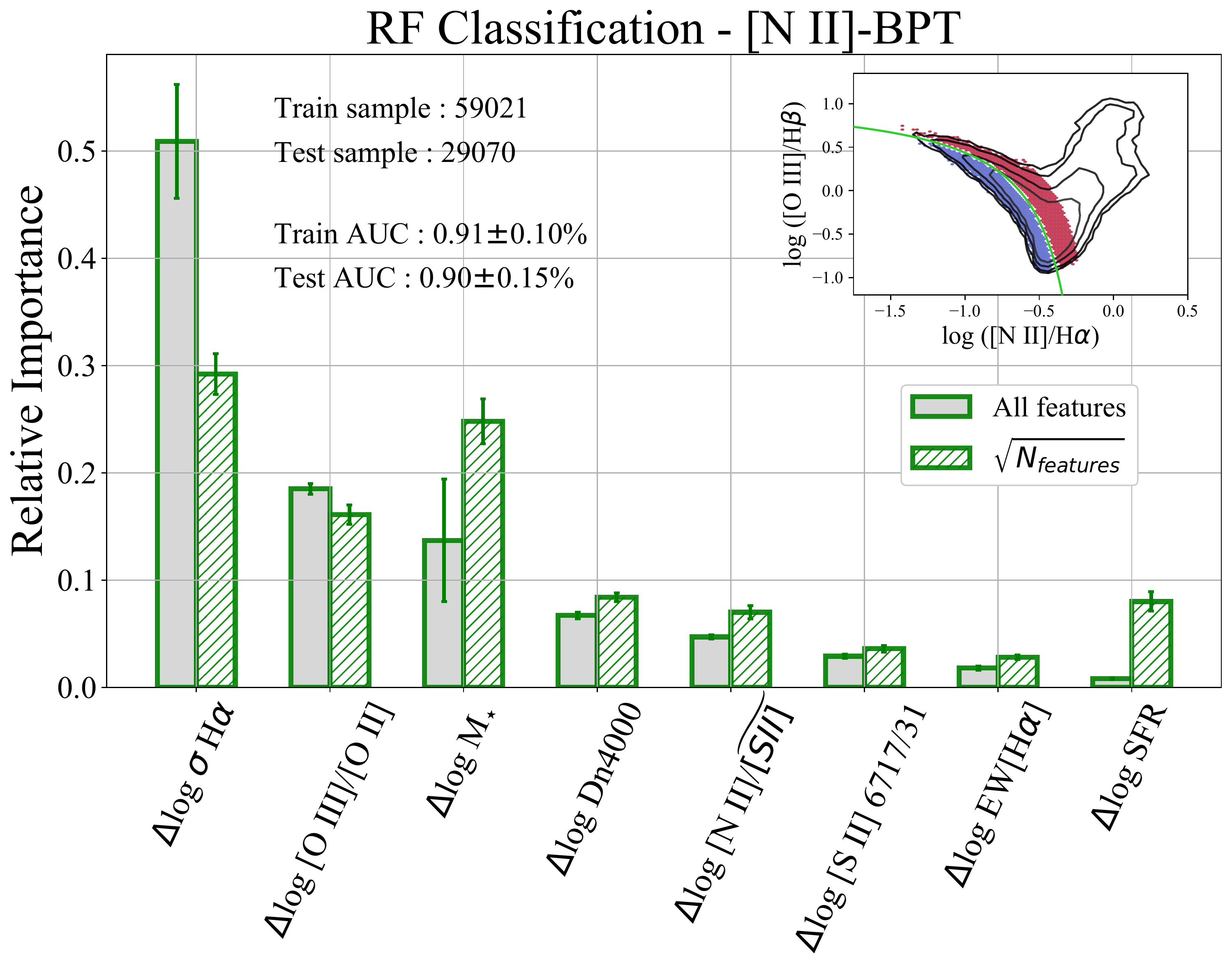}
    \includegraphics[width=0.48\textwidth]{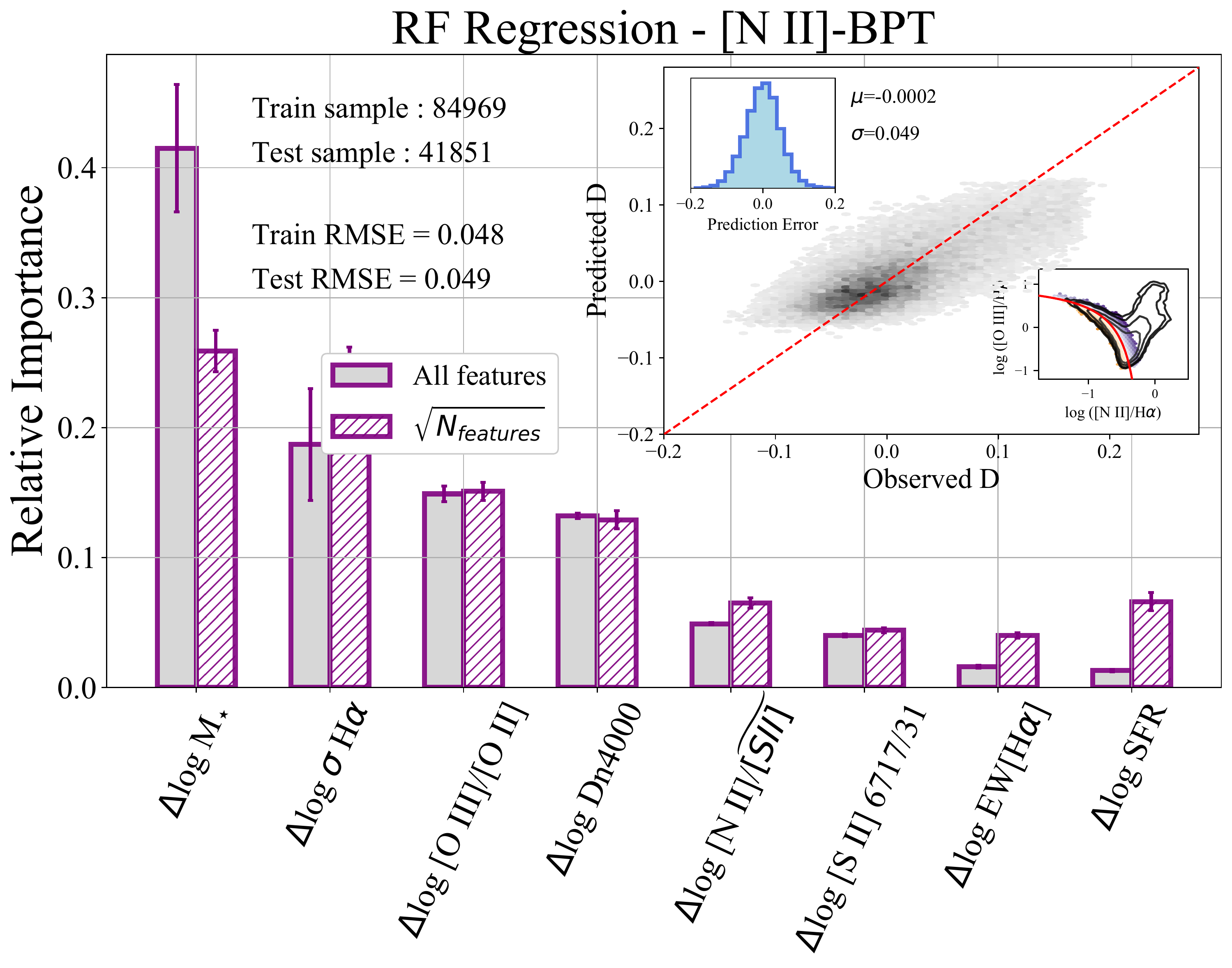}
    
    \caption{RF analysis of the [\ion{N}{ii}]-BPT diagram, where the [\ion{N}{ii}]/[\ion{S}{ii}] ratio has been replaced by a `pseudo-hybrid' parameter (i.e., [\ion{N}{ii}]/$\widetilde{[\ion{S}{ii}]}$) obtained by randomly shuffling the \sii\ fluxes among the full star-forming galaxy sample, while keeping the fluxes of the [\ion{N}{ii}]$\lambda6584$ line fixed. In this way, we can test to what extent the connection between our target labels and the \delNO\ parameter, as recovered by the ML algorithms, is trivially induced by sharing the [\ion{N}{ii}] emission line. 
    The results, which see the relative importance of [\ion{N}{ii}]/$\widetilde{[\ion{S}{ii}]}$ strongly suppressed compared to both our fiducial analysis and the rest of the features in the set, confirms that the information resides in the full [\ion{N}{ii}]/[\ion{S}{ii}] ratio, rather than just being driven by a trivial correlation between the nitrogen line fluxes, if not by $\sim 5$ per-cent.
    An equivalent conclusion can be drawn if we perform the same test on the [\ion{N}{ii}]/[\ion{O}{ii}] ratio, i.e., the other primary N/O tracer adopted in this work.}
    \label{fig:appendix_n2o2rnd}
\end{figure*}

The results of the RF classification and regression analysis are shown in the left and right panel of Fig.~\ref{fig:appendix_n2o2rnd}, respectively.
In both cases, the relative importance of $\upDelta$log([\ion{N}{ii}]/$\widetilde{[\ion{S}{ii}]}$) appears strongly suppressed (scoring less than $10$ per-cent), both compared to its fiducial value reported in Fig.~\ref{fig:Nii_RF_class}, and to that of the other parameters in the set, whose relative weight in the prediction of the target label are now instead increased. Interestingly, we note that the overall performances of the algorithm are clearly hampered, with a reduction in the AUC for classification and in RMSE for regression of $\sim 6$ and $\sim 20$ per cent, respectively, compared to the fiducial analysis presented in Section~\ref{sec:Nii_RF}.
These two observations, if taken together, confirms that, on the one hand, the relative importance of \delNO\ does not follow from trivial correlations with our target label as driven by line fluxes in common (if not for less than $10$ per-cent), whereas on the other, that the overall performance of our parameters set is significantly affected if we remove observables carrying direct information about the N/O abundances in galaxies, because even the combination of all the remaining parameters is not capable of providing the same level of predictive power.
This further corroborates the interpretation of N/O (and its relative variations compared to median behavior of galaxies) as the primary responsible for the observed scatter in the [\ion{N}{ii}]-BPT diagram.

\section{Partial correlation analysis}

\begin{figure*}
    \centering
    \includegraphics[width=0.48\textwidth]{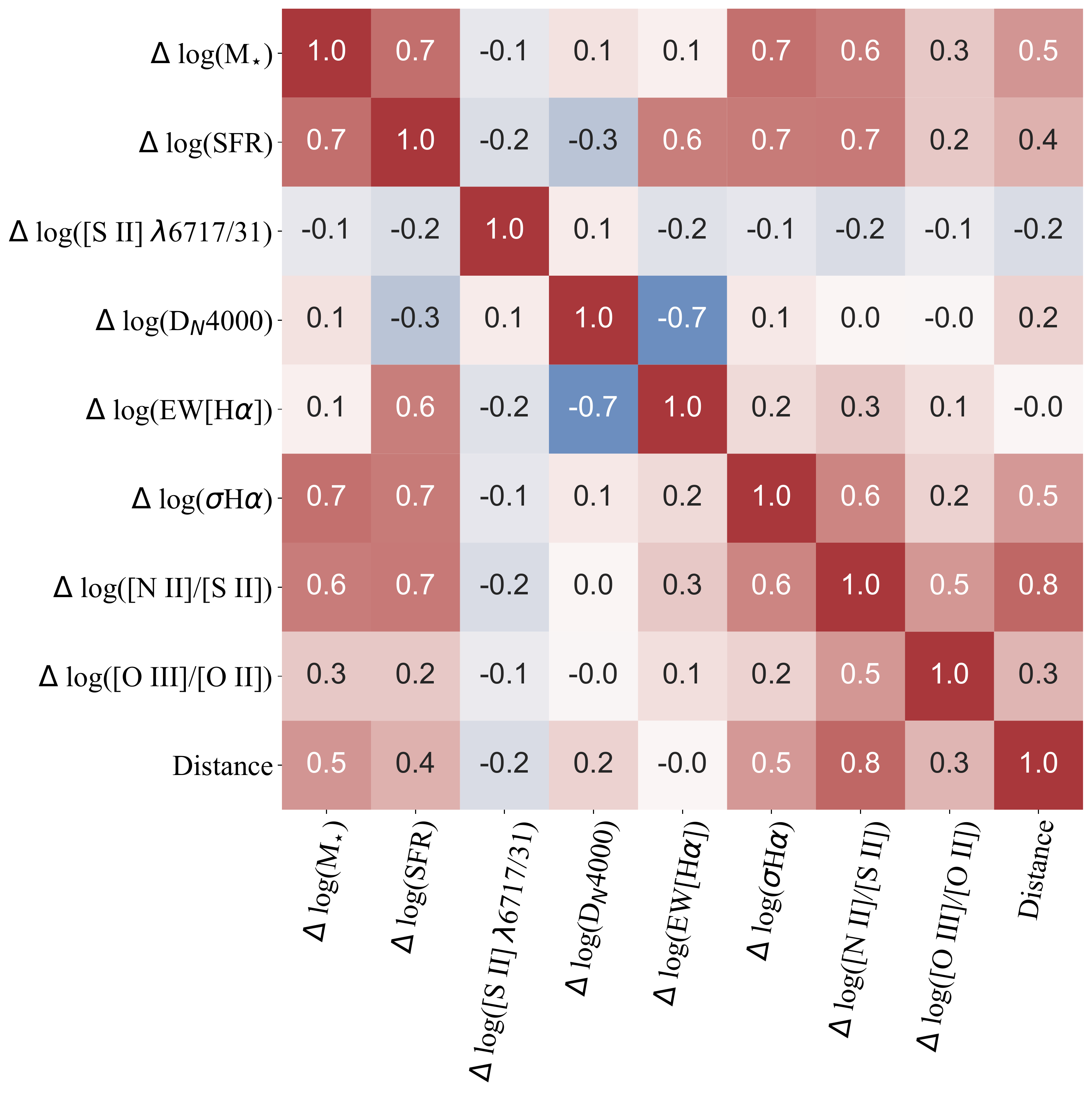}
    \includegraphics[width=0.48\textwidth]{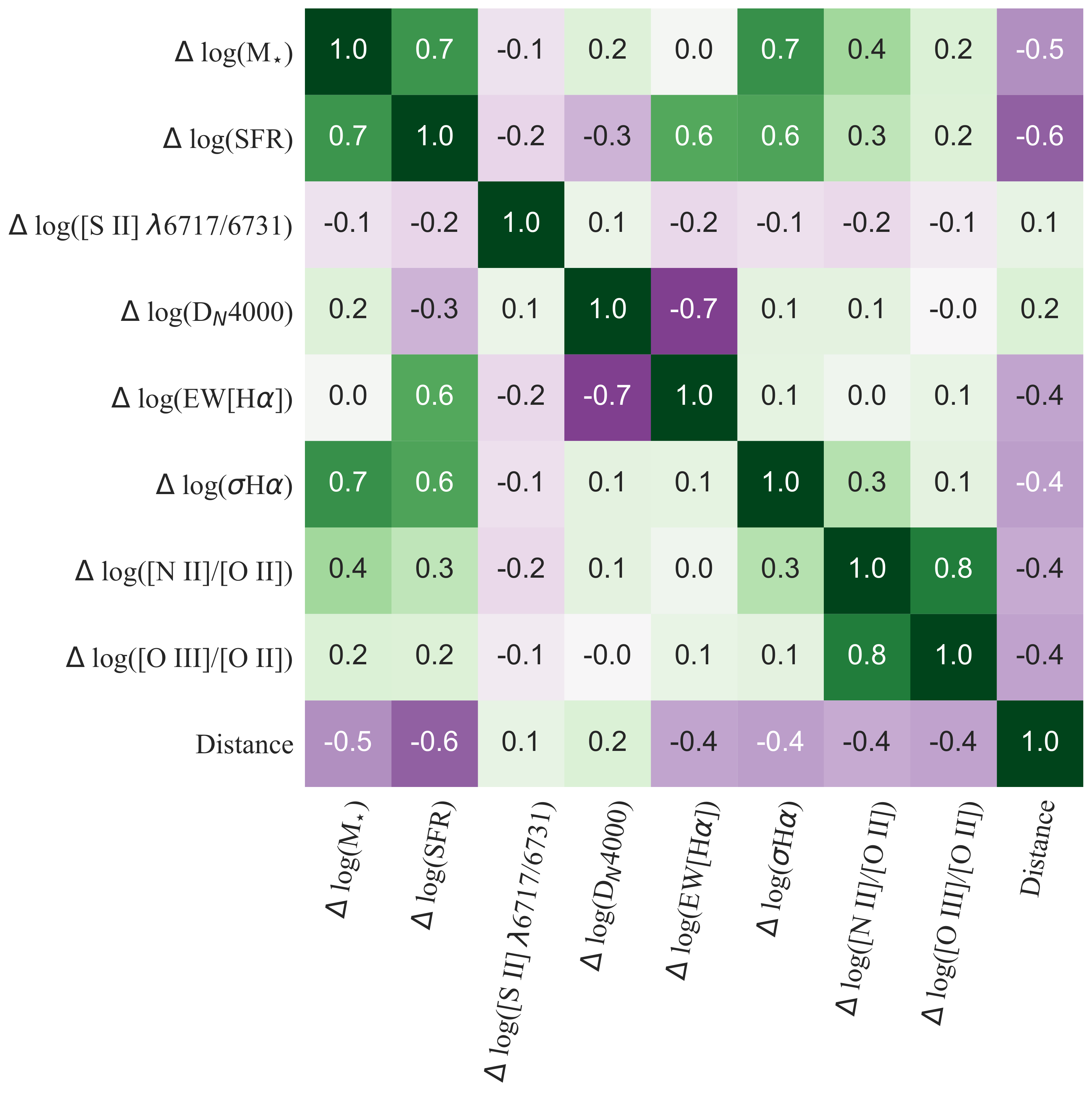}
    \caption{Matrix of Spearman correlation coefficients for the `multi-parameter' set of the [\ion{N}{ii}]-BPT (left panel) and [\ion{S}{ii}]-BPT (right panel), respectively. The target label for the ML regression problem, i.e, the distance \textbf{D} from the SF sequence, is also included.}
    \label{fig:appendix_correlation_matrix}
\end{figure*}

\begin{figure*}
    \centering
    \includegraphics[width=0.48\textwidth]{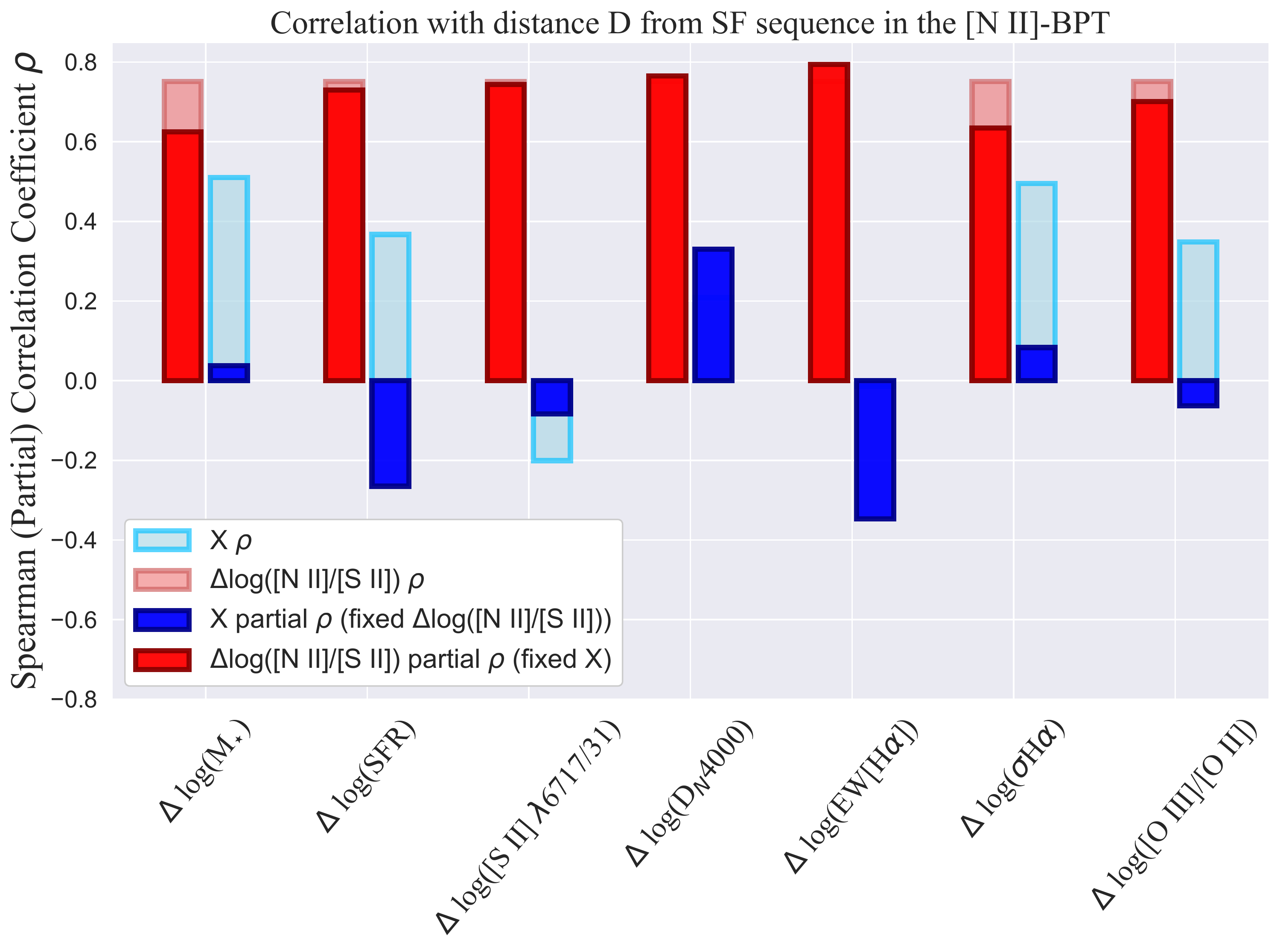}
    \includegraphics[width=0.48\textwidth]{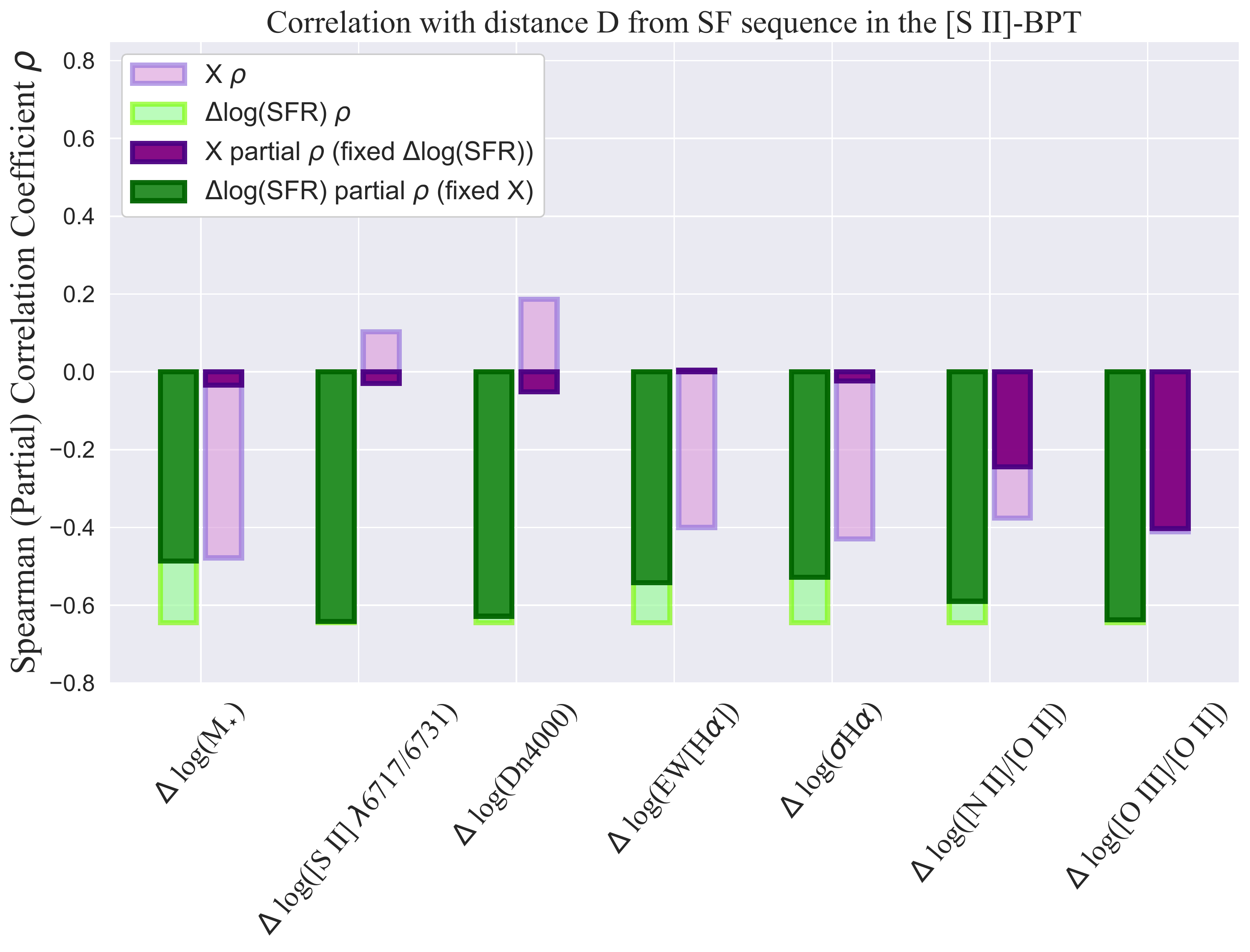}
    
    \caption{\textit{Upper panel}: Spearman (partial) correlation coefficients for our set of parameters with the distance \textbf{D} from the median SF sequence in the [\ion{N}{ii}]-BPT diagram. 
    The x-axis labels each parameter under consideration, and the bars are grouped into pairs for comparison with \delNO, taken as a reference because identified as the most relevant parameter in the ML analysis of Section~\ref{sec:ML}, as well as in the matrix of Spearman rank coefficients of Fig.~\ref{fig:appendix_correlation_matrix}. 
    Light shaded bars indicate the global Spearman rank correlation strength of each parameter with \textbf{D}, whereas solid coloured bars instead indicate the partial correlation strengths. 
    For blue bars, the partial correlation is computed by keeping fixed the value of \delNO, whereas the red bars report the partial correlation strength of \delNO\ at a fixed value of each of the other parameters, in turn. These results are fully consistent with our main ML analysis in showing that \delNO\ presents both the highest correlation and partial correlation coefficients with \textbf{D} than any other parameter. \newline
    \textit{Lower panel}: Same as \textit{upper panel}, for the [\ion{S}{ii}]-BPT. Here, the reference parameter is $\upDelta$log(SFR), which again shows both the highest correlation and partial correlation coefficients with \textbf{D} than any other parameter.}
    \label{fig:appendix_part_correlation}
\end{figure*}

In this appendix we present a rather different but complementary approach to our analysis of the connection between the offset from the SF sequence in the BPT diagrams and the set of physical parameters adopted in the paper, based on the evaluation of (partial) correlation coefficients.

In Fig.~\ref{fig:appendix_correlation_matrix} we present the matrix of Spearman correlation coefficients computed among the features in the `multi-parameter set' (in their `$\upDelta$log' form, and including the distance \textbf{D} from the SF sequence too), for both the \niibpt\ (left panel) and \siibpt\ (right panel). 
Each square in the matrix is colour-coded (on a diverging `blue-to-red' scheme) according to the value of Spearman correlation rank scored by the two parameters representing the `coordinates' of that element in the matrix.
In this way, it is readily immediate to visualise the amount of correlation between all the involved parameters, and between each parameter and our target label: \delNO\ is the quantity scoring the highest correlation rank with \textbf{D} in the \niibpt, whereas \delsfr\ scores the highest in the \siibpt, in agreement with the findings of the ML analysis presented in the main body of the paper.

Starting from these observations, and following e.g. \cite{bluck_how_2020}, in Fig~\ref{fig:appendix_part_correlation} we present a more detailed assessment of the correlation between \delNO\ and \textbf{D} (upper panel), and \delsfr\ and \textbf{D} (lower panel), but with a rather different way to visualise the results.
In the upper panel of Fig~\ref{fig:appendix_part_correlation} for instance, the \textit{total} Spearman correlation strength of \delNO\ with \textbf{D} is presented as light shaded red bars, and is reported adjacent to the Spearman rank correlation strengths of each other parameter in the set, as listed along the x-axis (in light shaded blue bars). The light shaded bars confirms what already shown by the correlation matrix in Fig.~\ref{fig:appendix_correlation_matrix}., i.e., that the correlation strengths of \delNO\ with \textbf{D} are higher than for any other variable.
For few parameters (e.g., $\upDelta$log(\mstar), $\upDelta$log(\sigha)) the correlations ranks are $\sim30-40$ per cent lower than for \delNO, whereas for the remaining parameters are even more suppressed.

The red, left-hand, solid shaded bars in the upper panel of Fig.~\ref{fig:appendix_part_correlation} represent instead the \textit{partial} correlation strengths of \delNO\ with \textbf{D}, at fixed values of each other parameter taken individually; the partial correlation strengths of each other variable with \textbf{D}, at a fixed \delNO\, are instead represented by the solid, blue, right-hand bars. 
Therefore, any subgroup of bars (and their relative x-axis labels) should be intended as representative of a pair of parameters, each constituted by \delNO\ and one of the other variables in the set, alternatively.
For instance, from the comparison of partial correlation coefficients, we observe that the strength of correlation between \delNO\ and \textbf{D} is only mildly reduced, at a fixed $\upDelta$log(\mstar) or $\upDelta$log(\sigha). 
However, at a fixed \delNO, the correlations between $\upDelta$log(\mstar), $\upDelta$log(\sigha) (which were the second and third ranked parameters in both RF and in terms of global correlation coefficients) and \textbf{D} are strongly affected and reduced in magnitude. 
Therefore, fixing the the \delNO\ parameter, which mainly traces variations in the N/O abundance, almost completely removes the correlations of \textbf{D} with both $\upDelta$log(\mstar) and $\upDelta$log(\sigha). 
Moreover, for a few parameters like \delsfr, the direction of the correlation is even inverted.

The same relationships between pairs of (partial) correlation coefficients are shown in the lower panel of Fig~\ref{fig:appendix_part_correlation} for the set of parameters adopted in the analysis of the \siibpt\ diagram; here, \delsfr\ is taken as the reference parameter to which all other variables should be compared to.
Again, \delsfr\ shows both the highest correlation coefficient and partial correlation coefficient with \textbf{D} than any other parameter, whose partial correlation ranks with our target label are, on the contrary, strongly suppressed when evaluated at fixed \delsfr.

In summary, the analysis based on (partial) correlation coefficients establish \delNO\ (tracing primarily deviations in the N/O abundance) as the most intrinsically connected parameter with the distance \textbf{D} from the SF sequence in the \niibpt\ diagram, and \delsfr\ as the most connected parameter with \textbf{D} in the \siibpt\ diagram,
in excellent agreement with the machine learning analysis presented in the main body of the paper.

\end{document}